\documentclass[11pt]{article}

\usepackage[final]{acl} 
\usepackage{booktabs} 
\usepackage{multirow}
\usepackage{siunitx}
\usepackage{xcolor}
\usepackage{comment}
\usepackage{epstopdf}
\usepackage{xspace}
\usepackage{graphicx}
\usepackage{subcaption}
\usepackage{amsmath}
\usepackage[inkscapelatex=false]{svg}

\usepackage[table]{xcolor} 
\usepackage{times}
\usepackage{latexsym}

\usepackage{color, colortbl}
\usepackage{graphicx}
\usepackage{xcolor}
\usepackage{float}
\definecolor{Gray}{gray}{0.9}
\definecolor{LightCyan}{rgb}{0.88,1,1}
\definecolor{Magenta}{rgb}{1,0.88,1}

\newcommand*{\yoruba}{Yor\`ub\'a\xspace}

\usepackage[T1]{fontenc}

\usepackage[utf8]{inputenc}

\usepackage{microtype}

\newcommand{\datasetname}{NaijaS2ST\xspace}

\newcommand{\datasetnamebf}{\textbf{NaijaS2ST}\xspace}

\usepackage{inconsolata}
\usepackage{graphicx}
\usepackage{textgreek}
\usepackage{tipa}
\usepackage{placeins}

\usepackage{tcolorbox}
\tcbuselibrary{listingsutf8}
\tcbuselibrary{skins,breakable}

\newenvironment{promptbox}{%
    \tcblisting{
        enhanced,
        listing only,
        colback=gray!10,
        colframe=black,
        top=2mm,
        bottom=2mm,
        boxrule=0.5pt,
        before skip=10pt,
        after skip=10pt,
        after={\par\vspace{0.5\baselineskip}\noindent},
    }
}
{\endtcblisting}

\title{NaijaS2ST: A Multi-Accent Benchmark for Speech-to-Speech Translation in Low-Resource Nigerian Languages}

\author{First Author \\
  Affiliation / Address line 1 \\
  Affiliation / Address line 2 \\
  Affiliation / Address line 3 \\
  \texttt{email@domain} \\\And
  Second Author \\
  Affiliation / Address line 1 \\
  Affiliation / Address line 2 \\
  Affiliation / Address line 3 \\
  \texttt{email@domain} \\}

\author{
 \textbf{Marie Maltais\textsuperscript{1,2}},
 \textbf{Yejin Jeon\textsuperscript{1,2}},
 \textbf{Min Ma\textsuperscript{3}},
 \textbf{Shamsuddeen Hassan Muhammad\textsuperscript{4,5}},
\\
 \textbf{Idris Abdulmumin\textsuperscript{4,6}},
 \textbf{Maryam Ibrahim Mukhtar\textsuperscript{4}},
 \textbf{Daud Abolade\textsuperscript{7}},
 \textbf{Joel Okepefi\textsuperscript{8}},
\\
 \textbf{Johnson Sewedo\textsuperscript{8}},
 \textbf{David Ifeoluwa Adelani\textsuperscript{1,2,9}}
\\
\\
 \textsuperscript{1}Mila - Quebec AI Institute,
 \textsuperscript{2}McGill University, Canada
 \textsuperscript{3}Google DeepMind,
 \textsuperscript{4}Hausa NLP, \\
 \textsuperscript{5}Imperial College, United Kingdom, 
 \textsuperscript{6}University of Pretoria, South Africa, 
 \textsuperscript{7}Masakhane NLP, \\
 \textsuperscript{8}Naij\'a Wikipedia Community, and 
 \textsuperscript{9}Canada CIFAR AI Chair
\\
 \texttt{\{marie.maltais, yejin.jeon, david.adelani\}@mila.quebec}
 }

\begin{document}
\maketitle
\begin{abstract}
Speech translation for low-resource languages is constrained by the scarcity of high-quality parallel speech data, which is particularly pronounced in African linguistic contexts. To address this, we introduce \datasetnamebf, a parallel speech translation dataset covering Igbo, Hausa, \yoruba, and Nigerian Pidgin paired with English. The dataset contains more than 40 hours of speech per language and includes several speaker and accent variations to reflect realistic settings.
Using NaijaS2ST, we benchmark cascaded, end-to-end (E2E), and AudioLLM-based approaches across bidirectional translation directions (XX$\leftrightarrow$English). Our results show that few-shot audio LLMs are more effective than cascaded and E2E methods finetuned for speech-to-text translation. For speech-to-speech translation, however, cascaded and audio LLM paradigms perform comparatively, highlighting the need for stronger task-specific models. \datasetname offers a strong foundation towards advancing research on low-resource multilingual speech translation. We open source the dataset here: \href{https://huggingface.co/datasets/McGill-NLP/NaijaS2ST}{NaijaS2ST Huggingface Dataset}; and the code here: \href{https://github.com/McGill-NLP/naijas2st}{NaijaS2ST Code}.
\end{abstract}

\section{Introduction}
Translation technologies play a central role in enabling equitable access to information and facilitating communication regardless of language boundaries \cite{MT-society}. In recent years, advances in neural architectures have led to substantial improvements in both speech-to-text translation (S2TT) and speech-to-speech translation (S2ST). More importantly, these improvements have not been driven by architectural enhancements alone, but have historically depended on the availability of large-scale parallel corpora. 

These advancements, however, have been disproportionately concentrated on a small set of high-resource languages. As a result, what started as a means to reduce language barriers and enable broader access to information and communication has not fully realized its goals, as the benefits of modern translation systems remain inaccessible to a large portion of the world’s population. This disparity is especially evident for African languages, which account for a significant portion of global linguistic diversity.  In fact, among the roughly 2,000 languages in the widely used Flores-101 \cite{flores} dataset, only 0.85\% are African. The more recent FLORES-200 added Hausa, Igbo and \yoruba, among other African languages, though not Nigerian Pidgin \citep{Costajuss2024ScalingNM}. 

\begin{figure}
    \centering
    \includegraphics[width=1\linewidth]{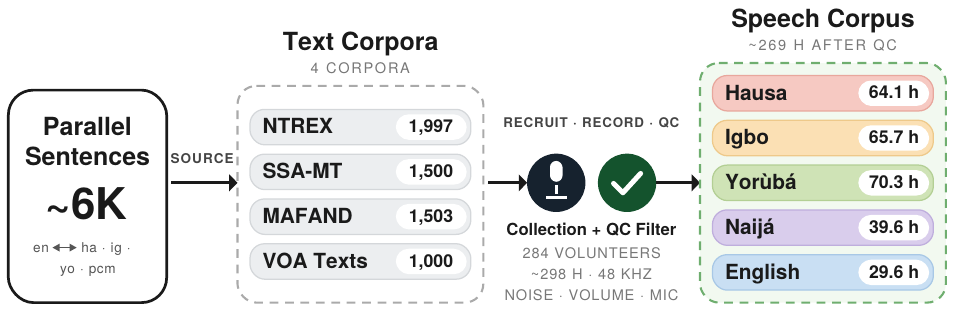}
\caption{\textbf{Construction of the NaijaS2ST corpus.} 6K
  English-aligned sentences are drawn from four parallel text corpora
  and read by 284 participants, resulting in \textasciitilde269 hours
  of speech after quality control (QC).}
\label{fig:naijas2tt_dataset}
\end{figure}

Recent efforts have led to the development of large-scale speech datasets for Nigerian languages~\citep{meyer22c_interspeech,ogunremi-etal-2024-iroyinspeech,emezue25_interspeech,adebara2026african}. However, these resources primarily target Automatic Speech Recognition (ASR) and Text-to-Speech (TTS) tasks. While there are some prior works for text-based translation~\cite{adelani-etal-2022-thousand,Costajuss2024ScalingNM}, it does not address the speech modality, and general-domain datasets capturing African–English accents remain particularly scarce~\citep{olatunji-etal-2023-afrispeech}. Moreover, speech-based translation, especially in bidirectional settings, remains significantly underexplored, as many existing evaluations are limited to the ``XX–English'' direction in a \textit{single} accent \citep{wang2021covost, conneau22_interspeech}.

This lack of data coverage reflects a deeper issue: under-representation of African languages constitutes a fundamental bottleneck for progress in speech translation research. Modern S2TT and S2ST systems, whether based on cascaded pipelines or end-to-end (E2E) architectures, depend on large-scale parallel speech data. 
In the absence of such datasets, it becomes difficult to (i) perform standardized and reproducible benchmarking, (ii) develop robust and generalizable models, and (iii) conduct meaningful comparisons across different modeling approaches. 

We address these challenges by introducing the \datasetnamebf speech dataset of actual spoken recordings across the most populous Nigerian languages of Hausa, Igbo, Nigerian Pidgin (Naij\'a\footnote{Naij\'a is a popular nickname for Nigeria, and also an alternative name for Nigerian-Pidgin (Naij\'a).}), and \yoruba. This dataset is designed to support both S2TT and S2ST, and includes parallel speech aligned with English. The speech data were collected from 284 participants, who were asked to read 6,000 sentences sourced from various news outlets. 
\autoref{fig:naijas2tt_dataset} illustrates the overall data collection process.

In addition to dataset construction, a systematic benchmarking study of various speech translation models is conducted. Specifically, we evaluate both cascaded approaches and E2E models across different translation directions (English$\leftrightarrow$African languages). This comprehensive evaluation framework enables a direct comparison of modeling paradigms under low-resource conditions and provides insights into their relative strengths and limitations. We hope that this work will serve as a foundation for future research in speech translation for African languages and contribute to more inclusive multilingual technologies.
\begin{table*}[t]
  \centering
  \resizebox{\textwidth}{!}{%
    \begin{tabular}{llr|llrr}
    \toprule[1pt]
     &  & \textbf{No. of } & &  & \textbf{No.} & \textbf{No.} \\ 
    \textbf{Language} & \textbf{Language Family} & \textbf{Speakers} & \textbf{Existing Data (Exist.)} & \textbf{Newly Added (Added)} & \textbf{Exist.} & \textbf{Added} \\ \midrule
    Hausa (\texttt{ha}) & Afro-Asiatic/Chadic & 94M & NTREX, SSA-MT \& MAFAND & VOA Texts & 5,000 & 1,000 \\
    Igbo (\texttt{ig})& Niger-Congo/Volta-Niger & 34M  & NTREX \& SSA-MT & VOA Texts \& MAFAND & 3,497 &  2,503\\
    Naij\'a (\texttt{pcm}) & English-based Creole & 121M  & -- & VOA Texts, MAFAND \& SSA-MT &  -- &  4,000 \\
    \yoruba (\texttt{yo}) & Niger-Congo/Volta-Niger & 54M & NTREX, SSA-MT \& MAFAND & VOA Texts & 5,000 & 1,000  \\
    \bottomrule[1pt]
    \end{tabular}%
  }
  \vspace{-2mm}
  \caption{\textbf{NaijaS2ST languages and data source}, including language families, and number of speakers. Existing dataset and their number of instances include: NTREX (1,997)~\cite{federmann-etal-2022-ntrex}, SSA-MT (1,500)~\cite{li-etal-2025-ssa}, and MAFAND (1503 out of 1925 parallel ha-yo sentences)~\citep{adelani-etal-2022-thousand}.} 
  \label{tab:lang-overview}
\end{table*}

\section{Related Work}
\label{sec:lit}
\vspace{-5pt}
The evolution of speech translation research has transitioned from cascaded architectures toward unified, E2E multimodal paradigms. Yet, such architectural progress has simultaneously exposed structural bottlenecks regarding data equity, accent robustness, and cross-lingual generalizability in low-resource settings \cite{sarim2025direct}.

In particular, the rapid advancement of speech translation architectures is fundamentally rooted in the availability of large-scale datasets. As such, the foundation of modern systems relies heavily on self-supervised learning (SSL) frameworks, such as wav2vec 2.0 \cite{baevski2020wav2vec} and HuBERT \cite{hsu2021hubert}, which utilize massive unannotated corpora like VoxPopuli \cite{wang2021voxpopuli} and Common Voice \cite{ardila2020common} for cross-lingual acoustic representations. For parallel translation tasks, monumental benchmarks such as CoVoST 2 \cite{wang2021covost} and FLEURS \cite{fleurs_paper} established standard evaluation protocols, while the CVSS corpus \cite{jia-etal-2022-cvss} and the mined SpeechMatrix \cite{duquenne-etal-2023-speechmatrix} expanded bidirectional S2ST capabilities. 
However, current corpora are predominantly focused on high-resource Indo-European languages. Recent speech datasets for Nigerian languages \cite{meyer22c_interspeech, ogunremi-etal-2024-iroyinspeech, emezue25_interspeech}, are also primarily restricted to ASR or TTS. 

Moreover, as the S2TT research paradigm has increasingly integrated Large Language Models (LLMs), this has further amplified resource requirements \cite{rubenstein1904audiopalm, dong2024polyvoice, zhang-etal-2024-streamspeech, seamless2025joint, alastruey2026omnilingual}. 
Concurrently, direct speech-to-speech translation has shifted toward textless modeling. While early systems mapped continuous spectrograms \cite{jia2019direct}, later work introduced discrete acoustic units \cite{lee2022direct, lee2022textless} that enabled models to bypass intermediate text generation and support applications in unwritten languages \cite{zhang2021uwspeech, chen2023speech}.
Despite this, because these approaches still require substantial parallel speech data, strategies such as cross-lingual pseudo-labeling \cite{dong22b_interspeech, popuri22_interspeech} and monolingual data utilization \cite{Translatotron} have been utilized to reduce this dependency. Regardless, their LLM reliance results in hallucinations when processing noisy and accented inputs \cite{sarim2025direct}.

Despite significant architectural advances, progress in speech translation remains limited by the absence of diverse, multi-accent, and bidirectional parallel speech corpora for low-resource languages. Moreover, evaluating whether discrete-unit or LLM-based E2E models truly outperform cascaded pipelines requires rigorous benchmarking on authentic low-resource data, rather than high-resource standardized proxies. Such work has been presented in high-resource language scenarios, as in \citet{bentivogli-etal-2021-cascade}, but still lack evaluation in low-resource African languages. To address this, we introduce \datasetnamebf, a multi-accent parallel speech translation benchmark for Nigerian languages, providing the data foundation necessary for the evaluation of speech translation paradigms under realistic low-resource conditions.

\section{\datasetname Dataset}
\vspace{-5pt}

\datasetnamebf is a parallel speech–text benchmark for training and evaluating \textit{speech-to-text} translation and \textit{speech-to-speech} translation models across the five most populous Nigerian languages and their accents. The languages include English (British \& Nigerian accent), Hausa, Igbo, Nigerian Pidgin (Naij\'a), and \yoruba,  each with at least 30 million native speakers. Unlike the other Nigerian languages, Naij\'a is an English-based Creole language that combines \textit{simplification} for broader accessibility (e.g. less educated people) with lexical influence from English and some local languages. This simplified and mixed linguistic structure poses unique challenges for automatic text and speech translation systems. Further information on language characteristics are described in  (\S\ref{sec:language_char})

Overall, \datasetnamebf enables multi-way S2ST and S2TT for a population of over 300 million native speakers. Details on the languages, their speaker population, and the original data source for data curation are shown in ~\autoref{tab:lang-overview}.

\subsection{Language Characteristics}
\label{sec:language_char}

All languages in NaijaS2ST make use of the Latin script and strictly follow the same Subject-Verb-Object word order as in English. The language specific characteristics are provided below. 
\vspace{-5pt}
\paragraph{Hausa} (\texttt{ha}) is spoken by more than 94M people across West and Central Africa and is native to Nigeria and the Niger Republic. It employs a Latin-based orthography consisting of 44 letters, including special characters such as \texthtb, \texthtd, and \texthtk. Hausa is a tonal language with high and low tones, and exhibits agglutinative morphological structure. It belongs to the Afro-Asiatic language family, which includes Arabic and Hebrew. Although multiple dialects exist, this work focuses on the dominant Kano dialect spoken in Nigeria, with all recordings collected in the city of Kano. 

\vspace{-5pt}
\paragraph{Igbo} (\texttt{ig}) is spoken by more than 34M people in the South-Eastern part of Nigeria. Igbo makes use of 34 Latin-based letters, including diacritics (grave (``\textbackslash'', acute (``/'') accents, and underdots on ``\d{i}'', ``\d{o}'', and ``\d{u}''). However, in modern times, upper diacritics are often ignored. Igbo is both tonal and agglutinative, and from the Naij\'a-Congo/Volta-Niger language family. 

\vspace{-5pt}
\paragraph{Nigerian-Pidgin (Naij\'a)} (\texttt{pcm}) is an English-based creole language native to Nigeria and part of the broader West African Pidgin English (WAPE) continuum. Naij\'a is spoken by over 121M people and is among the top 20 most spoken languages in the world. Despite this, it remains poorly represented in existing corpora. Owing to its similarity to WAPE, it is often broadly labeled as WAPE in prior datasets, leading to inconsistencies in representation. In this paper, the orthographic conventions used in Wikipedia are adopted, following the recommendations in \citet{adelani-etal-2025-generative}. 

\vspace{-5pt}
\paragraph{\yoruba} (\texttt{yo}) is native to South-Western Nigeria, Benin and Togo, and spoken by more than 54M people. \yoruba has 25 Latin letters, excluding the Latin characters (c, q, v, x and z), and instead includes additional characters ({\d e}, gb, {\d s} , {\d o}). \yoruba is a highly isolating language and belongs to the Naij\'a-Congo/Volta-Niger like Igbo. \yoruba is a tonal language with three tones: low (``\textbackslash
''), middle (``$-$'', optional) and high (``/'').  The tonal marks and underdots are very important in pronunciation of words and generating correct sounds; their absence often lead to worse results on downstream speech tasks such as TTS~\citep{ogunremi-etal-2024-iroyinspeech}.

\subsection{Text Data Collection}
\label{sec:text_collect}

Statistics of the text data are shown in \autoref{tab:lang-overview}. To obtain parallel sentences across language pairs, we use three existing resources: NTREX~\cite{federmann-etal-2022-ntrex} for Hausa–Igbo, SSA-MT~\citep{li-etal-2025-ssa} for Igbo–Yorùbá, and MAFAND~\citep{adelani-etal-2022-thousand} for Hausa–Yorùbá.

To mitigate data contamination, an additional set of 1,000 sentences is collected for English from the VOA website. These sentences are not parallel to any of the training data and are equally balanced across Nigerian and United Kingdom contexts. This is to enable assessment of S2ST models in handling named entity pronunciation across diverse geographical settings. Moreover, 1,000 evaluation sentences are collected each for Hausa and \yoruba. For Igbo and Naij\'a, we supplemented the evaluation data by translating the missing portions (i.e., MAFAND for Igbo and the full set for Naij\'a). In total, 5,000 sentences are collected for training and 1,000 are reserved for evaluation.

\begin{table}[t]
  \centering
  \resizebox{\columnwidth}{!}{%
    \begin{tabular}{lrr|rr|cc}
    \toprule[1pt]
     & \multicolumn{2}{c|}{\textbf{Train split (\# hours)}} &  \multicolumn{2}{c|}{\textbf{Dev-Test split}}  & \textbf{Gender ratio} \\ 
    \textbf{Language} & \textbf{\# Recorded} & \textbf{After QC} &\textbf{\# Dev} & \textbf{\# Test} & \textbf{Female:Male} \\ \midrule
    English (\texttt{en}) & 30.09 & 24.96 & 2.33 & 2.32 &  F:19, M:21\\
    Hausa (\texttt{ha}) & 58.34 & 55.06 & 3.89 & 5.11 & F:47, M:25 \\
    Igbo (\texttt{ig})& 65.57 & 56.89  & 4.43 & 4.38 & F:47, M:25\\
        Naij\'a (\texttt{pcm}) & 41.04 &  31.30 & 3.67 & 4.61 &  F:13, M:20  \\
    \yoruba (\texttt{yo}) & 63.39 & 61.38 & 4.38 & 4.49 &  F:46, M:26  \\
    \bottomrule[1pt]
    \end{tabular}%
  }
  \vspace{-2mm}
  \caption{\textbf{NaijaS2ST Speech information} before and after Quality Control (QC).}
  \label{tab:speech-stat}
\end{table}

\vspace{-5pt}
\subsection{Speech Data Collection}
\paragraph{Speaker Recruitment} We recruited language coordinators (LCs) who work in academic environments for each of the Nigerian languages. Each LC further recruits speakers within the same city, mostly from their home university, community, or closed family and friends. For Hausa, Igbo and \yoruba, we recruited 72 speakers mostly from Kano, Anambra/Imo and Lagos/Ogun states respectively to record the text data collected via the procedure detailed in \S\ref{sec:text_collect}. Each volunteer recorded 250 utterances, covering 6,000 sentences in total, with every sentence recorded three times.\footnote{Each speaker was paid \$15 per 250 utterances recorded.} Finally, for English (Naij\'a accent), speakers from two different regions in Nigeria were recruited across all data sources except NTREX, with half from the North (Kano) and half from the South (primarily Lagos and environs), involving 32 speakers in total. For NTREX, spoken English data was also collected in the same two accents with eight more speakers.\footnote{Due to budget constraints, we did not prioritize obtaining both accents for all NTREX sentences, as much of the content is more Western-oriented.} The train, development and test split speakers are completely distinct.

\begin{table*}[t]
\centering
\small
\setlength{\tabcolsep}{4.5pt}
\resizebox{\textwidth}{!}{%
\begin{tabular}{
    lcl|rrrcr|rrrcr
    }
\toprule
    & 
    & 
    & \multicolumn{5}{c|}{\textbf{XX $\rightarrow$ Eng}} 
    & \multicolumn{5}{c}{\textbf{Eng $\rightarrow$ XX}} \\
    
\cmidrule(lr){4-8} \cmidrule(lr){9-13}
        
\textbf{ Method} & \multicolumn{2}{c}{\textbf{Model}}
& \textbf{Hausa} & \textbf{Igbo} & \textbf{Naijá} & \textbf{\yoruba} & \textbf{Avg.} 
& \textbf{Hausa} & \textbf{Igbo} & \textbf{Naijá} & \textbf{\yoruba} & \textbf{Avg.} \\

\midrule
    \multirow{6}{*}{\begin{tabular}[c]{@{}c@{}} \textsc{Cascaded} \\ \texttt{(ASR + MT)}\end{tabular}}
    & \multirow{6}{*}{Omnilingual-ASR}
        & + NLLB 
        & 54.1 & 42.9 & N/A & 50.6 & -- & \underline{47.6} & 52.6 & N/A & 58.0 & -- \\ 
    & 
        & + Tiny Aya 
        & 52.5 & 40.7 &\textbf{ 54.2} & 35.5 & 45.7 & 39.0 & 40.0 & 15.7 & 26.3 & 30.3 \\
        \addlinespace
       &  & + Gemma 4 Zero-Shot
        & 44.6 & 30.2 & \underline{53.9} & 37.2 & 41.5 & 40.9 & 23.6 & 49.5 & 28.3 & 35.6 \\
    &  & + Gemma 4 Few-Shot
        & \textbf{63.3} & \underline{50.6} & 34.1 & 56.7 & \underline{51.2} & 42.1 & 26.8 & \underline{52.8} &  28.6 & 37.6 \\
        \midrule
        \midrule
    
\addlinespace
    \multirow{2}{*}{\textsc{End-to-End}} & \multirow{2}{*}{SeamlessM4T}
        & Zero-Shot 
        & 14.6 & 20.6 & 44.6 & \underline{57.0} & 34.2 & N/A & 53.1 &  N/A & 55.4 & -\\
    & 
        & Mono FT
        & \underline{54.9} & \textbf{52.4} & 50.7 & \textbf{60.3} & \textbf{54.6} & 12.1 & \underline{54.8} & \textbf{61.4} & \textbf{68.6} & \underline{49.2} \\ 
   & 
       & Multi FT
       & 46.7 & 47.4 & 56.0 & 54.3 & 51.1 & \textbf{53.4} & \textbf{64.6} & 42.6 & \underline{68.5} & \textbf{57.3}\\

\bottomrule
\end{tabular}
}
\caption{\textbf{Speech-to-text translation results (SSA-COMET $\uparrow$)}.
\textit{Italics} indicate best within each method; \textbf{bold} indicates best overall while \underline{underlined} indicate second best result. Multilingual fine-tuning (FT) is a model fine-tuned across all the Nigerian languages data. Naij\'a is not supported by NLLB, we label it \texttt{N/A} }
\label{stt_results_ssa_comet}
\end{table*}

\begin{table*}[t]
\centering
\small
\setlength{\tabcolsep}{4.5pt}

\resizebox{\textwidth}{!}{%
\begin{tabular}{
    lcl|rrrrr|rrrrr
    }
\toprule
    & 
    & 
    & \multicolumn{5}{c|}{\textbf{XX $\rightarrow$ Eng}} 
    & \multicolumn{5}{c}{\textbf{Eng $\rightarrow$ XX}} \\
    
\cmidrule(lr){4-8} \cmidrule(lr){9-13}
        
\textbf{ Method} & \multicolumn{2}{c}{\textbf{Model}}
& \textbf{Hausa} & \textbf{Igbo} & \textbf{Naijá} & \textbf{\yoruba} & \textbf{Avg.} 
& \textbf{Hausa} & \textbf{Igbo} & \textbf{Naijá} & \textbf{\yoruba} & \textbf{Avg.} \\

\midrule
    \multirow{2}{*}{\begin{tabular}[c]{@{}c@{}} \textsc{Cascaded} \\ \texttt{(ASR + MT)}\end{tabular}}
    & \multirow{4}{*}{Omnilingual-ASR}
        & + NLLB 
        & \textbf{42.0} & \underline{33.8} & N/A & 38.1 & -- & \textbf{50.0} & 45.2 & N/A & 29.1 & -- \\ 
    & 
        & + Gemma 4 Zero-Shot
        & 34.7 & 23.1 & 35.4 & 26.9 & 30.0 & 36.3 & 24.9 & \underline{20.8} &  11.8  &  23.5\\
    & 
        & + Gemma 4 Few-Shot
        & 36.9 & 24.7 & 36.6 & 26.1 & 31.1 & \underline{37.2} & 26.8 & \textbf{22.8} & 11.3 & 24.5 \\
\midrule
\addlinespace
    \multirow{2}{*}{\textsc{End-to-End}} & \multirow{2}{*}{SeamlessM4T}
        & Zero-Shot 
        & 15.9 & 19.4 & \underline{31.1} & \underline{43.3} & 27.4 & {N/A} & 24.5 & N/A & 14.2 & - \\
    & 
        & Mono-FT
        & \underline{41.9} & \textbf{39.2} & 18.1 & \textbf{43.5} & \underline{35.7} & 13.7 & \underline{54.7} & 6.1 & \underline{37.0} & 27.9 \\ 
   & 
       & Multi-FT
       & 38.0 & 36.0 & \textbf{40.1} & 41.8  & \textbf{39.0} & 14.5 & \textbf{55.3} & 15.3 & \textbf{37.1} & \underline{30.6} \\

\bottomrule
\end{tabular}
}
\caption{\textbf{Speech-to-text translation results (ChrF++ $\uparrow$)}.
\textbf{Bold} indicates best overall while \underline{underlined} indicate second best result. Multilingual fine-tuning (Multi FT) uses all Nigerian data. NLLB does not support Naij\'a.}
\label{stt_results_chrf}
\label{stt_results_chrf}
\end{table*}

\vspace{-5pt}
\paragraph{Recording Tool} Since annotators were geographically dispersed, the Telegram mobile app was used to record utterances for practical reasons. All coordination and project management was conducted via WhatsApp as it is a widely used communication platform in Nigeria.  Audio files are recorded at a sample rate of 48 kHz and a signal-to-noise ratio of at least 30 dB. 
\autoref{tab:speech-stat} provides the details on the number of recordings, number of hours and gender distribution. See Appendix~\ref{sec:recording} for details.

\vspace{-5pt} 
\paragraph{Quality Control}
Quality control (QC) was conducted systematically for each reader and language for the dev and test sets. Specifically, 3 to 5 utterances per reader were sampled to assess both speech quality (e.g., naturalness, repetitions, loudness) and recording conditions (e.g., background noise, microphone quality). The most commonly identified problems included excessive background noise, poor microphone quality, and low volume. Given this, utterances from readers who consistently exhibited such issues were discarded and re-recorded by new participants. All problematic recordings in the development and test sets were fully re-recorded. For budgetary reasons, re-recording was not conducted for the train set. Instead, priority was given to ensuring completely clean dev and test set curation. \autoref{tab:speech-stat} summarizes the speech dataset statistics before and after QC.

\section{Experimental Settings}
\vspace{-5pt}
\subsection{Models}
\paragraph{Cascaded Methods}
Cascaded S2TT and S2ST systems involve two steps: (1) \textit{ASR transcription} and  (2) \textit{Translation} into the target language. In S2ST, there is a \textit{third stage of ``audio synthesis,''} where  translated text is subsequently converted into speech via a TTS model. 
For ASR, we employ the Omnilingual-ASR 1B LLM model \cite{omnilingual2025}. For translation, we evaluate two systems: NLLB-200 3.3B \cite{Costajuss2024ScalingNM} trained exclusively for MT, and Tiny-Aya-Global 3B \cite{salamanca2026tinyayabridgingscale}, which supports three of our Nigerian languages, but does not natively support Naijá. We make use of 5-examples from the \texttt{DEV} set for few-shot prompting of TinyAya, Gemma 3n-E2B-it (Gemma 3) and Gemma 4-E4B-it (Gemma 4) (see Appendix~\ref{sec:prompting} for the prompt). All S2ST experiments use Gemini 2.5 Flash TTS.

\begin{table*}[h]
\centering
\small
\setlength{\tabcolsep}{8pt}
\resizebox{\textwidth}{!}{%
\begin{tabular}{
    llrrrrr|rrrrr
    }
\toprule
    & 
    & \multicolumn{5}{c}{\textbf{XX $\rightarrow$ Eng}} 
    & \multicolumn{5}{c}{\textbf{Eng $\rightarrow$ XX}} \\
    
\cmidrule(lr){3-7} \cmidrule(lr){8-12}
        
\textbf{Model} & \textbf{Method}
& \textbf{Hausa} & \textbf{Igbo} & \textbf{Naijá} & \textbf{\yoruba} & \textbf{Avg.} 
& \textbf{Hausa} & \textbf{Igbo} & \textbf{Naijá} & \textbf{\yoruba} & \textbf{Avg.}  \\

\midrule
    \multirow{2}{*}{SeamlessM4T}
        & Mono FT
        & 54.9 & \underline{52.4} & 50.7 & 60.3 & 54.6 & 12.1 & 54.8 & 61.4 & 68.6 & 49.2\\ 
        & Multi FT
        & 46.7 & 47.4 & 56.0 & 54.3 & 51.1 & 53.4 & 64.6 & 42.6 & 68.5 & 57.3\\
\midrule
    \multirow{2}{*}{Gemma 4}
        & Zero-Shot 
        & 39.2 & 31.7 & 52.7 & 37.0 & 40.2 & 40.6 & 25.8 & 53.2 & 30.9 & 37.6  \\
        & Few-Shot 
        & 41.5 & 28.0 & 52.9 & 35.7 & 39.5 & 39.3 & 30.0 & 51.3 & 32.2 & 38.2 \\
        
\midrule
    \multirow{2}{*}{GPT-Audio}
        & Zero-Shot 
        & 34.0 & 31.1 & 60.3 & 36.0 & 40.4 & 38.2 & 40.2 & 42.1 & 40.6 & 40.3\\
        & Few-Shot 
        & 33.8 & 30.4 & 60.3 & 35.4 & 40.0 & 63.2 & 61.3 & 60.2 & 61.2 & 61.5 \\ 
\midrule      
    \multirow{2}{*}{Gemini 2.5}
        & Zero-Shot 
        & 60.9 & 37.0 & 59.1 & 49.4 & 51.6 & 63.8 & 64.6 & 58.2 & 67.2 & 63.5 \\
        & Few-Shot 
        & 63.1 & 41.0 & 62.2 & 51.2 & 54.4 & 64.2 & 64.4 & 54.3 & 68.9 & 63.0\\ 

\addlinespace
    \multirow{2}{*}{Gemini 3.1}
        & Zero-Shot 
        & \underline{68.1} & 51.3 & \underline{65.4} & \underline{63.6} & \underline{62.1} & \underline{68.3} & \underline{67.1} & \underline{61.5} & \underline{72.1} & \underline{67.3} \\
        & Few-Shot 
        & \textbf{69.4} & \textbf{56.2} & \textbf{66.7} & \textbf{65.9} & \textbf{64.6} & \textbf{68.3} & \textbf{67.4} & \textbf{62.8} & \textbf{72.3} & \textbf{67.7} \\ 
\bottomrule
\end{tabular}
 }
 \vspace{-2mm}
\caption{\textbf{Speech-to-text translation results (SSA-COMET $\uparrow$) for AudioLLM}, compared with fully supervised fine tuning (Seamless M4T).
\textit{Italics} indicate best within each method; \textbf{bold} is best overall. Second best is \underline{underlined}.}
\label{stt_results_ssa_comet_audio_llm}
\end{table*}

\paragraph{End-to-End Methods} 
Unlike cascading pipelines, E2E S2TT and S2ST jointly learn direct mappings from speech inputs to either text or speech outputs, thereby eliminating the need for an intermediate ASR. For the S2TT task, we evaluate SeamlessM4T-Large V2 (2.3B) in \textit{three settings:} (1) \textbf{Zero-shot inference on the test set} (2) \textbf{Monolingual fine-tuning (\texttt{Mono-FT})} per language to further improve generalization
(3) \textbf{Multilingual fine-tuning (\texttt{Multi-FT})} that combines all languages per translation direction (Eng$\leftrightarrow$low-resource languages (LRL)).
For \texttt{Mono-FT}, each language-specific model is trained with a learning rate (LR) of $1\text{e-}5$, 16 gradient accumulation (GA) steps and 3 epochs. 
LR of $5\text{e-}6$, 32 GA steps and 3 epochs is used for \texttt{Multi-FT}.\footnote{Hyperparameters selected via hyperparameter tuning.} 

For LRL$\leftrightarrow$English S2ST evaluation, we use SeamlessM4T-large 2.3B for both \texttt{zero-shot} and \texttt{Mono-FT} settings, as SeamlessM4T-Large V2 does not support S2ST finetuning. As Hausa is not supported in SeamlessM4T, it is mapped to a proxy language token from the same language family (Arabic), following \citet{adelani2022few}. Similarly, we map Naijá to French, as it is very similar to English and should therefore be linked to a European language.  The results for these two languages should therefore be interpreted as a fine-tuned extension rather than a reflection of the model's native multilingual coverage.
Training is conducted with a LR of $1\text{e-}5$, batch size 2 and a maximum of 10 epochs with early stopping. Note that the other direction in S2ST (i.e. English$\rightarrow$LRL) is omitted because SeamlessM4T only supports translation into high-resource languages.

\vspace{-2pt}
\paragraph{AudioLLMs} We evaluate proprietary LLMs that support low-resource languages for S2TT and S2ST tasks (i.e., Gemini 2.5, Gemini 3.1, GPT-Audio). For comparison, we also evaluate Gemma 4 for S2TT as an open-source option. As Gemini does not currently support native S2ST, we construct a cascaded S2ST pipeline by applying Gemini 2.5 TTS to the outputs of its S2TT system. Note that most of the other high-performing AudioLLMs/multimodal LLMs, such as Qwen3-Omni \citep{xu2025qwen3omnitechnicalreport} or DeSTA2 \citep{10889444} do not support our studied languages.

\subsection{Evaluation Metrics}
\label{evaluation_metrics}
Both lexical and embedding-based metrics are used to obtain a comprehensive assessment of model performance. 
Specifically, we report SSA-COMET (-MTL) \cite{li-etal-2025-ssa}, 
which is an extension of the COMET~\citep{rei-etal-2020-comet} metric for African languages. COMET incorporates semantic similarity between hypothesis and reference translations by leveraging pretrained multilingual encoders to compute contextualized sentence representations. We additionally use ChrF++ \cite{popovic-2015-chrf}, 
a character n-gram-based metric that computes an F-score over overlapping character sequences between the hypothesis and reference. By operating at the character level, ChrF++ is more sensitive to morphological variations and orthographic similarity than word-level metrics, but are sometimes unreliable for many LRLs~\citep{freitag-etal-2022-results,wang-etal-2024-afrimte}.

For S2ST metrics, we calculate ASR-COMET by performing ASR on the predicted audio translations using Omnilingual-ASR, followed by SSA-COMET on the transcriptions, as in the S2TT task. 

\section{Results}
\label{sec:results}
\vspace{-5pt}
\subsection{Speech-to-Text Translation Results}
\label{sec:results5p1}
\paragraph{Architectural Comparisons}
Overall, we observe a consistent advantage of E2E and Audio LLM-based approaches over traditional cascaded S2TT pipelines (Tables~\ref{stt_results_ssa_comet}–\ref{stt_results_ssa_comet_audio_llm}). In particular, the best E2E configuration (i.e., SeamlessM4T with \texttt{Mono-FT} or \texttt{Multi-FT})
and the strongest Audio LLM (Gemini 3.1) both surpass the cascaded Omnilingual ASR + MT pipeline. 
This trend aligns with prior findings that cascaded S2TT systems are susceptible to error propagation, where transcription errors from ASR compound downstream during MT \cite{app12031097}. To assess whether degradation in cascaded performance stems primarily from ASR errors or MT limitations, we pair a fixed Omnilingual ASR backbone with multiple MT models (Tables ~\ref{stt_results_ssa_comet}, \ref{stt_results_chrf}). The results show that for Hausa and Igbo, performance remains relatively similar across NLLB and TinyAya backends regardless of language direction. 
In contrast, \yoruba exhibits substantial sensitivity to the choice of MT model: replacing NLLB with Tiny Aya leads to large drops in SSA-COMET (e.g., 50.6 $\rightarrow$ 35.5 for XX$\rightarrow$Eng; 58.0 $\rightarrow$ 48.4 for Eng$\rightarrow$XX).\footnote{Qualitative inspection reveals that Tiny Aya frequently produces extraneous “chain-of-thought”-like tokens and meta-commentary in \yoruba outputs, suggesting a mismatch between instruction-tuned generative behavior and the constrained requirements of translation.}\textsuperscript{,}\footnote{See Appendix~\ref{Other_Metrics} for results with other metrics.}

\begin{table*}[h]
\centering
\small
\setlength{\tabcolsep}{2pt}
\begin{tabular}{
   lll|rrrr|rrrr
    }
\toprule
    & &
    & \multicolumn{4}{c|}{\textbf{XX $\rightarrow$ Eng}} 
    & \multicolumn{4}{c}{\textbf{Eng $\rightarrow$ XX}} \\
    
\cmidrule(lr){4-7} \cmidrule(lr){8-11}
        
\textbf{Method} & \multicolumn{2}{c|}{\textbf{Model}} & {Hausa} & {Igbo} & Naijá & {\yoruba} & {Hausa} & {Igbo} & Naijá & {\yoruba} \\

\midrule

\cellcolor[gray]{0.92}\texttt{Naij\'a-accent} \\
 \multirow{1}{*}{End-to-End}
    & SeamlessM4T & Multilingual
    & 23.5 & 26.4 & {N/A} & 36.3 & {N/A} & {N/A} & {N/A} & {N/A} \\
    \multirow{1}{*}{Cascaded}& \multirow{1}{*}{Omni + NLLB/TinyAya} & + Gemini 2.5 TTS Naija 
    & 50.4 & 40.6 & 40.8 & 46.6 & 37.1 & 35.0 & 47.2 & 39.8 \\
     \multirow{1}{*}{AudioLLM}
    & \multirow{1}{*}{Gemini 3.1 Few-Shot}& + Gemini 2.5 TTS Naija
     & \underline{61.8} & \underline{50.7} & \underline{60.0} & \underline{65.4} &\textbf{45.8} & \underline{40.0} & \underline{47.6} & {42.5} \\
    \midrule
    \addlinespace
\cellcolor[gray]{0.92}\texttt{British-accent} \\
 \multirow{1}{*}{Cascaded}& {Omni + NLLB/TinyAya}
    & + Gemini 2.5 TTS British
    & 51.5 & 41.8 & 51.4 & 48.6 & 40.3 & 38.3 & 41.2 & \underline{45.1} \\
 \multirow{1}{*}{AudioLLM}
    & \multirow{1}{*}{Gemini 3.1 Few-Shot} & + Gemini 2.5 TTS British
     & \textbf{63.3} & \textbf{51.2} & \textbf{62.4} & \underline{60.7} & \underline{44.4} & \textbf{40.4} & \textbf{47.9} & \textbf{51.1} \\
\bottomrule
\end{tabular}
\vspace{-2mm}
\caption{\textbf{Speech-to-speech translation results (ASR-COMET $\uparrow$)}.
\textit{Italics} indicate best within each method; \textbf{bold} indicates best overall while second best is \underline{underlined}. Best models per method are used as a base for TTS. Thus, we use Omni+NLLB for Hausa, Igbo and \yoruba, and Tiny Aya for Naijá, as it is not supported by NLLB.}
\label{sts_results_ssa_comet}
\end{table*}

\vspace{-2pt}
\paragraph{End-to-End Fine-tuning}
Table~\ref{stt_results_ssa_comet} reveals a highly asymmetric zero-shot behavior of SeamlessM4T across directions and languages. In the LRL$\rightarrow$English setting, performance fluctuates: while \yoruba and Naijá results are relatively high at 52.4 and 57.0 SSA-COMET scores, Igbo and Hausa lag significantly behind (20.6 and 14.6, respectively). The particularly low performance for Hausa and Naijá for English$\rightarrow$LRL is expected given their unsupported status in the base model. 

\vspace{-2pt}
\paragraph{Inconsistencies of metrics to wrong language output} In the reverse English$\rightarrow$LRL direction, zero-shot scores for Igbo and \yoruba appear to exhibit relatively high SSA-COMET scores (53.1 \& 55.4, respectively). However, this is not reflected in ChrF++ (Table~\ref{stt_results_chrf}), where scores remain substantially lower. This discrepancy highlights a key limitation of embedding-based metrics: SSA-COMET is more tolerant to lexical and surface-form mismatches, whereas ChrF++ penalizes such deviations more directly. Manual inspection of the outputs shows that a considerable portion of SeamlessM4T predictions remain in the source language. This indicates that the model often fails to fully perform the translation despite achieving moderate semantic similarity scores, and highlights the need to have multiple evaluation metrics. While ChrF++ is generally unreliable as it often achieves low scores for languages like \yoruba with extensive use of diacritics and penalizes progress (i.e. the language with the highest SSA-COMET score out of the three languages achieved lower score with ChrF++ in \autoref{stt_results_chrf}), it is able to catch simpler errors that more sophisticated metrics cannot. 

SeamlessM4T exhibits notably different behaviour on Naijá compared to the other language pairs. While Mono-FT improves SSA-COMET scores over the Zero-Shot setting, the opposite trend is observed for ChrF++. One likely factor is the relatively small amount of available Naijá training data, which makes fine-tuning, particularly through the use of a proxy language, more difficult.


\vspace{-6pt}
\paragraph{Fine-tuning helps wrong language generation}
Fine-tuning substantially mitigates these issues, but the gains depend critically on both directionality and training strategy. In the LRL$\rightarrow$Eng direction, \texttt{Multi-FT} yields large improvements for Hausa, Igbo and Naijá which suggests that shared language representations help compensate for limited per-language data. Yet, this comes at a slight cost for \yoruba, where zero-shot performance is already strong, and indicates potential cross-lingual interference. On the other hand, \texttt{Mono-FT} consistently improves performance across all languages and achieves the best overall results in this direction (e.g., 60.3 for \yoruba), with Naijá being the exception, likely due to the comparatively limited amount of available training data. This suggests that joint training could introduce cross-lingual trade-offs that are absent in language-specific adaptation.

The trend reverses for English$\rightarrow$LRL, where \texttt{Multi-FT} consistently outperforms \texttt{Mono-FT} across both SSA-COMET and ChrF for Hausa, Igbo and \yoruba, and still achieves relatively high results for Naijá. Given the lower-resource nature of this setting, this suggests that performance is more strongly influenced by overall data availability, with shared multilingual training providing greater benefit than language-specific fine-tuning.

\vspace{-6pt}
\paragraph{AudioLLM Evaluation} 
As shown in Table~\ref{stt_results_ssa_comet_audio_llm}, Audio LLMs establish the highest baseline for S2TT, with Gemini 3.1 emerging as the strongest model across all language pairs and directions. Specifically, under few-shot prompting, Gemini 3.1 achieves the best overall SSA-COMET scores, and surpasses both the best fine-tuned SeamlessM4T models (average margin of -8 points) and all cascaded systems. 
Moreover, it is consistently seen across Gemini models that few-shot prompting helps in providing modest but reliable gains over its zero-shot performance. This indicates that in-context learning helps stabilize cross-lingual alignment in low-resource settings. In contrast, GPT-Audio exhibits weaker and less stable performance particularly for LRL$\rightarrow$English, where few-shot prompting does not consistently improve results, and even leads to degradation. Similarly, Gemma 4 does not improve with few-shot prompting, and reports the lowest average scores with the AudioLLM category, and is lower than that of SeamlessM4T. See Appendix \ref{sec:accents_appendix_british_additional} for more information. 

\vspace{-5pt}
\subsection{Speech-to-Speech Translation Results}

\begin{table}[t]
\centering
\small
\setlength{\tabcolsep}{4.5pt}
\resizebox{\columnwidth}{!}{%
\begin{tabular}{
    l|rrcr|rrcr
    }
\toprule
    & \multicolumn{4}{c|}{\textbf{XX $\rightarrow$ Eng}} 
    & \multicolumn{4}{c}{\textbf{Eng $\rightarrow$ XX}} \\
    
\cmidrule(lr){2-5} \cmidrule(lr){6-9}
        
\textbf{ Method}
& \textbf{ha} & \textbf{ig} & \textbf{yo} & \textbf{Avg.} 
& \textbf{ha} & \textbf{ig} & \textbf{yo} & \textbf{Avg.} \\

\midrule
    {\textsc{Cascaded}}
        & \underline{75.0} & \underline{47.5} & \underline{74.7} & \underline{65.7} &\underline{82.1} & \underline{73.9} & \underline{76.0}  & \underline{77.3} \\ 
\addlinespace
    \textsc{E2E}
        & 4.1 & 29.0 & 59.3 & 30.8 & {N/A}  &{N/A} & {N/A}  & {N/A}\\
\addlinespace
    \textsc{AudioLLM}
        & \textbf{99.2} & \textbf{72.8} & \textbf{93.2} & \textbf{88.4 } &\textbf{88.5} & \textbf{83.4} & \textbf{94.0} & \textbf{ 88.6}  \\
\bottomrule
\end{tabular}
}
\vspace{-2mm}
\caption{\textbf{S2ST human evaluation results (0-100 average Likert scale $\uparrow$)}.
\textit{Italics} indicate best within each method; \textbf{bold} is best overall. \underline{Underlined} is second best.}
\label{sts_human_eval}
\vspace{-3mm}
\end{table}

\paragraph{Speech-to-speech systems}
Table~\ref{sts_results_ssa_comet} presents S2ST results across cascaded, E2E, and AudioLLM-based approaches. The findings reinforce a consistent pattern observed in S2TT; performance is largely driven by the quality of the translation component rather than speech generation. For LRL$\rightarrow$English, AudioLLM pipelines achieve the strongest results overall, with Gemini 3.1 (few-shot) + TTS reaching the highest score for Hausa (63.3), outperforming cascaded systems by a 12.9 margin. This advantage is reflected across the other languages, all outperforming the cascaded methods by a margin over 10. 
Note that accent variation in the TTS component (Received Pronunciation (RP) British vs.\ Nigerian English) has minimal impact in the LRL$\rightarrow$English direction, with only marginal differences across all languages. The primary exception is the cascaded approach on Naijá, where performance decreases noticeably under British-accented speech. This may be explained by the linguistic similarity between Naijá and Nigerian-accented English, which likely provides an advantage when the target speech aligns more closely with the source language context. Overall, these findings suggest that accent choice in the TTS component has limited influence on speech-to-speech quality when English is the target language. 

In contrast, the English$\rightarrow$LRL direction reveals a clearer advantage for AudioLLM-based approaches compared to cascading systems. Gemini 3.1 with Nigerian English (Naija) TTS achieves the best performance across most of the target languages (e.g., 45.8 for Hausa, 40.0 for Igbo and 47.6 for Naijá), while the cascaded approach remains more effective for \yoruba (39.8).
Furthermore, S2ST with SeamlessM4T (multilingual) performs significantly worse than both cascaded and AudioLLM approaches, which is likely due to weaker translation quality from the V1 model, and simply outputted English for Naijá. This highlights that translation quality remains the primary bottleneck in S2ST performance.  

\subsection{Human Evaluation}
\label{sec:translation_quality_human}
To assess translation quality, we conduct six human evaluation studies, corresponding to each English$\leftrightarrow$XX language pair. Each survey is completed by three bilingual annotators, who evaluate both the semantic adequacy of the translation and the pronunciation quality of the generated S2ST output on a 0--100 scale. For 100 randomly sampled utterances from the Nigerian-context test set, we construct one evaluation item per modeling paradigm, using the strongest system in each: Omnilingual-ASR + NLLB (cascaded), SeamlessM4T Multi-FT (E2E), and Gemini 3.1 + Gemini 2.5-TTS (AudioLLM). Since Eng$\rightarrow$XX lacks an E2E paradigm, we evaluate only the cascaded and AudioLLM settings, yielding 200 items for Eng$\rightarrow$XX and 300 for XX$\rightarrow$Eng. Further details can be found in Appendix \ref{sec:human_eval_ablations}.

Overall, the human evaluation results reveal a consistent hierarchy across language pairs and translation directions. AudioLLM systems obtain the highest translation quality ratings, with average scores exceeding 88\%, whereas cascaded systems achieve moderate performance (65--77\%) and E2E models receive the lowest ratings (Table~\ref{sts_human_eval}). Across the evaluated languages, Igbo receives the lowest human scores, mirroring the pattern observed in the automatic evaluation results in Table~\ref{sts_results_ssa_comet}. This outcome may reflect Igbo's comparatively lower-resource status relative to the other languages.

\vspace{-5pt}
\section{Further Analysis}
\vspace{-5pt}
\subsection{Sentence Length Distribution}
\label{sec:sentence_length_main}
\begin{figure}[t]
\centering
\includegraphics[width=0.45\textwidth, height=3cm]{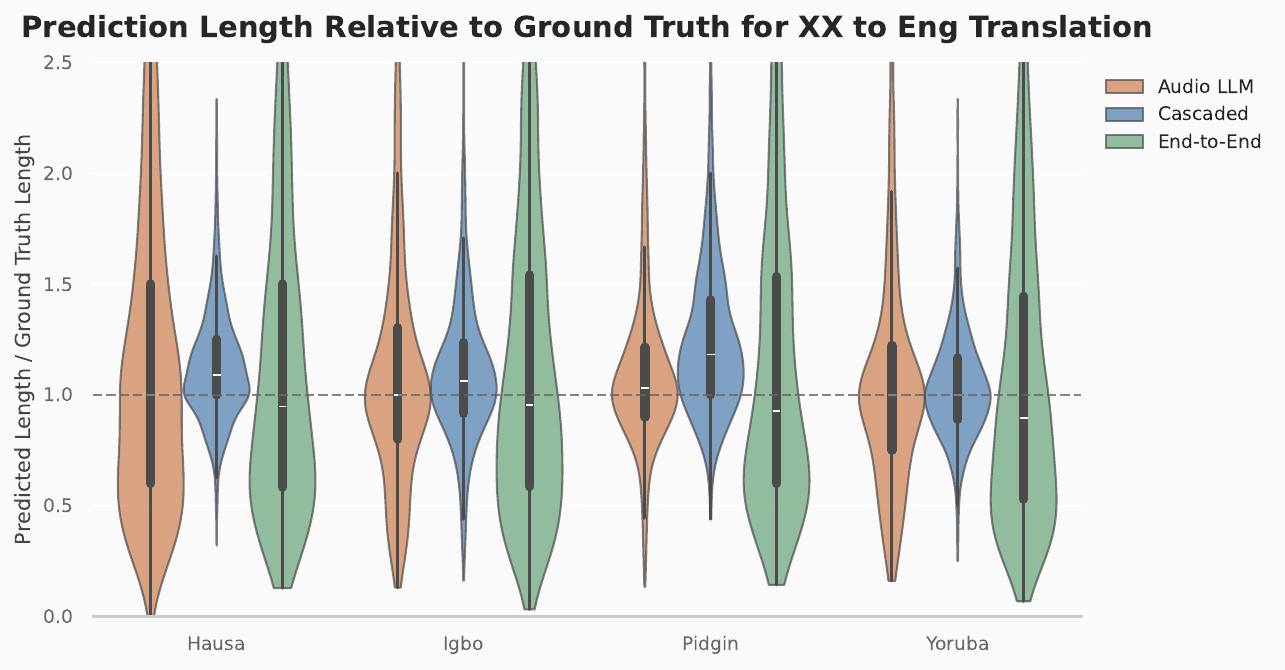}
\caption{Prediction and ground truth sentence length ratio for XX $\rightarrow$ English translation.}
\label{fig:violin1}
\end{figure}

\begin{figure}[t]
\centering
\includegraphics[width=0.45\textwidth]{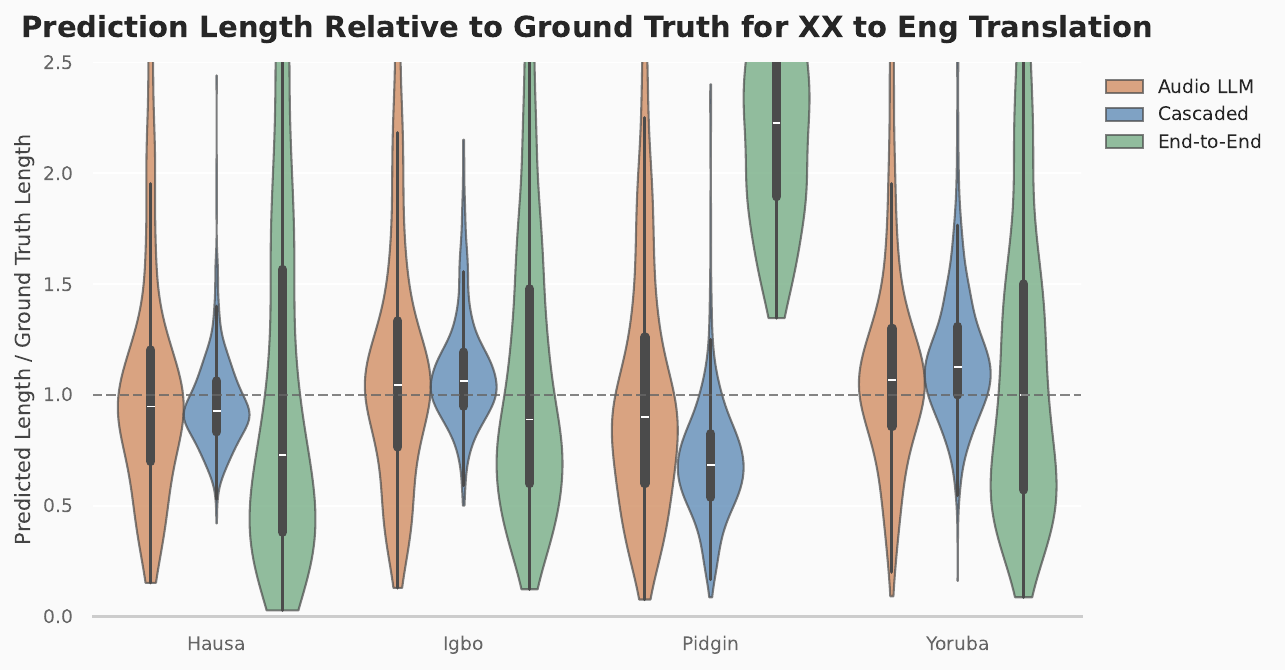}
\caption{Sentence length ratio for English$\rightarrow$XX.}
\label{fig:violin2}
\end{figure}

To further investigate translation quality, we analyze prediction-to-ground-truth sentence length ratios across architectures. Figures \ref{fig:violin1} and \ref{fig:violin2} show these distributions for the LRL$\rightarrow$Eng and Eng$\rightarrow$LRL directions, respectively. The cascaded approach yields the most stable and compact distribution, centering tightly around 1.0 with minimal variance. Although cascaded systems can propagate downstream ASR errors, their intermediate bottleneck helps enforce structural alignment and prevents runaway generation or severe truncation.

Conversely, SeamlessM4T shows severe instability, with bottom-heavy distributions that indicate systematic under-generation, especially for Hausa. This likely stems from Hausa being unsupported in the base SeamlessM4T model and handled through proxy Arabic tokens, leading to frequent generation failures, untranslated source-language outputs and repetitive words.

Audio LLMs exhibit the highest variance, with upper tails frequently exceeding a ratio of 2.0, indicating over-generation. Manual inspection of the 50 longest over-generated outputs revealed that all were valid, verbose paraphrases that unpack compact clauses into explicit compound sentences while preserving meaning. This accounts for the discrepancy between human preference for Audio LLM outputs and their apparent over-generation, as sentence-length ratio metrics penalize such verbosity, whereas human evaluators and semantic metrics (e.g., SSA-COMET) rate it positively, as translation accuracy remains unaffected. Further analyses are included in Appendix \ref{sec:sentence_analysis_ablations}.

\subsection{Part-of-Speech (POS) Analysis}
\begin{figure}[t]
\centering
\includegraphics[width=0.46\textwidth, height=3.3cm]{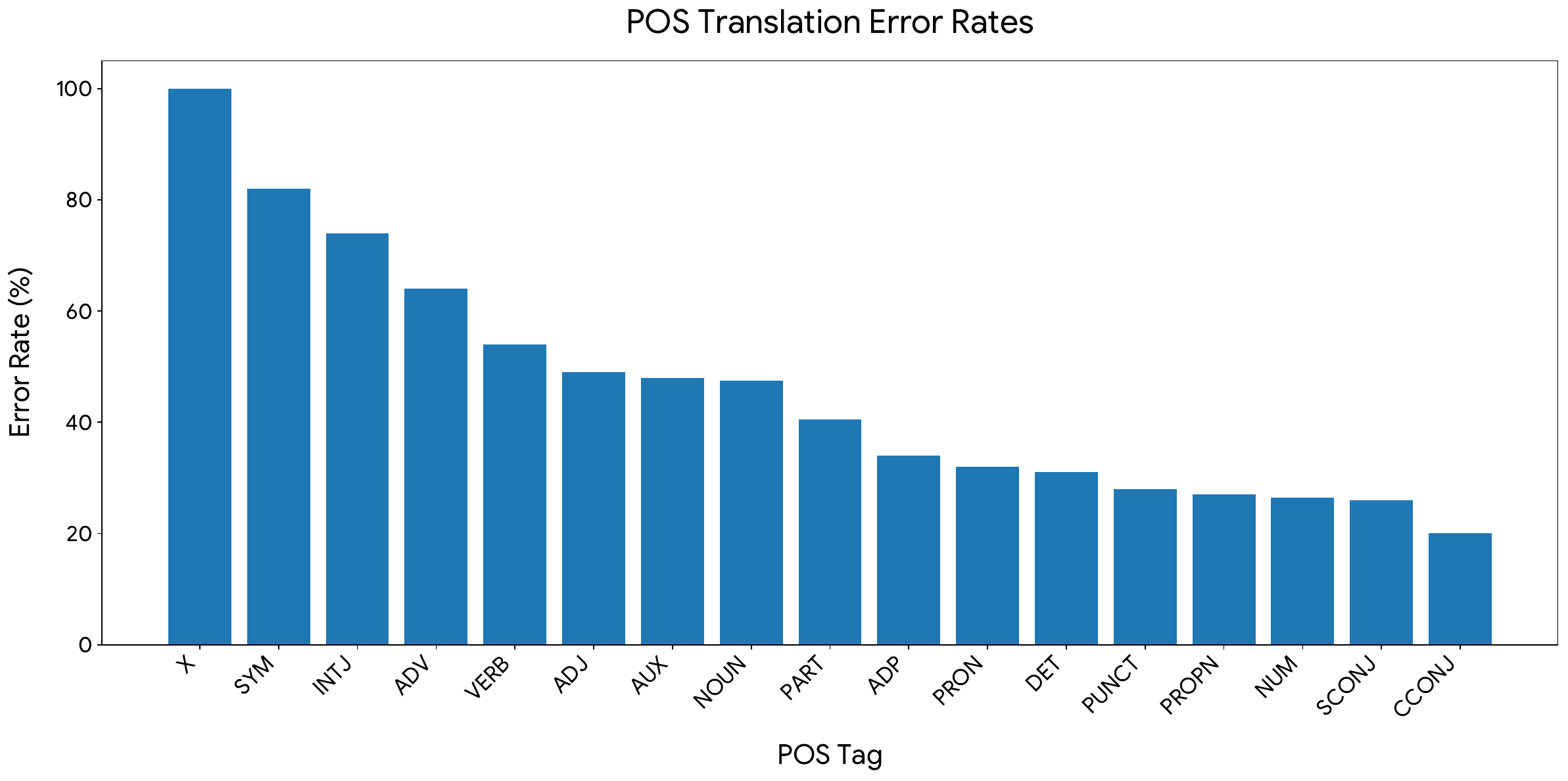}
\caption{POS error rates for English$\rightarrow$\yoruba S2TT using best-performing AudioLLM.}
\label{fig:eng_haus_pos_tag_main_body}
\end{figure}

To further understand the failure modes of different paradigms, we conduct POS error analysis on the outputs of best-performing AudioLLM and cascaded systems (Figure \ref{fig:eng_haus_pos_tag_main_body} and Appendix \ref{sec:POS_appendix}). POS tagging is performed automatically using open-source taggers, which are monolingually fine-tuned AfroXLMR-76L~\citep{adelani-etal-2024-sib} encoder models that are trained on MasakhaPOS \citep{dione-etal-2023-masakhapos}. Analysis is restricted to Eng$\rightarrow$XX, tagging both reference and predicted target-language text (with the same tagger) to isolate target-generation quality. 

It can be observed that aggregate scores mask systematic paradigm differences. POS-level breakdown shows AudioLLM errors concentrate on open-class, content-bearing tags (e.g., X, ADV, VERB, SYM) and are lowest on closed-class tags (e.g., CCONJ, PROPN, NUM). Cascaded systems show the opposite pattern, with dominant SYM, NUM, and PUNCT, and VERB remaining comparatively stable. We attribute this to differing inductive biases; AudioLLMs are prone to semantic drift on content words, while cascaded systems are more vulnerable on transcription-sensitive categories due to the intermediate textual bottleneck.

Moreover, \yoruba diverges from the other languages, showing elevated SYM and INTJ errors and a flatter error distribution, whereas Hausa, Igbo, and Naijá show broadly similar rankings across paradigms. This aligns with \yoruba's greater sensitivity to MT-model choice (Sec~\ref{sec:results5p1}), likely reflecting its profile as a highly isolating language reliant on tone marks and underdots~\cite{ogunremi-etal-2024-iroyinspeech}. Since \yoruba encodes POS prosodically and orthographically rather than morphologically, these distinctions are fragile under S2T transcription and harder for aggregate metrics to capture (see Appendix \ref{sec:POS_appendix}).

\section{Conclusion}
\vspace{-3pt}
In this paper, we introduced \textbf{NaijaS2ST}, a parallel speech–text dataset for Igbo, Hausa, \yoruba, and Nigerian Pidgin paired with English, designed to address the scarcity of high-quality, diverse speech resources for low-resource languages in African contexts. The dataset enables standardized evaluation of both S2TT and S2ST systems under realistic multi-accent and multilingual conditions, with carefully curated recordings spanning diverse speakers, dialects, and quality-controlled splits.

Using NaijaS2ST, comprehensive benchmarking studies were conducted across cascaded, E2E, and AudioLLM-based approaches in bidirectional translation settings. Results demonstrate a clear advantage of E2E and AudioLLM systems over traditional cascaded methods, with AudioLLMs achieving the strongest overall performance. Fine-tuning is essential for E2E models, but exhibits strong directionality effects; monolingual adaptation is most effective for LRL$\rightarrow$English, whereas multilingual training yields greater gains in the reverse direction. Moreover, translation quality remains the primary bottleneck for S2ST, while speech synthesis choices including accent variation, have comparatively limited impact in LRL$\rightarrow$English scenarios.
We hope this work will support future research toward more robust, scalable, and inclusive multilingual speech translation research.

\section{Limitations}
Despite the breadth of our study, several limitations remain. To start, our evaluations are conducted in a controlled offline setting and therefore does not capture practical deployment considerations such as inference latency, streaming constraints, or computational efficiency. In particular, although AudioLLM-based systems achieve strong translation performance, they may incur substantially higher inference cost and latency than cascaded or E2E approaches, potentially limiting their applicability in real-time or resource-constrained environments. Moreover, while we have benchmarked across varying cascaded, E2E, and AudioLLM-based paradigms, our exploration of model configurations is not exhaustive. For AudioLLMs especially, we evaluate only a limited set of prompting and decoding strategies, and do not systematically explore the broader design space of prompt engineering, in-context learning configurations, or alignment techniques. Given the sensitivity of AudioLLMs to prompt formulation, further performance gains may be achievable through more extensive optimization.

\section{Ethical considerations}
This project involved collecting speech data from more than 250 participants across Nigeria. All participants provided their informed consent for the use of their recordings in speech translation research and were fairly compensated for their contributions. Given the scale of the data collection effort, we did not rely on volunteer recordings. Instead, each participant recorded approximately 250 sentences and received \$15 in compensation (which is a fair rate in Nigeria). Similarly, language coordinators and translators were also adequately compensated for their work. Translators were paid \$0.50 per sentence, resulting in a total translation cost exceeding \$11,000 in all languages. Throughout the project, we ensured that contributors were compensated fairly while also allowing room for negotiation when necessary.

Regarding privacy and ethics, we did not collect private, personal, or sensitive content, nor did we ask participants to read such material. All recorded sentences were sourced from publicly available corpora developed for prior NLP research projects, including NTREX, SSA-MT, and VOA datasets. To further protect participant privacy, we will prohibit the use of the dataset for tasks intended to re-identify speakers. In addition, we do not intend to release any personally identifiable information (PII) associated with participants, such as names, email addresses, phone numbers, or other identifying metadata.

\section{Acknowledgments}
This research was supported by the Google grant via Mila, the Natural Sciences and Engineering Research Council
(NSERC) of Canada, and in part by the AI2050 program at Schmidt Sciences. This work was partially supported by Azure sponsorship credits granted by Microsoft’s AI for Good Research Lab and Google's Gemini Academic Program Award. 
We thank Elizabeth Salesky for her careful guidance and thoughtful feedback throughout this research. Finally, we thank Luel Hagos Beyene for his technical assistance with the speech-to-text model fine-tuning.


\bibliography{custom}

@inproceedings{emezue25_interspeech,
  title     = {{The NaijaVoices Dataset: Cultivating Large-Scale, High-Quality, Culturally-Rich Speech Data for African Languages}},
  author    = {Chris Emezue and  {The NaijaVoices Community} and Busayo Awobade and Abraham Toluwase Owodunni and Handel Emezue and Gloria Monica Tobechukwu Emezue and Nefertiti Nneoma Emezue and Sewade Ogun and Bunmi Akinremi and David Ifeoluwa Adelani and Chris Pal},
  year      = {2025},
  booktitle = {{Interspeech 2025}},
  pages     = {1338--1342},
  doi       = {10.21437/Interspeech.2025-1104},
  issn      = {2958-1796},
}

@inproceedings{
adebara2026african,
title={African Voices Nigeria: 2500 hours of ethically sourced speech data for four Nigerian Languages},
author={Ife Adebara and Oluwaseun Nifemi and Rashidat Damilola Sikiru and Olanrewaju Israel Lawal and Ololade Anjuwon and Olubayo Adekanmbi and Anthony Soronnadi and John Emeka Eze and Ewezu Ngim Ngim},
booktitle={7th Workshop on African Natural Language Processing},
year={2026},
url={https://openreview.net/forum?id=sJJ5ajmHFx}
}

@inproceedings{ogunremi-etal-2024-iroyinspeech,
    title = "{{\`I}}r{\`o}y{\`i}n{S}peech: A Multi-purpose {Y}or{\`u}b{\'a} Speech Corpus",
    author = "{\`O}g{\'u}nr\d{\'e}m{\'i}, Tol{\'u}l\d{o}p\d{e}  and
      T{\'u}b\d{o}s{\'u}n, K\d{\'o}l{\'a}  and
      Aremu, Anuoluwapo  and
      Orife, Iroro  and
      Adelani, David Ifeoluwa",
    editor = "Calzolari, Nicoletta  and
      Kan, Min-Yen  and
      Hoste, Veronique  and
      Lenci, Alessandro  and
      Sakti, Sakriani  and
      Xue, Nianwen",
    booktitle = "Proceedings of the 2024 Joint International Conference on Computational Linguistics, Language Resources and Evaluation (LREC-COLING 2024)",
    month = may,
    year = "2024",
    address = "Torino, Italia",
    publisher = "ELRA and ICCL",
    url = "https://aclanthology.org/2024.lrec-main.812/",
    pages = "9296--9303"
}

@inproceedings{meyer22c_interspeech,
  title     = {{BibleTTS: a large, high-fidelity, multilingual, and uniquely African speech corpus}},
  author    = {Josh Meyer and David Adelani and Edresson Casanova and Alp Öktem and Daniel Whitenack and Julian Weber and Salomon {KABONGO KABENAMUALU} and Elizabeth Salesky and Iroro Orife and Colin Leong and Perez Ogayo and Chris {Chinenye Emezue} and Jonathan Mukiibi and Salomey Osei and Apelete AGBOLO and Victor Akinode and Bernard Opoku and Olanrewaju Samuel and Jesujoba Alabi and Shamsuddeen Hassan Muhammad},
  year      = {2022},
  booktitle = {{Interspeech 2022}},
  pages     = {2383--2387},
  doi       = {10.21437/Interspeech.2022-10850},
  issn      = {2958-1796},
}

@article{fleurs_paper,
  title={FLEURS: FEW-Shot Learning Evaluation of Universal Representations of Speech},
  author={Alexis Conneau and Min Ma and Simran Khanuja and Yu Zhang and Vera Axelrod and Siddharth Dalmia and Jason Riesa and Clara Rivera and Ankur Bapna},
  journal={2022 IEEE Spoken Language Technology Workshop (SLT)},
  year={2022},
  pages={798-805},
  url={https://ieeexplore.ieee.org/abstract/document/10023141}
}

@article{olatunji-etal-2023-afrispeech,
    title = "{A}fri{S}peech-200: Pan-{A}frican Accented Speech Dataset for Clinical and General Domain {ASR}",
    author = "Olatunji, Tobi  and
      Afonja, Tejumade  and
      Yadavalli, Aditya  and
      Emezue, Chris Chinenye  and
      Singh, Sahib  and
      Dossou, Bonaventure F. P.  and
      Osuchukwu, Joanne  and
      Osei, Salomey  and
      Tonja, Atnafu Lambebo  and
      Etori, Naome  and
      Mbataku, Clinton",
    journal = "Transactions of the Association for Computational Linguistics",
    volume = "11",
    year = "2023",
    address = "Cambridge, MA",
    publisher = "MIT Press",
    url = "https://aclanthology.org/2023.tacl-1.93/",
    doi = "10.1162/tacl_a_00627",
    pages = "1669--1685"
}

@article{sarim2025direct,
  title={Direct speech to speech translation: A review},
  author={Sarim, Mohammad and Shakeel, Saim and Javed, Laeeba and Nadeem, Mohammad and others},
  journal={arXiv preprint arXiv:2503.04799},
  year={2025}
}

@inproceedings{baevski2020wav2vec,
author = {Baevski, Alexei and Zhou, Henry and Mohamed, Abdelrahman and Auli, Michael},
title = {wav2vec 2.0: a framework for self-supervised learning of speech representations},
year = {2020},
isbn = {9781713829546},
publisher = {Curran Associates Inc.},
address = {Red Hook, NY, USA},
booktitle = {Proceedings of the 34th International Conference on Neural Information Processing Systems},
url = {https://dl.acm.org/doi/abs/10.5555/3495724.3496768},
articleno = {1044},
numpages = {12},
location = {Vancouver, BC, Canada},
series = {NIPS '20}
}

@article{hsu2021hubert,
  title={Hubert: Self-supervised speech representation learning by masked prediction of hidden units},
  author={Hsu, Wei-Ning and Bolte, Benjamin and Tsai, Yao-Hung Hubert and Lakhotia, Kushal and Salakhutdinov, Ruslan and Mohamed, Abdelrahman},
  journal={IEEE/ACM transactions on audio, speech, and language processing},
  volume={29},
  pages={3451--3460},
  year={2021},
  publisher={IEEE},
  url={https://ieeexplore.ieee.org/document/9585401}
}

@inproceedings{wang2021voxpopuli,
    title = "{V}ox{P}opuli: A Large-Scale Multilingual Speech Corpus for Representation Learning, Semi-Supervised Learning and Interpretation",
    author = "Wang, Changhan  and
      Riviere, Morgane  and
      Lee, Ann  and
      Wu, Anne  and
      Talnikar, Chaitanya  and
      Haziza, Daniel  and
      Williamson, Mary  and
      Pino, Juan  and
      Dupoux, Emmanuel",
    editor = "Zong, Chengqing  and
      Xia, Fei  and
      Li, Wenjie  and
      Navigli, Roberto",
    booktitle = "Proceedings of the 59th Annual Meeting of the Association for Computational Linguistics and the 11th International Joint Conference on Natural Language Processing (Volume 1: Long Papers)",
    month = aug,
    year = "2021",
    address = "Online",
    publisher = "Association for Computational Linguistics",
    url = "https://aclanthology.org/2021.acl-long.80/",
    doi = "10.18653/v1/2021.acl-long.80",
    pages = "993--1003"
}

@inproceedings{ardila2020common,
    title = "Common Voice: A Massively-Multilingual Speech Corpus",
    author = "Ardila, Rosana  and
      Branson, Megan  and
      Davis, Kelly  and
      Kohler, Michael  and
      Meyer, Josh  and
      Henretty, Michael  and
      Morais, Reuben  and
      Saunders, Lindsay  and
      Tyers, Francis  and
      Weber, Gregor",
    editor = "Calzolari, Nicoletta  and
      B{\'e}chet, Fr{\'e}d{\'e}ric  and
      Blache, Philippe  and
      Choukri, Khalid  and
      Cieri, Christopher  and
      Declerck, Thierry  and
      Goggi, Sara  and
      Isahara, Hitoshi  and
      Maegaard, Bente  and
      Mariani, Joseph  and
      Mazo, H{\'e}l{\`e}ne  and
      Moreno, Asuncion  and
      Odijk, Jan  and
      Piperidis, Stelios",
    booktitle = "Proceedings of the Twelfth Language Resources and Evaluation Conference",
    month = may,
    year = "2020",
    address = "Marseille, France",
    publisher = "European Language Resources Association",
    url = "https://aclanthology.org/2020.lrec-1.520/",
    pages = "4218--4222",
    language = "eng",
    ISBN = "979-10-95546-34-4"
}

@inproceedings{wang2021covost,
  title={CoVoST 2 and massively multilingual speech translation.},
  author={Wang, Changhan and Wu, Anne and Gu, Jiatao and Pino, Juan},
  booktitle={Interspeech},
  volume={2021},
  pages={2247--2251},
  year={2021}
}

@inproceedings{jia-etal-2022-cvss,
    title = "{CVSS} Corpus and Massively Multilingual Speech-to-Speech Translation",
    author = "Jia, Ye  and
      Tadmor Ramanovich, Michelle  and
      Wang, Quan  and
      Zen, Heiga",
    editor = "Calzolari, Nicoletta  and
      B{\'e}chet, Fr{\'e}d{\'e}ric  and
      Blache, Philippe  and
      Choukri, Khalid  and
      Cieri, Christopher  and
      Declerck, Thierry  and
      Goggi, Sara  and
      Isahara, Hitoshi  and
      Maegaard, Bente  and
      Mariani, Joseph  and
      Mazo, H{\'e}l{\`e}ne  and
      Odijk, Jan  and
      Piperidis, Stelios",
    booktitle = "Proceedings of the Thirteenth Language Resources and Evaluation Conference",
    month = jun,
    year = "2022",
    address = "Marseille, France",
    publisher = "European Language Resources Association",
    url = "https://aclanthology.org/2022.lrec-1.720/",
    pages = "6691--6703"
}

@inproceedings{duquenne-etal-2023-speechmatrix,
    title = "{S}peech{M}atrix: A Large-Scale Mined Corpus of Multilingual Speech-to-Speech Translations",
    author = "Duquenne, Paul-Ambroise  and
      Gong, Hongyu  and
      Dong, Ning  and
      Du, Jingfei  and
      Lee, Ann  and
      Goswami, Vedanuj  and
      Wang, Changhan  and
      Pino, Juan  and
      Sagot, Beno{\^i}t  and
      Schwenk, Holger",
    editor = "Rogers, Anna  and
      Boyd-Graber, Jordan  and
      Okazaki, Naoaki",
    booktitle = "Proceedings of the 61st Annual Meeting of the Association for Computational Linguistics (Volume 1: Long Papers)",
    month = jul,
    year = "2023",
    address = "Toronto, Canada",
    publisher = "Association for Computational Linguistics",
    url = "https://aclanthology.org/2023.acl-long.899/",
    doi = "10.18653/v1/2023.acl-long.899",
    pages = "16251--16269"
}

@article{seamless2025joint,
	title = {Joint speech and text machine translation for up to 100 languages},
	volume = {637},
	copyright = {2025 Meta},
	issn = {1476-4687},
	url = {https://www.nature.com/articles/s41586-024-08359-z},
	doi = {10.1038/s41586-024-08359-z},
	language = {en},
	number = {8046},
	urldate = {2026-03-03},
	journal = {Nature},
	publisher = {Nature Publishing Group},
	author = {Barrault, Loïc and Chung, Yu-An and Meglioli, Mariano Coria and Dale, David and Dong, Ning and Duquenne, Paul-Ambroise and Elsahar, Hady and Gong, Hongyu and Heffernan, Kevin and Hoffman, John and Klaiber, Christopher and Li, Pengwei and Licht, Daniel and Maillard, Jean and Rakotoarison, Alice and Sadagopan, Kaushik Ram and Wenzek, Guillaume and Ye, Ethan and Akula, Bapi and Chen, Peng-Jen and El Hachem, Naji and Ellis, Brian and Gonzalez, Gabriel Mejia and Haaheim, Justin and Hansanti, Prangthip and Howes, Russ and Huang, Bernie and Hwang, Min-Jae and Inaguma, Hirofumi and Jain, Somya and Kalbassi, Elahe and Kallet, Amanda and Kulikov, Ilia and Lam, Janice and Li, Daniel and Ma, Xutai and Mavlyutov, Ruslan and Peloquin, Benjamin and Ramadan, Mohamed and Ramakrishnan, Abinesh and Sun, Anna and Tran, Kevin and Tran, Tuan and Tufanov, Igor and Vogeti, Vish and Wood, Carleigh and Yang, Yilin and Yu, Bokai and Andrews, Pierre and Balioglu, Can and Costa-jussà, Marta R. and Çelebi, Onur and Elbayad, Maha and Gao, Cynthia and Guzmán, Francisco and Kao, Justine and Lee, Ann and Mourachko, Alexandre and Pino, Juan and Popuri, Sravya and Ropers, Christophe and Saleem, Safiyyah and Schwenk, Holger and Tomasello, Paden and Wang, Changhan and Wang, Jeff and Wang, Skyler and {SEAMLESS Communication Team}},
	month = jan,
	year = {2025},
	pages = {587--593},
}

@article{alastruey2026omnilingual,
  title={Omnilingual MT: Machine Translation for 1,600 Languages},
  author={Belen Alastruey and Niyati Bafna and Andrea Caciolai and Kevin Heffernan and Artyom Kozhevnikov and Christophe Ropers and Eduardo S{\'a}nchez and Charles-'Eric Saint-James and Yiannis (Ioannis) Tsiamas and Chierh Cheng and Joe Chuang and Paul-Ambroise Duquenne and Mark Duppenthaler and Nate Ekberg and Cynthia Gao and Pere-Llu{\'i}s Huguet Cabot and Jo{\~a}o Janeiro and Jean Maillard and Gabriel Mejia Gonzalez and Holger Schwenk and Edan Toledo and Arina Turkatenko and Albert Ventayol-boada and Rashel Moritz and Alexandre Mourachko and Surya Sairam Parimi and Mary Williamson and Shireen Yates and David Dale and Marta Ruiz Costa-juss{\`a}},
  journal={ArXiv},
  year={2026},
  volume={abs/2603.16309},
  url={https://arxiv.org/abs/2603.16309}
}

@article{omnilingual2025,
  title={{Omnilingual ASR: Open-Source Multilingual Speech Recognition for 1600+ Languages}},
  author={Omnilingual ASR Team and Gil Keren and Artyom Kozhevnikov and Yen Meng and Christophe Ropers and Matthew Setzler and Skyler Wang and Ifeoluwanimi Adebara and Michael Auli and Can Balioglu and Kevin Chan and Chierh Cheng and Joe Chuang and Caley Droof and Mark Duppenthaler and Paul-Ambroise Duquenne and Alexander Erben and Cynthia Gao and Gabriel Mejia Gonzalez and Kehan Lyu and Sagar Miglani and Vineel Pratap and Kaushik Ram Sadagopan and Safiyyah Saleem and Arina Turkatenko and Albert Ventayol-boada and Zheng-Xin Yong and Yu-An Chung and Jean Maillard and Rashel Moritz and Alex Mourachko and Mary Williamson and Shireen Yates},
  journal={ArXiv},
  year={2025},
  volume={abs/2511.09690},
  url={https://arxiv.org/abs/2511.09690}
}

@article{app12031097,
	article-number = {1097},
	author = {Etchegoyhen, Thierry and Arzelus, Haritz and Gete, Harritxu and Alvarez, Aitor and Torre, Iv{\'a}n G. and Mart{\'\i}n-Do{\~n}as, Juan Manuel and Gonz{\'a}lez-Docasal, Ander and Fernandez, Edson Benites},
	doi = {10.3390/app12031097},
	issn = {2076-3417},
	journal = {Applied Sciences},
	number = {3},
	title = {Cascade or Direct Speech Translation? A Case Study},
	url = {https://www.mdpi.com/2076-3417/12/3/1097},
	volume = {12},
	year = {2022}}

@article{rubenstein1904audiopalm,
  title={AudioPaLM: A Large Language Model That Can Speak and Listen},
  author={Paul K. Rubenstein and Chulayuth Asawaroengchai and Duc Dung Nguyen and Ankur Bapna and Zal{\'a}n Borsos and F{\'e}lix de Chaumont Quitry and Peter Chen and Dalia El Badawy and Wei Han and Eugene Kharitonov and Hannah Muckenhirn and Dirk Ryan Padfield and James Qin and Daniel Rozenberg and Tara N. Sainath and Johan Schalkwyk and Matthew Sharifi and Michelle D. Tadmor and Ramanovich and Marco Tagliasacchi and Alexandru Tudor and Mihajlo Velimirovi'c and Damien Vincent and Jiahui Yu and Yongqiang Wang and Victoria Zayats and Neil Zeghidour and Yu Zhang and Zhishuai Zhang and Luk{\'a}s Zilka and Christian Havn{\o} Frank},
  journal={ArXiv},
  year={2023},
  volume={abs/2306.12925},
  url={https://arxiv.org/abs/2306.12925}
}

@inproceedings{
dong2024polyvoice,
title={PolyVoice: Language Models for Speech to Speech Translation},
author={Qian qian Dong and Zhiying Huang and Qiao Tian and Chen Xu and Tom Ko and yunlong zhao and Siyuan Feng and Tang Li and Kexin Wang and Xuxin Cheng and Fengpeng Yue and Ye Bai and Xi Chen and Lu Lu and Zejun MA and Yuping Wang and Mingxuan Wang and Yuxuan Wang},
booktitle={The Twelfth International Conference on Learning Representations},
year={2024},
url={https://openreview.net/forum?id=hCrFG9cyuC}
}

@inproceedings{jia2019direct,
  title={Direct Speech-to-Speech Translation with a Sequence-to-Sequence Model},
  author={Jia, Ye and Weiss, Ron J and Biadsy, Fadi and Macherey, Wolfgang and Johnson, Melvin and Chen, Zhifeng and Wu, Yonghui},
  booktitle={Proc. Interspeech 2019},
  pages={1123--1127},
  year={2019}
}

@inproceedings{lee2022direct,
    title = "Direct Speech-to-Speech Translation With Discrete Units",
    author = "Lee, Ann  and
      Chen, Peng-Jen  and
      Wang, Changhan  and
      Gu, Jiatao  and
      Popuri, Sravya  and
      Ma, Xutai  and
      Polyak, Adam  and
      Adi, Yossi  and
      He, Qing  and
      Tang, Yun  and
      Pino, Juan  and
      Hsu, Wei-Ning",
    editor = "Muresan, Smaranda  and
      Nakov, Preslav  and
      Villavicencio, Aline",
    booktitle = "Proceedings of the 60th Annual Meeting of the Association for Computational Linguistics (Volume 1: Long Papers)",
    month = may,
    year = "2022",
    address = "Dublin, Ireland",
    publisher = "Association for Computational Linguistics",
    url = "https://aclanthology.org/2022.acl-long.235/",
    doi = "10.18653/v1/2022.acl-long.235",
    pages = "3327--3339"
}

@inproceedings{lee2022textless,
    title = "Textless Speech-to-Speech Translation on Real Data",
    author = "Lee, Ann  and
      Gong, Hongyu  and
      Duquenne, Paul-Ambroise  and
      Schwenk, Holger  and
      Chen, Peng-Jen  and
      Wang, Changhan  and
      Popuri, Sravya  and
      Adi, Yossi  and
      Pino, Juan  and
      Gu, Jiatao  and
      Hsu, Wei-Ning",
    editor = "Carpuat, Marine  and
      de Marneffe, Marie-Catherine  and
      Meza Ruiz, Ivan Vladimir",
    booktitle = "Proceedings of the 2022 Conference of the North American Chapter of the Association for Computational Linguistics: Human Language Technologies",
    month = jul,
    year = "2022",
    address = "Seattle, United States",
    publisher = "Association for Computational Linguistics",
    url = "https://aclanthology.org/2022.naacl-main.63/",
    doi = "10.18653/v1/2022.naacl-main.63",
    pages = "860--872"
}

@inproceedings{zhang2021uwspeech,
  title={Uwspeech: Speech to speech translation for unwritten languages},
  author={Zhang, Chen and Tan, Xu and Ren, Yi and Qin, Tao and Zhang, Kejun and Liu, Tie-Yan},
  booktitle={Proceedings of the AAAI Conference on Artificial Intelligence},
  volume={35},
  number={16},
  pages={14319--14327},
  year={2021}
}

@inproceedings{chen2023speech,
    title = "Speech-to-Speech Translation for a Real-world Unwritten Language",
    author = "Chen, Peng-Jen  and
      Tran, Kevin  and
      Yang, Yilin  and
      Du, Jingfei  and
      Kao, Justine  and
      Chung, Yu-An  and
      Tomasello, Paden  and
      Duquenne, Paul-Ambroise  and
      Schwenk, Holger  and
      Gong, Hongyu  and
      Inaguma, Hirofumi  and
      Popuri, Sravya  and
      Wang, Changhan  and
      Pino, Juan  and
      Hsu, Wei-Ning  and
      Lee, Ann",
    editor = "Rogers, Anna  and
      Boyd-Graber, Jordan  and
      Okazaki, Naoaki",
    booktitle = "Findings of the Association for Computational Linguistics: ACL 2023",
    month = jul,
    year = "2023",
    address = "Toronto, Canada",
    publisher = "Association for Computational Linguistics",
    url = "https://aclanthology.org/2023.findings-acl.307/",
    doi = "10.18653/v1/2023.findings-acl.307",
    pages = "4969--4983"
}

@inproceedings{zhang-etal-2024-streamspeech,
    title = "{S}tream{S}peech: Simultaneous Speech-to-Speech Translation with Multi-task Learning",
    author = "Zhang, Shaolei  and
      Fang, Qingkai  and
      Guo, Shoutao  and
      Ma, Zhengrui  and
      Zhang, Min  and
      Feng, Yang",
    editor = "Ku, Lun-Wei  and
      Martins, Andre  and
      Srikumar, Vivek",
    booktitle = "Proceedings of the 62nd Annual Meeting of the Association for Computational Linguistics (Volume 1: Long Papers)",
    month = aug,
    year = "2024",
    address = "Bangkok, Thailand",
    publisher = "Association for Computational Linguistics",
    url = "https://aclanthology.org/2024.acl-long.485/",
    doi = "10.18653/v1/2024.acl-long.485",
    pages = "8964--8986"
}

@inproceedings{dong22b_interspeech,
  title     = {{Leveraging Pseudo-labeled Data to Improve Direct Speech-to-Speech Translation}},
  author    = {Qianqian Dong and Fengpeng Yue and Tom Ko and Mingxuan Wang and Qibing Bai and Yu Zhang},
  year      = {2022},
  booktitle = {{Interspeech 2022}},
  pages     = {1781--1785},
  doi       = {10.21437/Interspeech.2022-10011},
  issn      = {2958-1796},
}

@inproceedings{popuri22_interspeech,
  title     = {{Enhanced Direct Speech-to-Speech Translation Using Self-supervised Pre-training and Data Augmentation}},
  author    = {Sravya Popuri and Peng-Jen Chen and Changhan Wang and Juan Pino and Yossi Adi and Jiatao Gu and Wei-Ning Hsu and Ann Lee},
  year      = {2022},
  booktitle = {{Interspeech 2022}},
  pages     = {5195--5199},
  doi       = {10.21437/Interspeech.2022-11032},
  issn      = {2958-1796},
}

@INPROCEEDINGS{Translatotron,
  author={Nachmani, Eliya and Levkovitch, Alon and Ding, Yifan and Asawaroengchai, Chulayuth and Zen, Heiga and Ramanovich, Michelle Tadmor},
  booktitle={ICASSP 2024 - 2024 IEEE International Conference on Acoustics, Speech and Signal Processing (ICASSP)}, 
  title={Translatotron 3: Speech to Speech Translation with Monolingual Data}, 
  year={2024},
  volume={},
  number={},
  pages={10686-10690},
  doi={10.1109/ICASSP48485.2024.10448426}}

@misc{salamanca2026tinyayabridgingscale,
      title={Tiny Aya: Bridging Scale and Multilingual Depth}, 
      author={Alejandro R. Salamanca and Diana Abagyan and Daniel D'souza and Ammar Khairi and David Mora and Saurabh Dash and Viraat Aryabumi and Sara Rajaee and Mehrnaz Mofakhami and Ananya Sahu and Thomas Euyang and Brittawnya Prince and Madeline Smith and Hangyu Lin and Acyr Locatelli and Sara Hooker and Tom Kocmi and Aidan Gomez and Ivan Zhang and Phil Blunsom and Nick Frosst and Joelle Pineau and Beyza Ermis and Ahmet Üstün and Julia Kreutzer and Marzieh Fadaee},
      year={2026},
      eprint={2603.11510},
      archivePrefix={arXiv},
      primaryClass={cs.CL},
      url={https://arxiv.org/abs/2603.11510}, 
}

@article{Costajuss2024ScalingNM,
  title={Scaling neural machine translation to 200 languages},
  author={Marta Ruiz Costa-juss{\`a} and James Cross and Onur Çelebi and Maha Elbayad and Kenneth Heafield and Kevin Heffernan and Elahe Kalbassi and Janice Lam and Daniel Licht and Jean Maillard and Anna Sun and Skyler Wang and Guillaume Wenzek and Al Youngblood and Bapi Akula and Lo{\"i}c Bar-rault and Gabriel Mejia Gonzalez and Prangthip Hansanti and John Hoffman and Semarley Jarrett and Kaushik Ram Sadagopan and Dirk Rowe and Shannon Spruit and C. Tran and Pierre Andrews and Necip Fazil Ayan and Shruti Bhosale and Sergey Edunov and Angela Fan and Cynthia Gao and Vedanuj Goswami and Francisco Guzm{\'a}n and Philipp Koehn and Alex Mourachko and Christophe Ropers and Safiyyah Saleem and Holger Schwenk and Jeff Wang},
  journal={Nature},
  year={2024},
  volume={630},
  pages={841 - 846},
  url={https://www.nature.com/articles/s41586-024-07335-x}
}

@article{flores,
    title = "The {F}lores-101 Evaluation Benchmark for Low-Resource and Multilingual Machine Translation",
    author = "Goyal, Naman  and
      Gao, Cynthia  and
      Chaudhary, Vishrav  and
      Chen, Peng-Jen  and
      Wenzek, Guillaume  and
      Ju, Da  and
      Krishnan, Sanjana  and
      Ranzato, Marc{'}Aurelio  and
      Guzm{\'a}n, Francisco  and
      Fan, Angela",
    editor = "Roark, Brian  and
      Nenkova, Ani",
    journal = "Transactions of the Association for Computational Linguistics",
    volume = "10",
    year = "2022",
    address = "Cambridge, MA",
    publisher = "MIT Press",
    url = "https://aclanthology.org/2022.tacl-1.30/",
    doi = "10.1162/tacl_a_00474",
    pages = "522--538"
}

@article{MT-society,
title = {Overview and challenges of machine translation for contextually appropriate translations},
journal = {iScience},
volume = {27},
number = {10},
pages = {110878},
year = {2024},
issn = {2589-0042},
doi = {https://doi.org/10.1016/j.isci.2024.110878},
url = {https://www.sciencedirect.com/science/article/pii/S2589004224021035},
author = {Palanichamy Naveen and Pavel Trojovský}
}

@inproceedings{federmann-etal-2022-ntrex,
    title = "{NTREX}-128 {--} News Test References for {MT} Evaluation of 128 Languages",
    author = "Federmann, Christian  and
      Kocmi, Tom  and
      Xin, Ying",
    editor = "Ahuja, Kabir  and
      Anastasopoulos, Antonios  and
      Patra, Barun  and
      Neubig, Graham  and
      Choudhury, Monojit  and
      Dandapat, Sandipan  and
      Sitaram, Sunayana  and
      Chaudhary, Vishrav",
    booktitle = "Proceedings of the First Workshop on Scaling Up Multilingual Evaluation",
    month = nov,
    year = "2022",
    address = "Online",
    publisher = "Association for Computational Linguistics",
    url = "https://aclanthology.org/2022.sumeval-1.4/",
    doi = "10.18653/v1/2022.sumeval-1.4",
    pages = "21--24"
}

@inproceedings{adelani-etal-2022-thousand,
    title = "A Few Thousand Translations Go a Long Way! Leveraging Pre-trained Models for {A}frican News Translation",
    author = "Adelani, David Ifeoluwa  and
      Alabi, Jesujoba Oluwadara  and
      Fan, Angela  and
      Kreutzer, Julia  and
      Shen, Xiaoyu  and
      Reid, Machel  and
      Ruiter, Dana  and
      Klakow, Dietrich  and
      Nabende, Peter  and
      Chang, Ernie  and
      Gwadabe, Tajuddeen  and
      Sackey, Freshia  and
      Dossou, Bonaventure F. P.  and
      Emezue, Chris  and
      Leong, Colin  and
      Beukman, Michael  and
      Muhammad, Shamsuddeen H.  and
      Jarso, Guyo D.  and
      Yousuf, Oreen  and
      Niyongabo Rubungo, Andre N.  and
      Hacheme, Gilles  and
      Wairagala, Eric Peter  and
      Nasir, Muhammad Umair  and
      Ajibade, Benjamin A.  and
      Ajayi, Tunde Oluwaseyi  and
      Gitau, Yvonne Wambui  and
      Abbott, Jade  and
      Ahmed, Mohamed  and
      Ochieng, Millicent  and
      Aremu, Anuoluwapo  and
      Ogayo, Perez  and
      Mukiibi, Jonathan  and
      Ouoba Kabore, Fatoumata  and
      Kalipe, Godson Koffi  and
      Mbaye, Derguene  and
      Tapo, Allahsera Auguste  and
      Memdjokam Koagne, Victoire M.  and
      Munkoh-Buabeng, Edwin  and
      Wagner, Valencia  and
      Abdulmumin, Idris  and
      Awokoya, Ayodele  and
      Buzaaba, Happy  and
      Sibanda, Blessing  and
      Bukula, Andiswa  and
      Manthalu, Sam",
    editor = "Carpuat, Marine  and
      de Marneffe, Marie-Catherine  and
      Meza Ruiz, Ivan Vladimir",
    booktitle = "Proceedings of the 2022 Conference of the North American Chapter of the Association for Computational Linguistics: Human Language Technologies",
    month = jul,
    year = "2022",
    address = "Seattle, United States",
    publisher = "Association for Computational Linguistics",
    url = "https://aclanthology.org/2022.naacl-main.223/",
    doi = "10.18653/v1/2022.naacl-main.223",
    pages = "3053--3070"
}

@inproceedings{li-etal-2025-ssa,
    title = "{SSA}-{COMET}: Do {LLM}s Outperform Learned Metrics in Evaluating {MT} for Under-Resourced {A}frican Languages?",
    author = "Li, Senyu  and
      Wang, Jiayi  and
      Ali, Felermino D. M. A.  and
      Cherry, Colin  and
      Deutsch, Daniel  and
      Briakou, Eleftheria  and
      Sousa-Silva, Rui  and
      Lopes Cardoso, Henrique  and
      Stenetorp, Pontus  and
      Adelani, David Ifeoluwa",
    editor = "Christodoulopoulos, Christos  and
      Chakraborty, Tanmoy  and
      Rose, Carolyn  and
      Peng, Violet",
    booktitle = "Proceedings of the 2025 Conference on Empirical Methods in Natural Language Processing",
    month = nov,
    year = "2025",
    address = "Suzhou, China",
    publisher = "Association for Computational Linguistics",
    url = "https://aclanthology.org/2025.emnlp-main.656/",
    doi = "10.18653/v1/2025.emnlp-main.656",
    pages = "12979--12998",
    ISBN = "979-8-89176-332-6"
}

@inproceedings{adelani-etal-2025-generative,
    title = "Does Generative {AI} speak {N}igerian-{P}idgin?: Issues about Representativeness and Bias for Multilingualism in {LLM}s",
    author = {Adelani, David Ifeoluwa  and
      Do{\u{g}}ru{\"o}z, A. Seza  and
      Shode, Iyanuoluwa  and
      Aremu, Anuoluwapo},
    editor = "Chiruzzo, Luis  and
      Ritter, Alan  and
      Wang, Lu",
    booktitle = "Findings of the Association for Computational Linguistics: NAACL 2025",
    month = apr,
    year = "2025",
    address = "Albuquerque, New Mexico",
    publisher = "Association for Computational Linguistics",
    url = "https://aclanthology.org/2025.findings-naacl.85/",
    doi = "10.18653/v1/2025.findings-naacl.85",
    pages = "1571--1583",
    ISBN = "979-8-89176-195-7"
}

@inproceedings{popovic-2015-chrf,
    title = "chr{F}: character n-gram {F}-score for automatic {MT} evaluation",
    author = "Popovi{\'c}, Maja",
    editor = "Bojar, Ond{\v{r}}ej  and
      Chatterjee, Rajan  and
      Federmann, Christian  and
      Haddow, Barry  and
      Hokamp, Chris  and
      Huck, Matthias  and
      Logacheva, Varvara  and
      Pecina, Pavel",
    booktitle = "Proceedings of the Tenth Workshop on Statistical Machine Translation",
    month = sep,
    year = "2015",
    address = "Lisbon, Portugal",
    publisher = "Association for Computational Linguistics",
    url = "https://aclanthology.org/W15-3049/",
    doi = "10.18653/v1/W15-3049",
    pages = "392--395"
}

@inproceedings{rei-etal-2020-comet,
    title = "{COMET}: A Neural Framework for {MT} Evaluation",
    author = "Rei, Ricardo  and
      Stewart, Craig  and
      Farinha, Ana C  and
      Lavie, Alon",
    editor = "Webber, Bonnie  and
      Cohn, Trevor  and
      He, Yulan  and
      Liu, Yang",
    booktitle = "Proceedings of the 2020 Conference on Empirical Methods in Natural Language Processing (EMNLP)",
    month = nov,
    year = "2020",
    address = "Online",
    publisher = "Association for Computational Linguistics",
    url = "https://aclanthology.org/2020.emnlp-main.213/",
    doi = "10.18653/v1/2020.emnlp-main.213",
    pages = "2685--2702"
}

@inproceedings{wang-etal-2024-afrimte,
    title = "{A}fri{MTE} and {A}fri{COMET}: Enhancing {COMET} to Embrace Under-resourced {A}frican Languages",
    author = "Wang, Jiayi  and
      Adelani, David Ifeoluwa  and
      Agrawal, Sweta  and
      Masiak, Marek  and
      Rei, Ricardo  and
      Briakou, Eleftheria  and
      Carpuat, Marine  and
      He, Xuanli  and
      Bourhim, Sofia  and
      Bukula, Andiswa  and
      Mohamed, Muhidin  and
      Olatoye, Temitayo  and
      Adewumi, Tosin  and
      Mokayed, Hamam  and
      Mwase, Christine  and
      Kimotho, Wangui  and
      Yuehgoh, Foutse  and
      Aremu, Anuoluwapo  and
      Ojo, Jessica  and
      Muhammad, Shamsuddeen Hassan  and
      Osei, Salomey  and
      Omotayo, Abdul-Hakeem  and
      Chukwuneke, Chiamaka  and
      Ogayo, Perez  and
      Hourrane, Oumaima  and
      El Anigri, Salma  and
      Ndolela, Lolwethu  and
      Mangwana, Thabiso  and
      Mohamed, Shafie Abdi  and
      Hassan, Ayinde  and
      Awoyomi, Oluwabusayo Olufunke  and
      Alkhaled, Lama  and
      Al-Azzawi, Sana  and
      Etori, Naome A.  and
      Ochieng, Millicent  and
      Siro, Clemencia  and
      Njoroge, Samuel  and
      Muchiri, Eric  and
      Kimotho, Wangari  and
      Wamba Momo, Lyse Naomi  and
      Abolade, Daud  and
      Ajao, Simbiat  and
      Shode, Iyanuoluwa  and
      Macharm, Ricky  and
      Iro, Ruqayya Nasir  and
      Abdullahi, Saheed S.  and
      Moore, Stephen E.  and
      Opoku, Bernard  and
      Akinjobi, Zainab  and
      Afolabi, Abeeb  and
      Obiefuna, Nnaemeka  and
      Ogbu, Onyekachi Raphael  and
      Brian, Sam  and
      Otiende, Verrah Akinyi  and
      Mbonu, Chinedu Emmanuel  and
      Toadoum Sari, Sakayo  and
      Lu, Yao  and
      Stenetorp, Pontus",
    editor = "Duh, Kevin  and
      Gomez, Helena  and
      Bethard, Steven",
    booktitle = "Proceedings of the 2024 Conference of the North American Chapter of the Association for Computational Linguistics: Human Language Technologies (Volume 1: Long Papers)",
    month = jun,
    year = "2024",
    address = "Mexico City, Mexico",
    publisher = "Association for Computational Linguistics",
    url = "https://aclanthology.org/2024.naacl-long.334/",
    doi = "10.18653/v1/2024.naacl-long.334",
    pages = "5997--6023"
}

@inproceedings{freitag-etal-2022-results,
    title = "Results of {WMT}22 Metrics Shared Task: Stop Using {BLEU} {--} Neural Metrics Are Better and More Robust",
    author = "Freitag, Markus  and
      Rei, Ricardo  and
      Mathur, Nitika  and
      Lo, Chi-kiu  and
      Stewart, Craig  and
      Avramidis, Eleftherios  and
      Kocmi, Tom  and
      Foster, George  and
      Lavie, Alon  and
      Martins, Andr{\'e} F. T.",
    editor = {Koehn, Philipp  and
      Barrault, Lo{\"i}c  and
      Bojar, Ond{\v{r}}ej  and
      Bougares, Fethi  and
      Chatterjee, Rajen  and
      Costa-juss{\`a}, Marta R.  and
      Federmann, Christian  and
      Fishel, Mark  and
      Fraser, Alexander  and
      Freitag, Markus  and
      Graham, Yvette  and
      Grundkiewicz, Roman  and
      Guzman, Paco  and
      Haddow, Barry  and
      Huck, Matthias  and
      Jimeno Yepes, Antonio  and
      Kocmi, Tom  and
      Martins, Andr{\'e}  and
      Morishita, Makoto  and
      Monz, Christof  and
      Nagata, Masaaki  and
      Nakazawa, Toshiaki  and
      Negri, Matteo  and
      N{\'e}v{\'e}ol, Aur{\'e}lie  and
      Neves, Mariana  and
      Popel, Martin  and
      Turchi, Marco  and
      Zampieri, Marcos},
    booktitle = "Proceedings of the Seventh Conference on Machine Translation (WMT)",
    month = dec,
    year = "2022",
    address = "Abu Dhabi, United Arab Emirates (Hybrid)",
    publisher = "Association for Computational Linguistics",
    url = "https://aclanthology.org/2022.wmt-1.2/",
    doi = "10.18653/v1/2022.wmt-1.2",
    pages = "46--68"
}

@misc{xu2025qwen3omnitechnicalreport,
      title={Qwen3-Omni Technical Report}, 
      author={Jin Xu and Zhifang Guo and Hangrui Hu and Yunfei Chu and Xiong Wang and Jinzheng He and Yuxuan Wang and Xian Shi and Ting He and Xinfa Zhu and Yuanjun Lv and Yongqi Wang and Dake Guo and He Wang and Linhan Ma and Pei Zhang and Xinyu Zhang and Hongkun Hao and Zishan Guo and Baosong Yang and Bin Zhang and Ziyang Ma and Xipin Wei and Shuai Bai and Keqin Chen and Xuejing Liu and Peng Wang and Mingkun Yang and Dayiheng Liu and Xingzhang Ren and Bo Zheng and Rui Men and Fan Zhou and Bowen Yu and Jianxin Yang and Le Yu and Jingren Zhou and Junyang Lin},
      year={2025},
      eprint={2509.17765},
      archivePrefix={arXiv},
      primaryClass={cs.CL},
      url={https://arxiv.org/abs/2509.17765}, 
}

@INPROCEEDINGS{10889444,
  author={Lu, Ke-Han and Chen, Zhehuai and Fu, Szu-Wei and Yang, Chao-Han Huck and Balam, Jagadeesh and Ginsburg, Boris and Wang, Yu-Chiang Frank and Lee, Hung-Yi},
  booktitle={ICASSP 2025 - 2025 IEEE International Conference on Acoustics, Speech and Signal Processing (ICASSP)}, 
  title={Developing Instruction-Following Speech Language Model Without Speech Instruction-Tuning Data}, 
  year={2025},
  volume={},
  number={},
  pages={1-5},
  doi={10.1109/ICASSP49660.2025.10889444}}

@inproceedings{conneau22_interspeech,
  title     = {{XTREME-S: Evaluating Cross-lingual Speech Representations}},
  author    = {Alexis Conneau and Ankur Bapna and Yu Zhang and Min Ma and Patrick {von Platen} and Anton Lozhkov and Colin Cherry and Ye Jia and Clara Rivera and Mihir Kale and Daan {van Esch} and Vera Axelrod and Simran Khanuja and Jonathan Clark and Orhan Firat and Michael Auli and Sebastian Ruder and Jason Riesa and Melvin Johnson},
  year      = {2022},
  booktitle = {{Interspeech 2022}},
  pages     = {3248--3252},
  doi       = {10.21437/Interspeech.2022-10007},
  issn      = {2958-1796},
}

@inproceedings{bentivogli-etal-2021-cascade,
    title = "Cascade versus Direct Speech Translation: Do the Differences Still Make a Difference?",
    author = "Bentivogli, Luisa  and
      Cettolo, Mauro  and
      Gaido, Marco  and
      Karakanta, Alina  and
      Martinelli, Alberto  and
      Negri, Matteo  and
      Turchi, Marco",
    editor = "Zong, Chengqing  and
      Xia, Fei  and
      Li, Wenjie  and
      Navigli, Roberto",
    booktitle = "Proceedings of the 59th Annual Meeting of the Association for Computational Linguistics and the 11th International Joint Conference on Natural Language Processing (Volume 1: Long Papers)",
    month = aug,
    year = "2021",
    address = "Online",
    publisher = "Association for Computational Linguistics",
    url = "https://aclanthology.org/2021.acl-long.224/",
    doi = "10.18653/v1/2021.acl-long.224",
    pages = "2873--2887"
}

@inproceedings{dione-etal-2023-masakhapos,
    title = "{M}asakha{POS}: Part-of-Speech Tagging for Typologically Diverse {A}frican languages",
    author = "Dione, Cheikh M. Bamba  and
      Adelani, David Ifeoluwa  and
      Nabende, Peter  and
      Alabi, Jesujoba  and
      Sindane, Thapelo  and
      Buzaaba, Happy  and
      Muhammad, Shamsuddeen Hassan  and
      Emezue, Chris Chinenye  and
      Ogayo, Perez  and
      Aremu, Anuoluwapo  and
      Gitau, Catherine  and
      Mbaye, Derguene  and
      Mukiibi, Jonathan  and
      Sibanda, Blessing  and
      Dossou, Bonaventure F. P.  and
      Bukula, Andiswa  and
      Mabuya, Rooweither  and
      Tapo, Allahsera Auguste  and
      Munkoh-Buabeng, Edwin  and
      Memdjokam Koagne, Victoire  and
      Ouoba Kabore, Fatoumata  and
      Taylor, Amelia  and
      Kalipe, Godson  and
      Macucwa, Tebogo  and
      Marivate, Vukosi  and
      Gwadabe, Tajuddeen  and
      Elvis, Mboning Tchiaze  and
      Onyenwe, Ikechukwu  and
      Atindogbe, Gratien  and
      Adelani, Tolulope  and
      Akinade, Idris  and
      Samuel, Olanrewaju  and
      Nahimana, Marien  and
      Musabeyezu, Th{\'e}og{\`e}ne  and
      Niyomutabazi, Emile  and
      Chimhenga, Ester  and
      Gotosa, Kudzai  and
      Mizha, Patrick  and
      Agbolo, Apelete  and
      Traore, Seydou  and
      Uchechukwu, Chinedu  and
      Yusuf, Aliyu  and
      Abdullahi, Muhammad  and
      Klakow, Dietrich",
    editor = "Rogers, Anna  and
      Boyd-Graber, Jordan  and
      Okazaki, Naoaki",
    booktitle = "Proceedings of the 61st Annual Meeting of the Association for Computational Linguistics (Volume 1: Long Papers)",
    month = jul,
    year = "2023",
    address = "Toronto, Canada",
    publisher = "Association for Computational Linguistics",
    url = "https://aclanthology.org/2023.acl-long.609/",
    doi = "10.18653/v1/2023.acl-long.609",
    pages = "10883--10900"
}

@inproceedings{adelani2022few,
  title={A few thousand translations go a long way! Leveraging pre-trained models for African news translation},
  author={Adelani, David Ifeoluwa and Alabi, Jesujoba and Fan, Angela and Kreutzer, Julia and Shen, Xiaoyu and Reid, Machel and Ruiter, Dana and Klakow, Dietrich and Nabende, Peter and Chang, Ernie and others},
  booktitle={Proceedings of the 2022 Conference of the North American Chapter of the Association for Computational Linguistics: Human Language Technologies},
  pages={3053--3070},
  year={2022}
}

@inproceedings{adelani-etal-2024-sib,
    title = "{SIB}-200: A Simple, Inclusive, and Big Evaluation Dataset for Topic Classification in 200+ Languages and Dialects",
    author = "Adelani, David Ifeoluwa  and
      Liu, Hannah  and
      Shen, Xiaoyu  and
      Vassilyev, Nikita  and
      Alabi, Jesujoba O.  and
      Mao, Yanke  and
      Gao, Haonan  and
      Lee, En-Shiun Annie",
    editor = "Graham, Yvette  and
      Purver, Matthew",
    booktitle = "Proceedings of the 18th Conference of the European Chapter of the Association for Computational Linguistics (Volume 1: Long Papers)",
    month = mar,
    year = "2024",
    address = "St. Julian{'}s, Malta",
    publisher = "Association for Computational Linguistics",
    url = "https://aclanthology.org/2024.eacl-long.14/",
    doi = "10.18653/v1/2024.eacl-long.14",
    pages = "226--245"
}

\appendix

\section{British English Dataset Addition}\label{sec:british_english}

\subsection{Dataset}
\label{British_dataset}
Using the same English data from our accented-English dataset, we record a new British English test set and synthesize the train set using a mixture of 10 Gemini 2.5 TTS voices. 
We provide our own recording tool to the British annotators, who were instructed the speakers to record on their laptops with a good microphone in a quiet environment. We recruited one female and three males speakers through our community, who are native Southern British English speakers. Each volunteer recorded 250 utterances, covering the 1000 necessary for the dev and test sets. The speakers were aware that their recordings would be made public, but used only for research purposes, as were the Nigerian languages participants. 

\subsection{Experiments with Accents}
\label{sec:accents_appendix_british_additional}
We further investigate the difference in performance between Nigerian and British accented English, the best cascaded and AudioLLM methods are evaluated (Table \ref{stt_results_ssa_comet_audio_llm}) on the newly created dataset mentioned in Section \ref{British_dataset}.

The results in Table \ref{stt_brit_vs_naija} highlight clear differences in robustness to British- and Naijá-accented English inputs across model architectures. The strongest cascaded approaches are substantially affected by accent variation, with performance generally improving under British-accented input compared to Naijá-accented speech. In contrast, the best-performing AudioLLM remains comparatively stable across accent conditions, achieving similar performance for Hausa, Igbo, and Naijá translation under both input varieties. The primary exception is \yoruba, where performance decreases notably for British-accented input despite achieving the strongest overall results under Naijá-accented speech. This finding suggests that accent alignment between the source speech and target language context may particularly benefit \yoruba translation. Additionally, the performance gap between zero-shot and few-shot prompting is smaller for British-accented inputs, indicating that in-context examples provide comparatively limited gains in this setting. Overall, these results suggest that end-to-end AudioLLM approaches exhibit substantially greater robustness to accent variation than cascaded ASR+MT pipelines. 

\begin{table}[h]
\centering
\small
\setlength{\tabcolsep}{4.5pt}
\resizebox{\columnwidth}{!}{
\begin{tabular}{lllrrrrr}
\toprule
    &&&\multicolumn{5}{c}{\textbf{Eng $\rightarrow$ XX}} \\
    
\cmidrule(lr){4-8}
        
\textbf{Model} & \textbf{Method}&
& \textbf{Hausa} & \textbf{Igbo} & \textbf{Naijá} & \textbf{\yoruba} & \textbf{Avg.}  \\

\midrule
\cellcolor[gray]{0.92}\texttt{British-accent} \\
    \multirow{2}{*}{\begin{tabular}[c]{@{}c@{}} \textsc{Cascaded} \\ \texttt{(ASR + MT)}\end{tabular}}
    & \multirow{2}{*}{Omnilingual-ASR}
        & + NLLB 
       & 52.6 & 58.4 & 10.9 & 63.2 & 46.3\\ 
    & 
        & + Tiny Aya 
        & 49.3 & 53.2 & 45.4 & 44.9 & 48.2 \\
    \midrule

\cellcolor[gray]{0.92}\texttt{Naij\'a-accent} \\
    \multirow{2}{*}{\begin{tabular}[c]{@{}c@{}} \textsc{Cascaded} \\ \texttt{(ASR + MT)}\end{tabular}}
    & \multirow{2}{*}{Omnilingual-ASR}
        & + NLLB 
       & 47.6 & 52.6 & 9.7 & 58.0 & 42.0 \\ 
    & 
        & + Tiny Aya 
        & 39.0 & 40.0 & 15.7 & 26.3 & 30.3\\
    \midrule

\cellcolor[gray]{0.92}\texttt{British-accent} \\
\multirow{2}{*}{\textsc{AudioLLM}}
& \multirow{2}{*}{Gemini 3.1}
    & Zero-Shot
    & 68.2 & 67.2 &  \underline{62.7} & 63.1 & 65.3 \\
    &
    & Few-Shot
        & \textbf{68.9} & \textbf{67.7} & 61.8 & 63.0 & 65.4\\

\cellcolor[gray]{0.92}\texttt{Naij\'a-accent} \\
\multirow{2}{*}{\textsc{AudioLLM}}
& \multirow{2}{*}{Gemini 3.1}
    & Zero-Shot
    & \underline{68.3} & 67.1 & 61.5 & \underline{72.1} & \underline{67.3}\\
    &
    & Few-Shot
        & \underline{68.3} & \underline{67.4} & \textbf{62.8} & \textbf{72.3} & \textbf{67.7}  \\

\bottomrule
\end{tabular}}
\caption{\textbf{Speech-to-text translation results (SSA-COMET $\uparrow$) for AudioLLM and Cascade evaluation} for British English $\rightarrow$ XX. \textit{Italics} indicate best within each method; \textbf{bold} indicates best overall while the second best is \underline{underlined}.}
\label{stt_brit_vs_naija}
\end{table}

\paragraph{Speech-to-Speech Evaluation}
Omnilingual-ASR 1B model is used to transcribe speech outputs, on which SSA-COMET and ChrF++ scores are computed. We also finetune Omnilingual on our train and dev sets, to create Naij\'a-Omni.
To analyze the impact of the ASR system on the evaluation, we also synthesize the text translation to RP British English, and compare results between British and Naij\'a-accented speech and the base Omnilingual and Naij\'a-Omni. As can be seen in Table \ref{sts_results_omni_ssa_comet}, Naij\'a-Omni generally improves results both for the Nigerian and the British accented speech translations. Interestingly, British TTS still gets better results than Naij\'a TTS using Naij\'a-Omni. These results indicate that biases introduced at the TTS stage propagate into evaluation metrics, motivating further investigation into accent-invariant evaluation strategies for S2ST.

\begin{table}[t]
\centering
\small
\setlength{\tabcolsep}{4.5pt}
\resizebox{\columnwidth}{!}{%
\begin{tabular}{
    l l l
    S[table-format=2.1]
    S[table-format=2.1]
    S[table-format=2.1]
    S[table-format=2.1]
    S[table-format=2.1]
    S[table-format=2.1]
    S[table-format=2.1]
    }
\toprule
    & &
    & \multicolumn{4}{c}{\textbf{XX $\rightarrow$ Eng}} \\
    
\cmidrule(lr){4-7} 
        
\textbf{Model} & \textbf{Output Accent} & \textbf{ASR model} & {Hausa} & {Igbo} & {Naijá} & {\yoruba} \\

\midrule
\addlinespace
    \multirow{4}{*}{Gemini 3.1}
    & \multirow{2}{*}{Naij\'a} & + Naij\'a-Omni
     & 61.8 & 50.7 & 60.0 & 65.4 \\
    & & + Omni
    & 60.8 & 50.0 & 60.7 & 58.6 \\
 \addlinespace

    &\multirow{2}{*}{British} 
    & + Naij\'a-Omni
    &  63.2 & 51.4 & 62.6 & 60.3 \\
    && + Omni 
    & 63.3 & 51.2 & 62.4 & 60.7  \\

\bottomrule
\end{tabular}
}
\caption{S2ST ASR-COMET ($\uparrow$) results computed with finetuned Naij\'a-Omni ASR and the original Omnilingual-LLM-1B (Omni).}
\label{sts_results_omni_ssa_comet}
\end{table}

\section{Recording Instructions}
\label{sec:recording}
Participants were instructed to record in quiet environments, speak clearly, and avoid word repetitions. Following the initial recordings, feedback was provided on audio quality, along with guidance on segments requiring re-recording.

\section{License}
\vspace{-4pt}
All resources released as part of this work are distributed under the CC-BY license.

\section{Prompt Templates}
\label{sec:prompting}

\begin{promptbox}
[PROMPT CONTENTS]
You are a translation expert. Here are {len(fewshot_files)} examples of {language} audio transcribed, then translated into English. Following these examples, transcribe the last given audio, and use the transcription to provide its exact English translation. Return only the English translation without any additional text.
[AUDIO] .wav
\end{promptbox}
\captionof{figure}{Prompt template used for LRL$\rightarrow$English speech-to-text translation in Gemini 2.5 and Gemini 3.1.}
\label{fig:prompt2}

\begin{promptbox}
[USER]
You are a professional translator. Here are {number_of_few_shot_examples} examples of {language} transcriptions and their English translations:\n{examples_str}\n\n"
Following these examples, translate the following {language} transcription to English.
Only output the English translation without any additional text or formatting.{input_text}
\end{promptbox}
\captionof{figure}{Prompt template used for LRL $\rightarrow$ Eng MT for Tiny Aya.}
\label{fig:prompt3}

\label{prompting}
\begin{promptbox}
[SYSTEM]
You are a translation expert. Here are {len(fewshot_examples)} examples of {language} audio transcribed, then translated into English.Following these examples, transcribe the last given audio, and use the transcription to provide its exact English translation. Return only the English translation without any additional text."
[USER]
Translate this {language} audio into English.
[AUDIO]: .wav
\end{promptbox}
\captionof{figure}{Prompt template used for LRL$\rightarrow$English speech-to-text translation for GPT-Audio}
\label{fig:prompt1}

\begin{table*}[t]
\centering
\small
\setlength{\tabcolsep}{2.8pt}
\begin{tabular}{
    lcl|rrcrr|rrcrr
    }
\toprule
    & 
    & 
    & \multicolumn{5}{c|}{\textbf{XX $\rightarrow$ Eng}} 
    & \multicolumn{5}{c}{\textbf{Eng $\rightarrow$ XX}} \\
    
\cmidrule(lr){4-8} \cmidrule(lr){9-12}
        
\textbf{ Method} & \multicolumn{2}{c}{\textbf{Model}}
& \textbf{Hausa} & \textbf{Igbo} & \textbf{Naijá} & \textbf{\yoruba} & \textbf{Avg.} 
& \textbf{Hausa} & \textbf{Igbo} & \textbf{Naijá} & \textbf{\yoruba} & \textbf{Avg.} \\

\midrule
    \multirow{2}{*}{\begin{tabular}[c]{@{}c@{}} \textsc{Cascaded} \\ \texttt{(ASR + MT)}\end{tabular}}
    & \multirow{2}{*}{Omnilingual-ASR}
        & + NLLB 
        & \underline{17.3} & 11.0 & 16.1 & 17.0 & 11.1 & \textbf{16.7} & 29.2 & 0.0 & 23.4 & 17.32\\ 
    & 
        & + Tiny Aya 
        & 12.4 & 6.2 & 4.3 & 10.9 & 8.45 & \underline{16.2} & 23.8 & 1.1 & 18.7 & 15.0\\
        \midrule
    
\addlinespace
    \multirow{3}{*}{\textsc{End-to-End}} & \multirow{3}{*}{SeamlessM4T}
          & Zero-Shot 
        & 1.3 & 4.0 & {N/A} & 19.5 & 12.4 & {N/A} & 7.5 & {N/A} & 2.5 & 5.0\\
        && Mono FT
        & \textbf{18.6} & \textbf{17.6} & \textbf{1.2} & 21.1 & 14.7 & 1.1 & \underline{37.2} & 0.3 & \underline{30.0} & 22.9 \\ 
        && Multi FT
        & 13.9 & \underline{14.8} & \textbf{18.5} & 15.8 & 15.8 & 1.5 & \textbf{40.0} & 1.2 & \textbf{30.3} & 23.8\\

\bottomrule
\end{tabular}
\caption{\textbf{Speech-to-text translation results (SpBLEU $\uparrow$)}.
\textit{Italics} indicate best within each method; \textbf{bold} indicates best overall while \underline{underlined} indicate second best result. Multilingual fine-tuning (FT) is a model fine-tuned across all the Nigerian languages data.}
\label{stt_results_spbleu}
\end{table*}

\section{ChrF++ and SpBLEU Results}
\label{Other_Metrics}

In addition to the SSA-COMET results reported previously, we also provide ChrF++ scores for other S2TT experiments. Furthermore, we include \textbf{SpBLEU}, which is a metric designed for machine translation that evaluates translation quality. We also report a variant, \textbf{ASR-SpBLEU}, which applies the same evaluation procedure on ASR-transcribed hypotheses, thereby capturing the compounded effects of recognition and translation errors in cascaded or speech-based pipelines. Table \ref{stt_results_spbleu} through Table \ref{sts_results_chrf} explicitly detail the comprehensive SpBLEU, ChrF++, and ASR-SpBLEU metrics. They serve to provide a highly granular breakdown of the evaluation settings discussed in Section \ref{sec:results}, showing the performance variations across both Naij\'a-accented and British-accented evaluations.

\begin{table*}[htbp!]
\centering
\small
\setlength{\tabcolsep}{6pt}
\begin{tabular}{
    llrrrrrrrrrr
    }
\toprule
    & 
    & \multicolumn{5}{c}{\textbf{XX $\rightarrow$ Eng}} 
    & \multicolumn{5}{c}{\textbf{Eng $\rightarrow$ XX}} \\
    
\cmidrule(lr){3-7} \cmidrule(lr){8-12}
        
\textbf{Model} & \textbf{Method}
& \textbf{Hausa} & \textbf{Igbo} & \textbf{Naijá} & \textbf{\yoruba} & \textbf{Avg.} 
& \textbf{Hausa} & \textbf{Igbo} & \textbf{Naijá} & \textbf{\yoruba} & \textbf{Avg.}  \\

\midrule
    \multirow{3}{*}{SeamlessM4T}
        & Zero-Shot 
        & 1.3 & 4.0 & 7.4 & 8.3 & 5.3 & {N/A} & 7.5 & {N/A} & 2.5 & -\\
        & Mono FT
        & 18.6 & 17.6 & 1.2 & 21.1 & 14.7 & 1.1 & \underline{37.2} & 0.3 & \underline{30.0} & 17.2 \\ 
        & Multi FT
        & 13.9 & 14.8 & 18.5 & 15.8 & 16.1 & 1.5 & \textbf{40.0} & 1.5& \textbf{30.3} & 18.3\\
    
\addlinespace 
    \multirow{2}{*}{Gemini 2.5}
        & Zero-Shot 
        & 19.2 & 7.2 & 12.5 & 13.5 & 13.2 & 26.7 & 37.4 & 5.7 & 31.2 & 25.3 \\
        & Few-Shot 
        & 23.3 & 10.9 & 16.5 & 17.0 & 16.9 & 36.1 & 5.2 & \textbf{25.2} & 29.1 & 23.9\\ 

\addlinespace
    \multirow{2}{*}{Gemini 3.1}
        & Zero-Shot 
        & \underline{30.0} & \underline{19.6} & \underline{28.4} & 21.1 & \underline{24.8} & \underline{30.3} & 39.0 & 6.6 & \underline{35.6} & \underline{27.9} \\
        & Few-Shot 
        & \textbf{35.6} & \textbf{25.2} & \textbf{33.3} & \textbf{28.4} & \textbf{38.7} & \textbf{32.4} & \textbf{40.6} & \underline{20.4} & \textbf{36.3} & \textbf{32.4} \\ 

\addlinespace
    \multirow{2}{*}{GPT-Audio}
        & Zero-Shot 
        & 4.4 & 4.6 & 16.2 & \underline{24.7} & 12.5 & 7.2 & 8.1 & 1.9 & 6.5 & 5.9\\
        & Few-Shot 
        & 4.3 & 4.4 & 16.3 & 5.6 & 7.7 & 25.2 & 29.8 & 5.8 & 13.6 & 18.6 \\

\bottomrule
\end{tabular}
\caption{\textbf{Speech-to-text translation results (SpBLEU $\uparrow$) for AudioLLM evaluation}, with a comparison to fully supervised fine tuning (SFT) (Seamless M4T).
\textbf{Bold} indicates best overall while the second best is \underline{underlined}.}
\label{stt_results_spbleu_audio_llm}
\end{table*}

\begin{table*}[h]
\centering
\small
\setlength{\tabcolsep}{5pt}
\begin{tabular}{
    llrrrrrrrrrr
    }
\toprule
    & 
    & \multicolumn{5}{c}{\textbf{XX $\rightarrow$ Eng}} 
    & \multicolumn{5}{c}{\textbf{Eng $\rightarrow$ XX}} \\
    
\cmidrule(lr){3-7} \cmidrule(lr){8-12}
        
\textbf{Model} & \textbf{Method}
& \textbf{Hausa} & \textbf{Igbo} & \textbf{Naijá} & \textbf{\yoruba} & \textbf{Avg.} 
& \textbf{Hausa} & \textbf{Igbo} & \textbf{Naijá} & \textbf{\yoruba} & \textbf{Avg.}  \\

\midrule
    \multirow{2}{*}{SeamlessM4T}
         & Zero-Shot 
        & 15.9 & 19.4 & 31.1 & 43.3 & 27.4 & {N/A} & 24.5 & N/A & 14.2 & - \\
        & Mono-FT
        & 41.9 & 39.2 & 18.1 & 43.5 & 35.7 & 13.7 & 54.7 & 6.1 & 37.0 & 27.9 \\ 
       & Multi-FT
       & 38.0 & 36.0 & 40.1 & 41.8  & 39.0 & 14.5 & 55.3 & 15.3 & 37.1 & 30.6 \\
    
\addlinespace 
    \multirow{2}{*}{Gemini 2.5}
        & Zero-Shot 
        & 46.5 & 30.0 & 38.9 & 37.4 & 38.2 & 49.3 & 53.1 & 28.5 & 32.4 & 40.8\\
        & Few-Shot 
        & 49.3 & 33.5 & 41.2 & 39.4 & 40.9 & 48.9 & 53.8 & 27.0 & 37.6 & 41.8\\ 

\addlinespace
    \multirow{2}{*}{Gemini 3.1}
        & Zero-Shot 
        & \underline{54.9} & \underline{41.5} & \underline{44.9} & \underline{50.8} & \underline{48.0} & \underline{53.0} & \underline{56.3} & \underline{28.8} & \underline{40.7}  &\underline{44.7}\\
        & Few-Shot 
        & \textbf{58.1} & \textbf{46.0} & \textbf{49.7} &\textbf{ 54.5} & \textbf{52.1} & \textbf{53.6} & \textbf{57.0} & \textbf{41.0} & \textbf{42.0} & \textbf{48.4} \\ 

\addlinespace 
    \multirow{2}{*}{Gemma 4}
        & Zero-Shot 
        & 17.0 & 15.6 & 29.9 & 17.9 & 20.1 & 33.2 & 21.8 & 22.0 & 12.8 & 22.5 \\
        & Few-Shot 
        & 24.7 & 15.9 & 30.2 & 18.0 & 22.2 & 33.5 & 22.1 & 22.2 & 13.0 & 22.7\\ 

\addlinespace
    \multirow{2}{*}{GPT-Audio}
        & Zero-Shot 
        & 24.2 & 22.3 & 39.9 & 24.7 & 27.8 & 23.9 & 22.7 & 18.9 & 14.1 & 19.9 \\
        & Few-Shot 
        & 24.1 & 22.0 & 40.1 & 24.4 & 27.7 & 48.1 & 47.2 & 28.0 & 24.1 & 36.9 \\

\bottomrule
\end{tabular}
\caption{\textbf{Speech-to-text translation results (ChrF++ $\uparrow$) for AudioLLM evaluation}, with a comparison to fully supervised fine tuning (SFT) (Seamless M4T).
\textbf{Bold} indicates best overall while the second best is \underline{underlined}.}
\label{stt_results_chrf_audio_llm}
\end{table*}

\begin{table*}[h]
\centering
\small
\setlength{\tabcolsep}{1.9pt}
\begin{tabular}{
   lll|rrrr|rrrr
    }
\toprule
    & &
    & \multicolumn{4}{c|}{\textbf{XX $\rightarrow$ Eng}} 
    & \multicolumn{4}{c}{\textbf{Eng $\rightarrow$ XX}} \\
    
\cmidrule(lr){4-7} \cmidrule(lr){8-10}
        
\textbf{Method} & \multicolumn{2}{c|}{\textbf{Model}} & {Hausa} & {Igbo} & Naijá & {\yoruba} & {Hausa} & {Igbo} & Naijá & {Yoruba} \\

\midrule

\cellcolor[gray]{0.92}\texttt{Naij\'a-accent} \\

 \multirow{1}{*}{End-to-End}
    & SeamlessM4T & Multilingual
    & 1.2 & 1.9 & {N/A} & 4.4 & & & \\
    
    \multirow{1}{*}{Cascaded}& \multirow{1}{*}{Omni + NLLB/TinyAya} & + Gemini 2.5 TTS Naij\'a 
    & 11.9 &\underline{7.8} & 0.1 & \underline{9.5} & {11.0} & \underline{6.4} & {0.3} & \underline{12.8} \\
   
    \multirow{1}{*}{AudioLLM}
    & \multirow{1}{*}{Gemini 3.1 Few-Shot}& + Gemini 2.5 TTS Naijá
    & \underline{18.8} & \underline{12.4} & 0.3 & \textbf{32.7} & \textbf{13.6} &\textbf{ 6.6} & \underline{1.2} & \textbf{14.2} \\
    \midrule
    \addlinespace
    
\cellcolor[gray]{0.92}\texttt{British-accent} \\

 \multirow{1}{*}{Cascaded}& \multirow{1}{*}{Omni + NLLB/TinyAya}
    & + Gemini 2.5 TTS British
    & 12.6 & 8.5 & \underline{0.4} & 10.5 & 11.4 & 4.6 & 0.2 & 9.6 \\
    
 \multirow{1}{*}{AudioLLM}
    & \multirow{1}{*}{Gemini 3.1 Few-Shot} & + Gemini 2.5 TTS British
     & \textbf{19.3} & \textbf{ 12.7} & \textbf{14.1} & \underline{17.7} & \textbf{12.2} & 5.6 &  \textbf{2.1} & \underline{14.0} \\
     
\bottomrule
\end{tabular}
\caption{\textbf{Speech-to-speech translation results (ASR-SpBLEU $\uparrow$)}.
\textbf{Bold} indicates best overall while the second best is \underline{underlined}. }
\label{sts_results_spbleu}
\end{table*}

\begin{table*}[h]
\centering
\small
\setlength{\tabcolsep}{1.5pt}
\begin{tabular}{
   lll|rrrr|rrrr
    }
\toprule
    & &
    & \multicolumn{4}{c|}{\textbf{XX $\rightarrow$ Eng}} 
    & \multicolumn{4}{c}{\textbf{Eng $\rightarrow$ XX}} \\
    
\cmidrule(lr){4-6} \cmidrule(lr){7-9}
        
\textbf{Method} & \multicolumn{2}{c|}{\textbf{Model}} & {Hausa} & {Igbo} & {Naijá} &{\yoruba} & {Hausa} & {Igbo} & {Naijá} & {Yoruba} \\

\midrule

\cellcolor[gray]{0.92}\texttt{Naij\'a-accent} \\

 \multirow{1}{*}{End-to-End}
    & SeamlessM4T & Multilingual
    & 21.4 & 22.5 & {N/A} & 27.0 & & & \\
    
    \multirow{1}{*}{Cascaded}& \multirow{1}{*}{Omni + NLLB/TinyAya} & + Gemini 2.5 TTS Naij\'a 
    & 41.7 &\underline{34.5} & 14.08 & 36.3 & \underline{}{35.7} & \underline{26.0}& \textbf{23.9} & {20.8} \\
   
    \multirow{1}{*}{AudioLLM}
    & \multirow{1}{*}{Gemini 3.1 Few-Shot}& + Gemini 2.5 TTS Naija
    & \textbf{47.9} & \textbf{38.0} & 18.7 &\textbf{53.7} & \textbf{40.5} & \textbf{27.3}& {22.1} & \underline{21.6} \\
    \midrule
    \addlinespace
    
\cellcolor[gray]{0.92}\texttt{British-accent} \\

 \multirow{1}{*}{Cascaded}& \multirow{1}{*}{Omni + NLLB}
    & + Gemini 2.5 TTS British
    & 38.2 & 31.8 & \underline{19.7} & 34.1 & 35.4 & 24.2 & 13.3 & {20.1} \\
    
 \multirow{1}{*}{AudioLLM}
    & \multirow{1}{*}{Gemini 3.1 Few-Shot} & + Gemini 2.5 TTS British
     & \underline{47.6} & \underline{37.8} & \textbf{40.3} & \underline{44.7} & \underline{37.0} & 25.5 & \underline{22.3} & \textbf{26.3} \\
     
\bottomrule
\end{tabular}
\caption{\textbf{Speech-to-speech translation results (ASR-ChrF $\uparrow$)}.
\textbf{Bold} indicates best overall while the second best is \underline{underlined}. }
\label{sts_results_chrf}
\end{table*}



\clearpage
\vspace{-10pt}
\clearpage

\section{POS Tags}
\label{sec:POS_appendix}

\begin{figure*}[!htbp]
  \centering
  \begin{subfigure}[b]{0.5\linewidth}
    \centering
    \includegraphics[width=\linewidth]{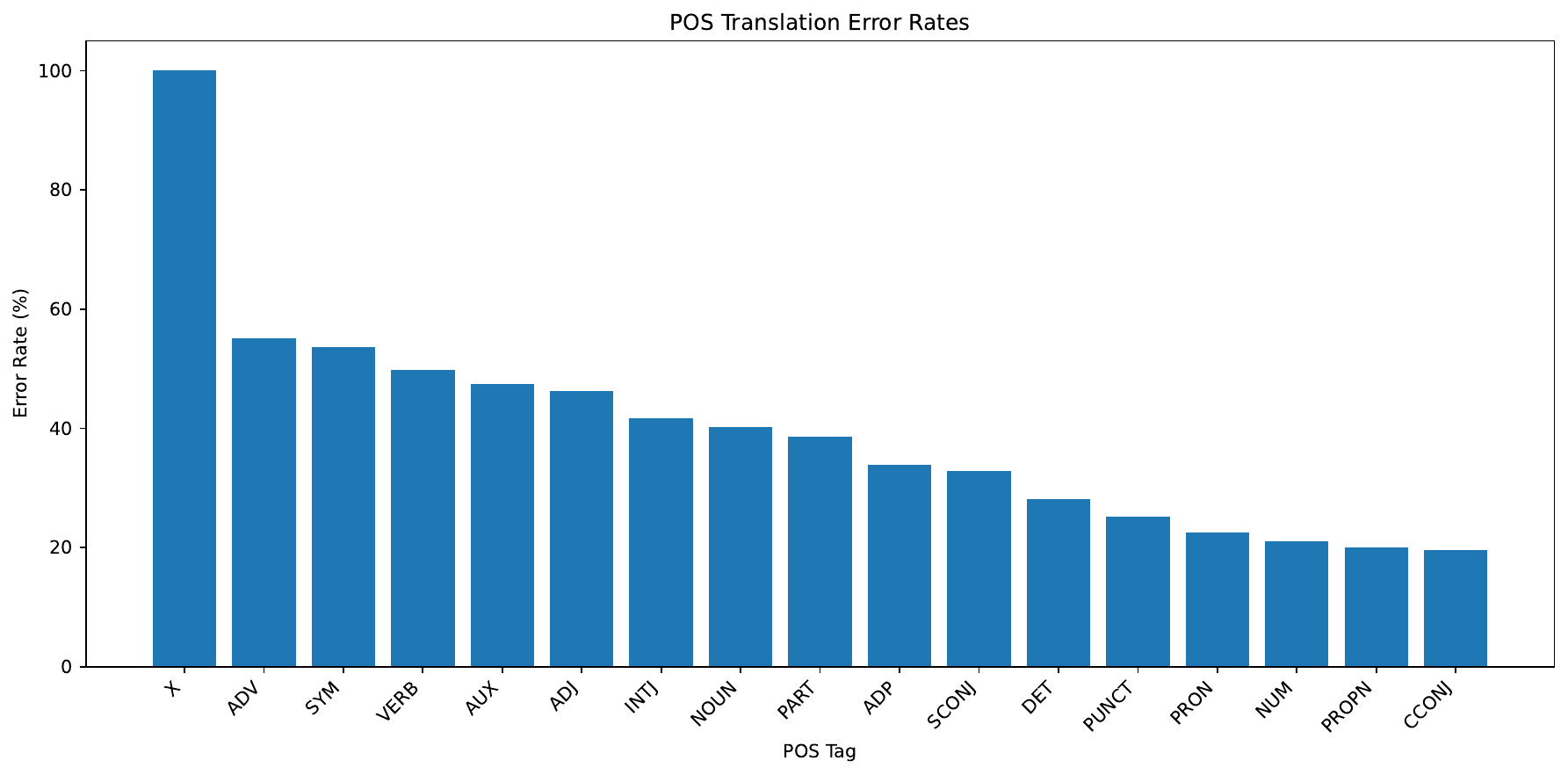}
    \caption{Hausa$\rightarrow$English (AudioLLM)}
    \label{fig:eng_haus_pos_tag}
  \end{subfigure}\hfill
  \begin{subfigure}[b]{0.5\linewidth}
    \centering
    \includegraphics[width=\linewidth]{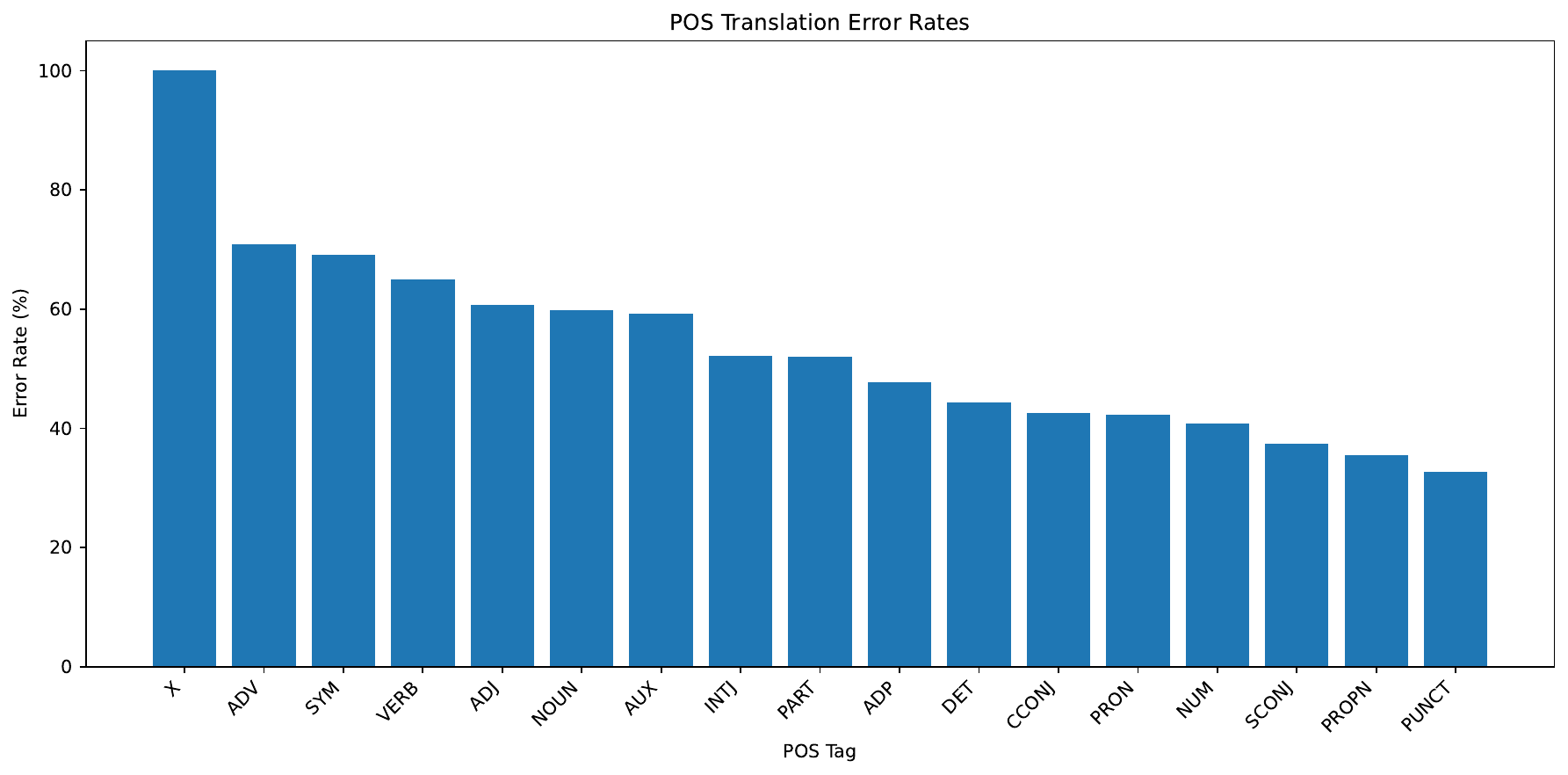}
    \caption{Igbo$\rightarrow$English (AudioLLM)}
    \label{fig:eng_ibo_pos_tag}
  \end{subfigure}

  \vspace{2em}

  \begin{subfigure}[b]{0.5\linewidth}
    \centering
    \includegraphics[width=\linewidth]{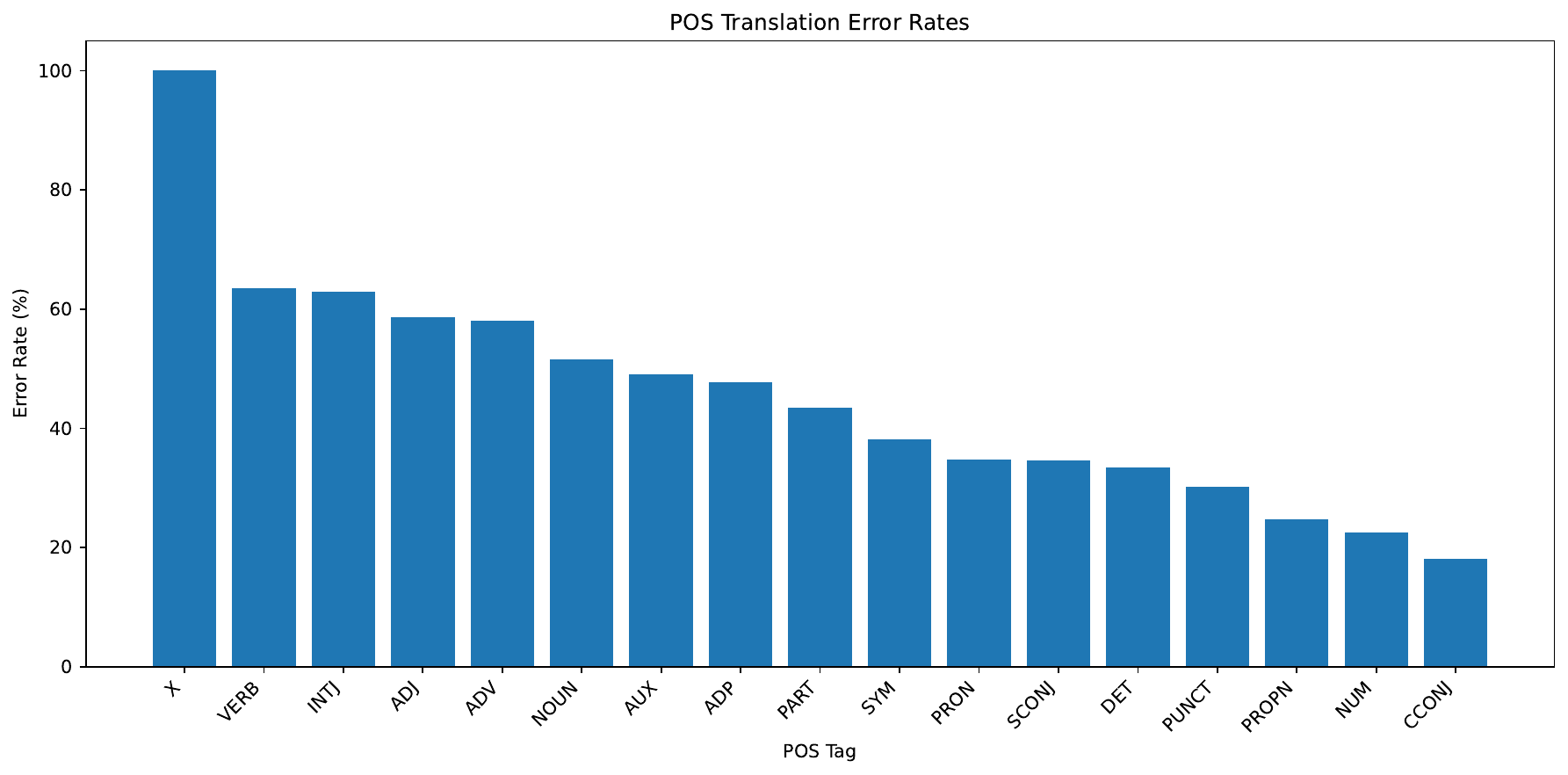}
    \caption{Naijá$\rightarrow$English (AudioLLM)}
    \label{fig:eng_pcm_pos_tag}
  \end{subfigure}\hfill
  \begin{subfigure}[b]{0.5\linewidth}
    \centering
    \includegraphics[width=\linewidth]{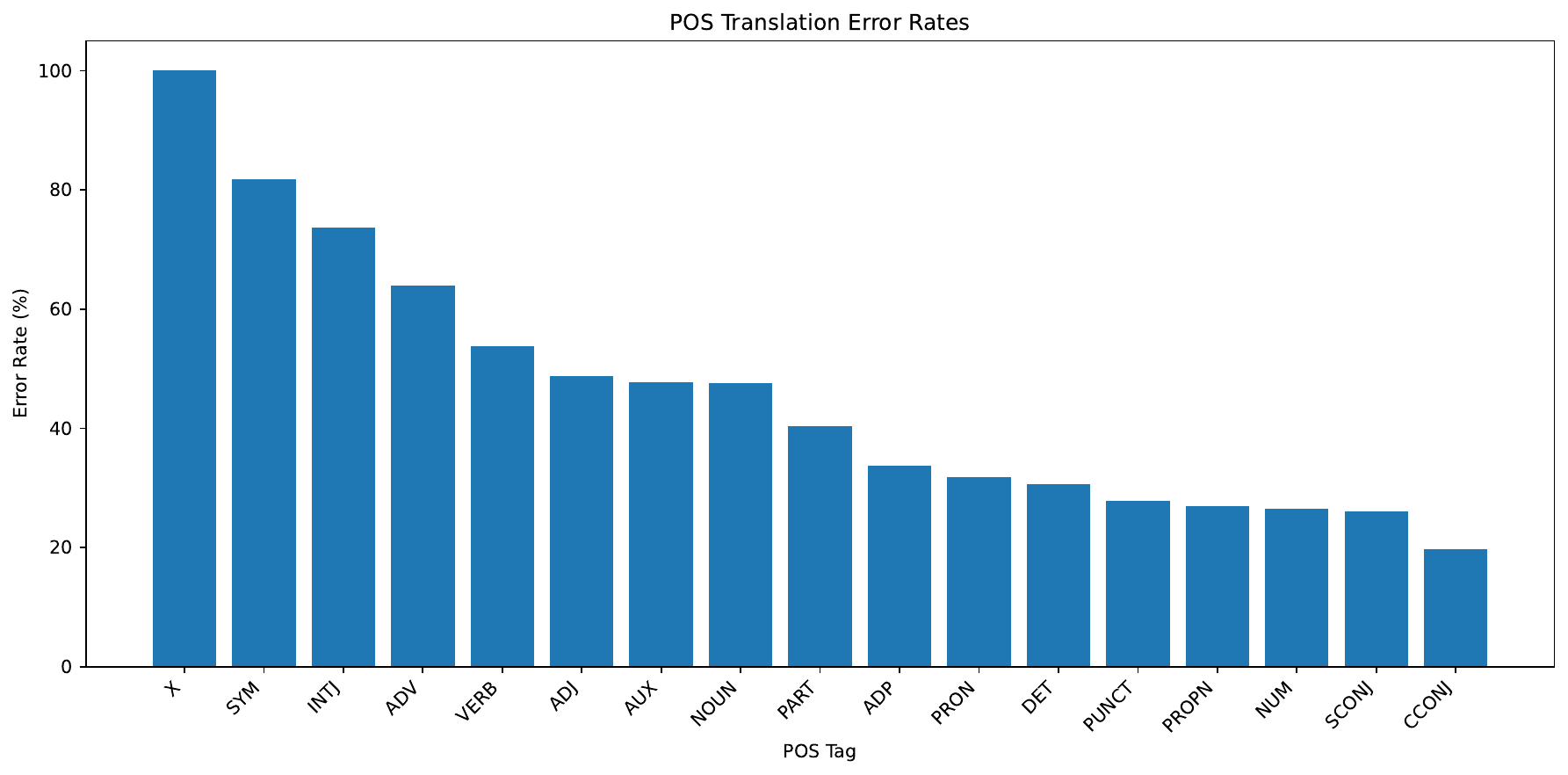}
    \caption{\yoruba$\rightarrow$English (AudioLLM)}
    \label{fig:eng_yor_pos_tag}
  \end{subfigure}

  \vspace{2em}

  \begin{subfigure}[b]{0.5\linewidth}
    \centering
    \includegraphics[width=\linewidth]{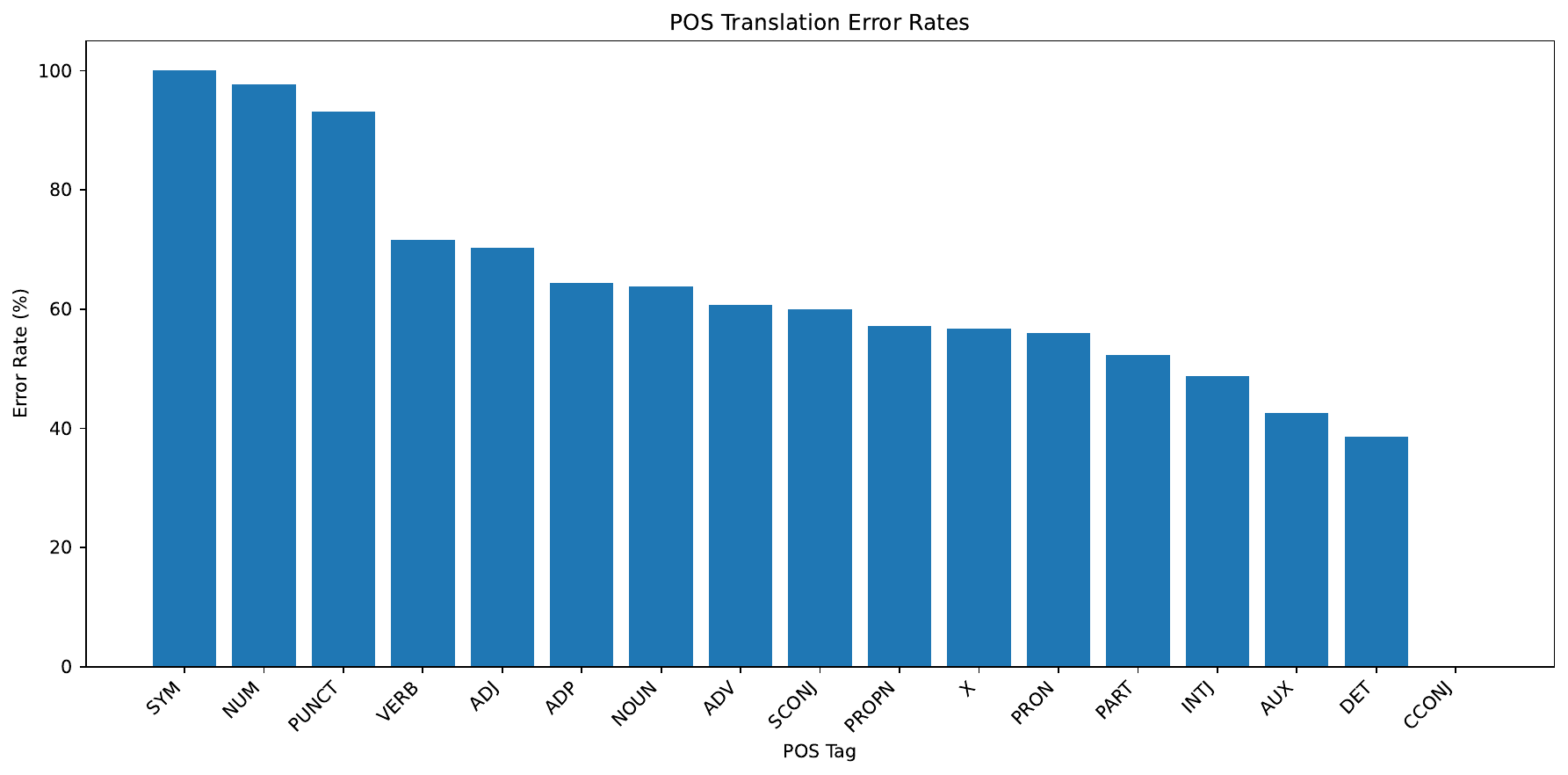}
    \caption{English$\rightarrow$Hausa (Cascaded)}
    \label{fig:eng_hau_cas_pos_tag}
  \end{subfigure}\hfill
  \begin{subfigure}[b]{0.5\linewidth}
    \centering
    \includegraphics[width=\linewidth]{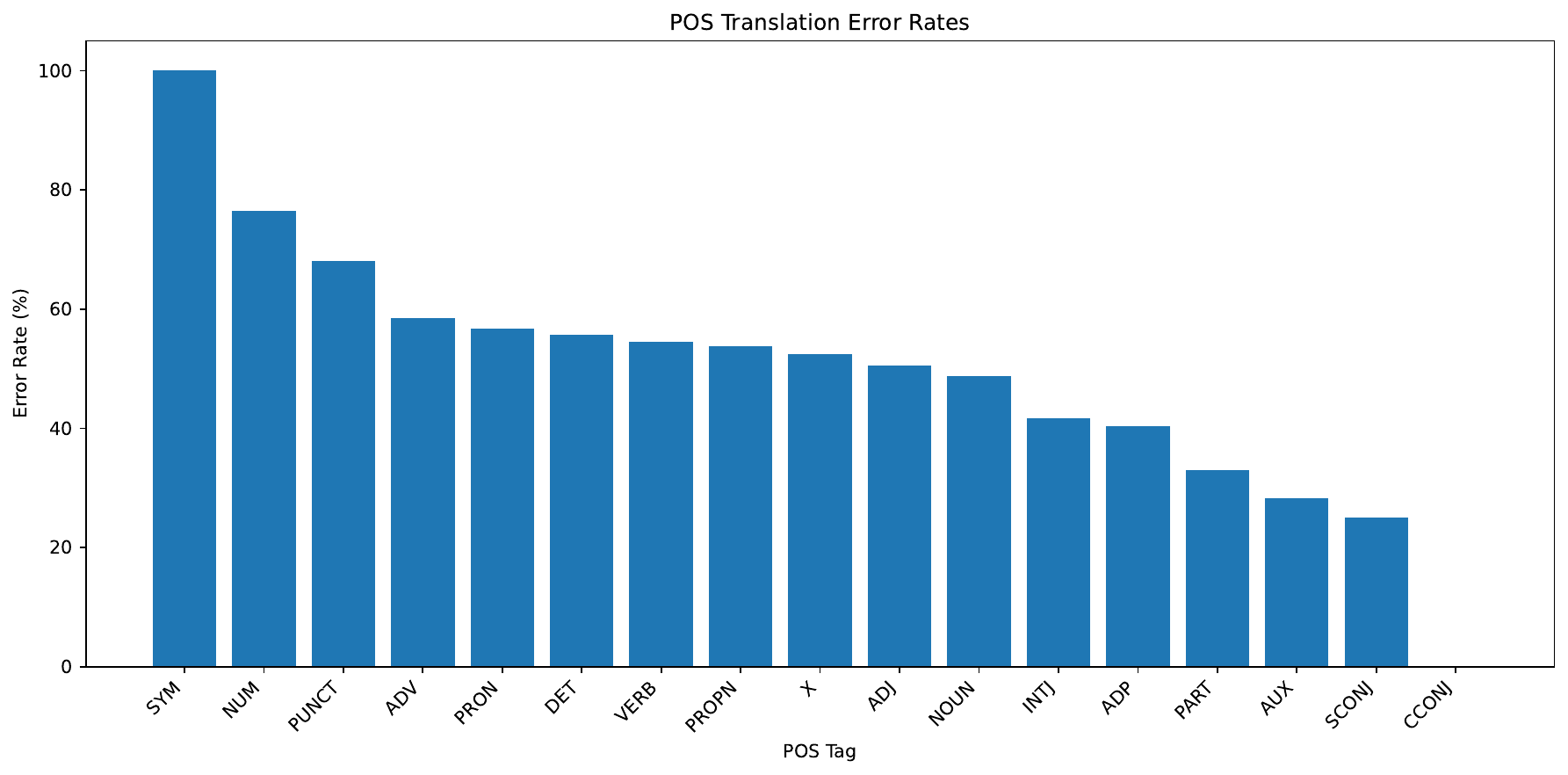}
    \caption{English$\rightarrow$Igbo (Cascaded)}
    \label{fig:eng_ibo_cas_pos_tag}
  \end{subfigure}

  \vspace{2em}

  \begin{subfigure}[b]{0.5\linewidth}
    \centering
    \includegraphics[width=\linewidth]{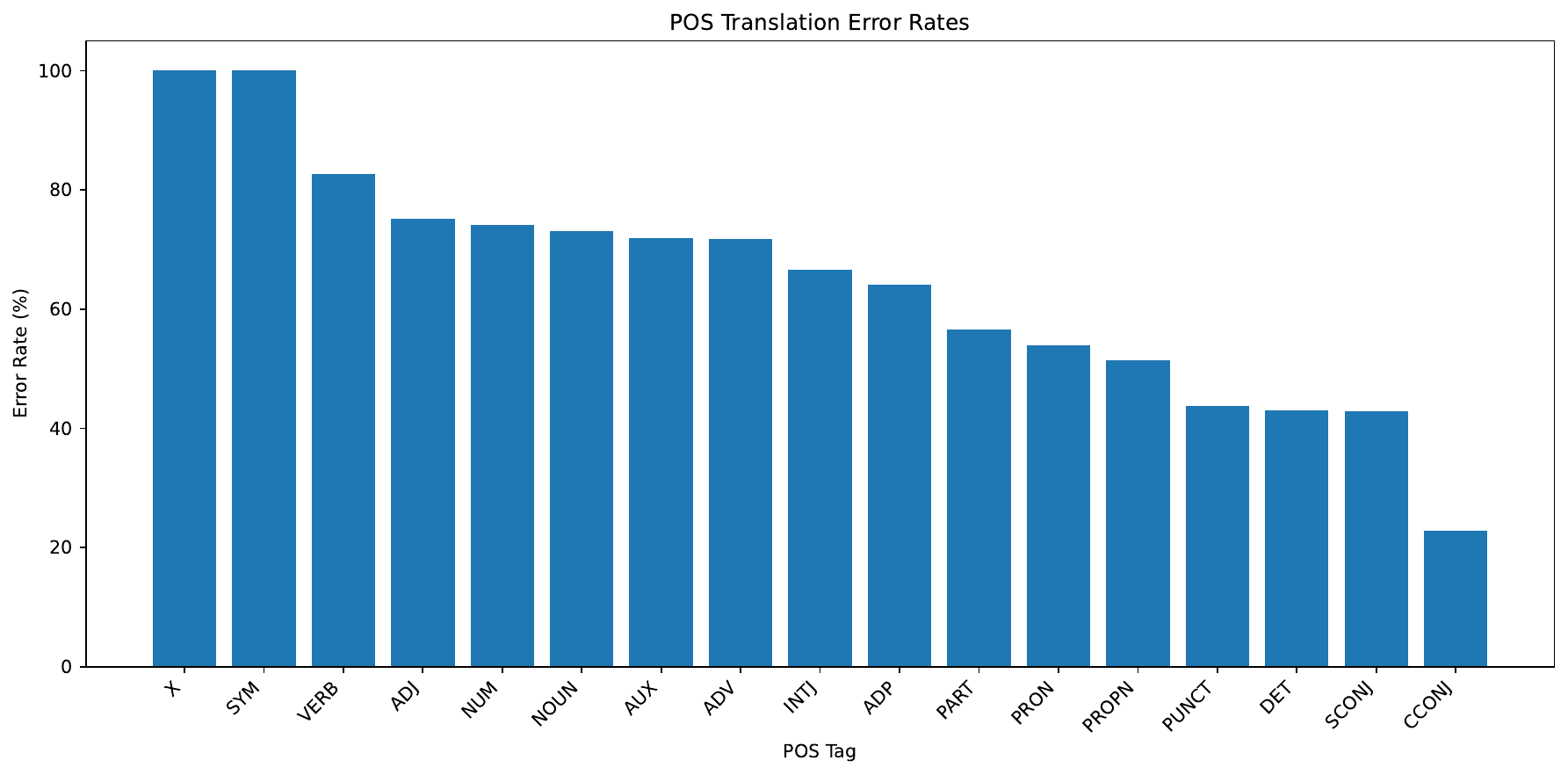}
    \caption{English$\rightarrow$Naijá (Cascaded)}
    \label{fig:eng_pcm_cas_pos_tag}
  \end{subfigure}\hfill
  \begin{subfigure}[b]{0.5\linewidth}
    \centering
    \includegraphics[width=\linewidth]{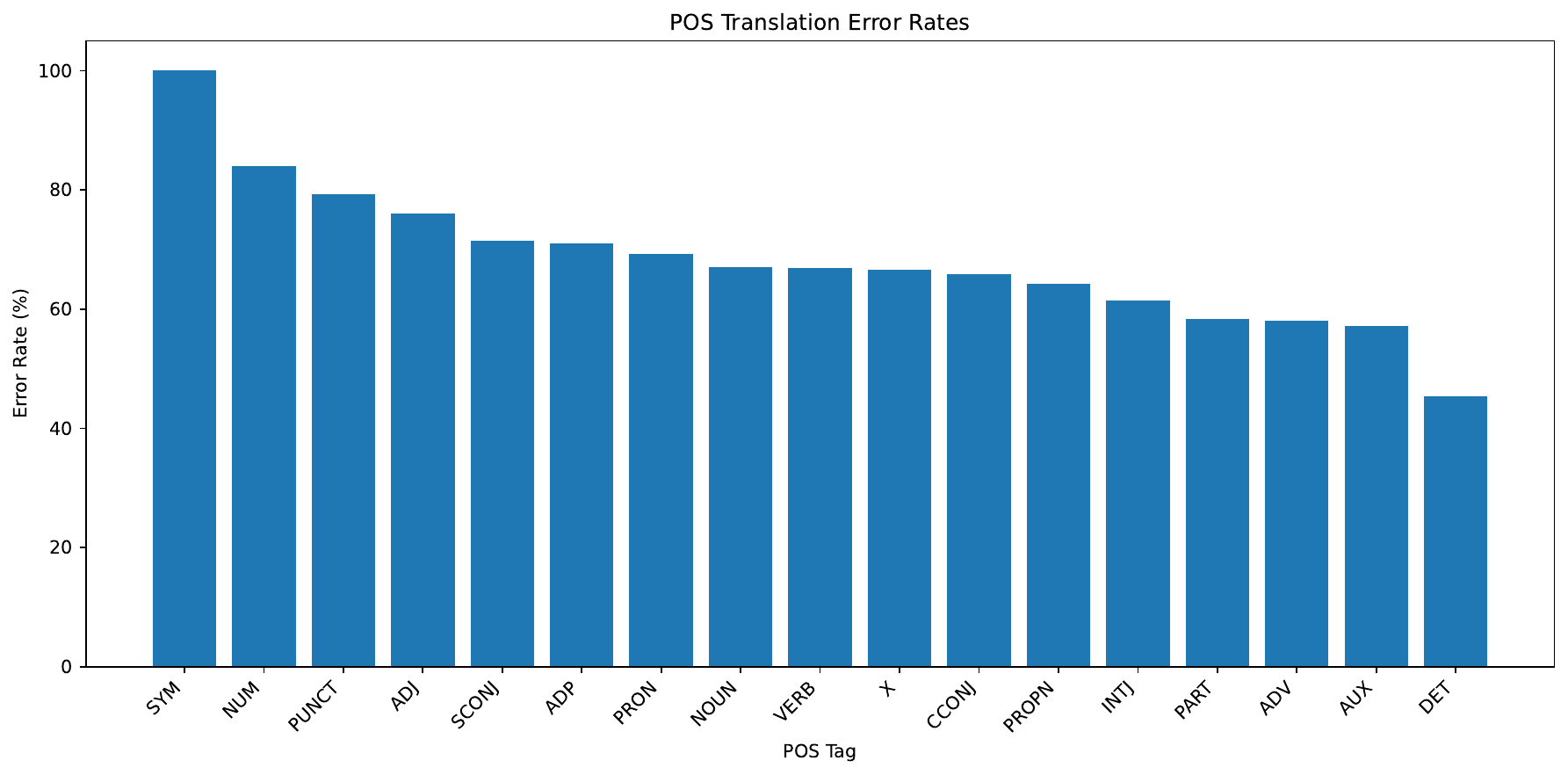}
    \caption{English$\rightarrow$\yoruba (Cascaded)}
    \label{fig:eng_yor_cas_pos_tag}
  \end{subfigure}

  \caption{POS translation error rates across language pairs for AudioLLM (top four) and Cascaded (bottom four) models.}
  \label{fig:all_pos_error_rates}
\end{figure*}

\subsection{POS Tagging Error Rate}

Since references and predictions may differ in word choice, order, or count, we align their tag sequences using Sequence Matcher-based token alignment rather than assuming 1:1 correspondence, and compute a per-tag error rate:

\begin{equation}
    \text{Error Rate}_p = \frac{S_p+D_p}{M_p+S_p+D_p} \times 100
\end{equation}

, where $M_p$, $S_p$ and $D_p$ are match, substitution, and deletion counts for tag $p$. This method highlights tag-level patterns (e.g. high PROPN error rates indicating dropped or mistranslated names) attributable to translation behavior rather than tagging noise.

\subsection{\yoruba POS Tagging Analysis}

Because \yoruba lacks the extensive morphological inflection found in agglutinative languages, syntactic roles (POS) are heavily dependent on exact lexical identification, which in turn is strictly dictated by tones. For example, a minor ASR transcription error in tone can change the word ti (a preposition/ADP meaning 'from' or 'of') into tì (a verb meaning 'to push' or 'to lock'), or bí (a conjunction/SCONJ meaning 'if') into bi (a verb meaning 'to ask').

Therefore, when speech models fail to capture these phonological nuances (a common vulnerability in S2T systems), it does not merely result in orthographic typos; it leads to severe lexical substitutions that fundamentally alter the grammatical structure (POS) of the generated sentence.

\section{Sentence Analysis Ablations}\label{sec:sentence_analysis_ablations}

To address the over-generation in AudioLLMs, we ran a quantitative breakdown across all eight translation directions, measuring word-count differences between predictions and references. While over-generation is consistent, its impact is generally low in that the average excess length ranges from 2.46 to 4.19 words per sentence (Table \ref{tab:audiollm_overgeneration}), with minimum differences as small as one word. Larger deviations (15-52 words), driven by meta-commentary, ``reasoning-like" continuations or more verbose outputs, such as the Igbo$\rightarrow$English example below, are outliers, occurring in only a small fraction of cases relative to near-reference-length outputs.

\begin{table}[ht]
    \centering
    \small
    \setlength{\tabcolsep}{4.5pt}
    \resizebox{\columnwidth}{!}{%
    \label{tab:audiollm_overgeneration}
    \begin{tabular}{lrrrr}
        \toprule
        \textbf{Direction} &
        \textbf{Samples} &
        \textbf{Min.} &
        \textbf{Max.} &
        \textbf{Avg.} \\
        \midrule
        Eng $\rightarrow$ Hausa   & 167/500   & 1 & 15 & 2.53 \\
        Eng $\rightarrow$ Igbo    & 288/500   & 1 & 15 & 3.39 \\
        Eng $\rightarrow$ Naijá  & 124/500   & 1 & 30 & 3.41 \\
        Eng $\rightarrow$ \yoruba  & 333/500   & 1 & 16 & 4.19 \\
        Hausa $\rightarrow$ Eng    & 765/1500  & 1 & 20 & 3.19 \\
        Igbo $\rightarrow$ Eng     & 711/1500  & 1 & 52 & 3.95 \\
        Naijá $\rightarrow$ Eng  & 767/1500  & 1 & 20 & 3.49 \\
        \yoruba $\rightarrow$ Eng   & 554/1500  & 1 & 9  & 2.46 \\
        \bottomrule
    \end{tabular}
    }
    \caption{AudioLLM (Gemini 3.1) over-generation statistics. 
    Difference is defined as the prediction word count minus the reference word count. \textbf{Samples} refers to the total number of ``over-generated" samples, Min. to the minimum difference, Max. to the maximum difference and Avg. to the average word count difference. }\label{tab:audiollm_overgeneration}
\end{table}

\textbf{An Example from the Test Set}

\textit{Prediction} (31 words): It is the home of golf. The Old Course of the Royal and Ancient Golf Club is located by the ocean. Some of the most famous images in the sporting world.

\textit{Reference} (ground truth) (7 words): St. Andrews is famous for many things.

This suggests the length variance is driven by a long tail of infrequent, severe over-generations rather than systematic verbosity, and that the strong automatic/human scores are broadly consistent with generation behaviour. 

\clearpage
\section{Human Evaluation Ablations}\label{sec:human_eval_ablations}

\subsection{Pronunciation Quality}
As with translation quality in Section \ref{sec:translation_quality_human}, a similar trend is observed for pronunciation quality (Table \ref{sts_pronunciation_human_eval}). AudioLLMs again achieve the best scores overall, followed by Cascaded systems, while End-to-End models consistently perform worst. These results suggest that AudioLLM approaches produce both more accurate translations and more natural speech outputs across low-resource language settings.

\begin{table}[h]
\centering
\small
\setlength{\tabcolsep}{4.5pt}
\resizebox{\columnwidth}{!}{%
\begin{tabular}{
    l|rrcr|rrcr
    }
\toprule
    & \multicolumn{4}{c|}{\textbf{XX $\rightarrow$ Eng}} 
    & \multicolumn{4}{c}{\textbf{Eng $\rightarrow$ XX}} \\
    
\cmidrule(lr){2-5} \cmidrule(lr){6-9}
        
\textbf{ Method}
& \textbf{Hausa} & \textbf{Igbo} & \textbf{\yoruba} & \textbf{Avg.} 
& \textbf{Hausa} & \textbf{Igbo} & \textbf{\yoruba} & \textbf{Avg.} \\

\midrule
    {\begin{tabular}[c]{@{}c@{}} \textsc{Cascaded} \\ \texttt{(ASR + MT)}\end{tabular}}
        & \underline{97.3} & \underline{59.3} & \underline{88.4} & 81.7 &\underline{77.1} & \underline{62.4} & \underline{55.9} & 65.1 \\ 
\addlinespace
    \textsc{End-to-End}
        & 78.6 & 41.4 & 72.3 & 64.1 & {N/A} & {N/A} & {N/A}  & \\
\addlinespace
    \textsc{AudioLLM}
        & \textbf{98.5} & \textbf{69.8} & \textbf{93.6} & 87.3 &\textbf{83.4}& \textbf{78.6} & \textbf{58.5}&  73.5  \\
\bottomrule
\end{tabular}
}
\caption{S2ST \textbf{pronunciation} human evaluation results (0-100 average Likert scale $\uparrow$).}
\label{sts_pronunciation_human_eval}
\end{table}

\subsection{Source Text Addition}
\label{sec:source_text_addition}
To examine the impact of source-text availability, we further duplicate the 300 evaluation items and augment the duplicated versions with the corresponding source transcription, resulting in 600 items in total. In other words, annotators first evaluated 300 target–source audio pairs and later evaluated the same 300 pairs with additional source text provided.  
We find that for all languages and directions, adding source text had little effect on annotator scores, as shown in Tables \ref{tab:quality_method_text_stats_e_h} 
\textasciitilde{}
\ref{tab:quality_method_text_stats_y_e}.

\begin{table}[h]
\setlength{\tabcolsep}{4.5pt}
\resizebox{\columnwidth}{!}{%
\begin{tabular}{llrrrr}
\toprule
 &  & N & Mean & SD & Median \\
Method & Has Source Text &  &  &  &  \\
\midrule
\multirow[t]{2}{*}{AudioLLM} & No text & 105 & 85.29 & 9.54 & 87.67 \\
 & With text & 105 & 85.19 & 7.40 & 88.00 \\
\cline{1-6}
\multirow[t]{2}{*}{Cascaded} & No text & 105 & 71.12 & 13.96 & 70.67 \\
 & With text & 105 & 74.78 & 11.60 & 73.33 \\
\cline{1-6}
\bottomrule
\end{tabular}}
\caption{Audio Quality by method and text condition (English to Hausa).}
\label{tab:quality_method_text_stats_e_h}
\end{table}

\begin{table}
\setlength{\tabcolsep}{4.5pt}
\resizebox{\columnwidth}{!}{%
\begin{tabular}{llrrrr}
\toprule
 &  & N & Mean & SD & Median \\
Method & Has Source Text &  &  &  &  \\
\midrule
\multirow[t]{2}{*}{AudioLLM} & No text & 105 & 68.94 & 9.65 & 69.67 \\
 & With text & 105 & 73.70 & 8.36 & 75.67 \\
\cline{1-6}
\multirow[t]{2}{*}{Cascaded} & No text & 105 & 67.66 & 8.66 & 68.33 \\
 & With text & 105 & 71.66 & 8.37 & 72.00 \\
\cline{1-6}
\bottomrule
\end{tabular}}
\caption{Audio Quality by method and text condition (English to Igbo).}
\label{tab:quality_method_text_stats_e_i}
\end{table}

\begin{table}
\setlength{\tabcolsep}{4.5pt}
\resizebox{\columnwidth}{!}{%
\begin{tabular}{llrrrr}
\toprule
 &  & N & Mean & SD & Median \\
Method & Has Source Text &  &  &  &  \\
\midrule
\multirow[t]{2}{*}{AudioLLM} & No text & 105 & 76.31 & 10.47 & 77.33 \\
 & With text & 105 & 77.31 & 8.52 & 77.33 \\
\cline{1-6}
\multirow[t]{2}{*}{Cascaded} & No text & 105 & 58.89 & 16.52 & 61.33 \\
 & With text & 105 & 54.56 & 15.23 & 52.67 \\
\cline{1-6}
\bottomrule
\end{tabular}}
\caption{Audio Quality by method and text condition (English to \yoruba).}
\label{tab:quality_method_text_stats_e_y}
\end{table}

\begin{table}
\setlength{\tabcolsep}{4.5pt}
\resizebox{\columnwidth}{!}{%
\begin{tabular}{llrrrr}
\toprule
 &  & N & Mean & SD & Median \\
Method & Has Source Text &  &  &  &  \\
\midrule
\multirow[t]{2}{*}{AudioLLM} & No text & 105 & 99.54 & 2.85 & 100.00 \\
 & With text & 105 & 98.79 & 5.48 & 100.00 \\
\cline{1-6}
\multirow[t]{2}{*}{Cascaded} & No text & 105 & 78.70 & 27.05 & 91.67 \\
 & With text & 105 & 71.24 & 35.19 & 93.33 \\
\cline{1-6}
\multirow[t]{2}{*}{End-to-End} & No text & 105 & 6.50 & 16.08 & 0.00 \\
 & With text & 105 & 1.75 & 7.81 & 0.00 \\
\cline{1-6}
\bottomrule
\end{tabular}}
\caption{Audio Quality by method and text condition (Hausa to English).}
\label{tab:quality_method_text_stats_h_e}
\end{table}


\begin{table}
\setlength{\tabcolsep}{4.5pt}
\resizebox{\columnwidth}{!}{%
\begin{tabular}{llrrrr}
\toprule
 &  & N & Mean & SD & Median \\
Method & Has Source Text &  &  &  &  \\
\midrule
\multirow[t]{2}{*}{AudioLLM} & No text & 105 & 67.39 & 14.48 & 70.33 \\
 & With text & 105 & 76.55 & 13.24 & 81.67 \\
\cline{1-6}
\multirow[t]{2}{*}{Cascaded} & No text & 105 & 63.47 & 16.07 & 66.33 \\
 & With text & 105 & 64.37 & 18.11 & 69.67 \\
\cline{1-6}
\multirow[t]{2}{*}{End-to-End} & No text & 105 & 48.18 & 17.37 & 46.33 \\
 & With text & 105 & 48.87 & 17.01 & 45.67 \\
\cline{1-6}
\bottomrule
\end{tabular}}
\caption{Audio Quality by method and text condition (Igbo to English).}
\label{tab:quality_method_text_stats_i_e}
\end{table}

\begin{table}
\setlength{\tabcolsep}{4.5pt}
\resizebox{\columnwidth}{!}{%
\begin{tabular}{llrrrr}
\toprule
 &  & N & Mean & SD & Median \\
Method & Has Source Text &  &  &  &  \\
\midrule
\multirow[t]{2}{*}{AudioLLM} & No text & 105 & 87.14 & 11.09 & 90.67 \\
 & With text & 105 & 90.44 & 11.07 & 94.67 \\
\cline{1-6}
\multirow[t]{2}{*}{Cascaded} & No text & 105 & 70.96 & 17.76 & 73.33 \\
 & With text & 105 & 66.14 & 24.37 & 69.33 \\
\cline{1-6}
\multirow[t]{2}{*}{End-to-End} & No text & 105 & 44.53 & 12.58 & 40.00 \\
 & With text & 105 & 36.72 & 12.42 & 31.67 \\
\cline{1-6}
\bottomrule
\end{tabular}}
\caption{Audio Quality by method and text condition (Yorúbà to English).}
\label{tab:quality_method_text_stats_y_e}
\end{table}

\clearpage
\section{Annotator Reliability and Agreement}
\label{human_eval_disagreement}
Recall in Appendix \ref{sec:source_text_addition} that there were 300 target-source audio pairs, and an additional 300 pairs with additional source text provided, resulting in 600 evaluation items. Additionally, in order to assess whether annotators assigned consistent ratings to the same S2ST pair across different intervals and thereby estimate annotation reliability, five randomly selected repeated items are also inserted throughout each survey. Consequently, each survey contains 630 evaluation items.

Specifically, inter-annotator variance measures the degree to which human evaluators differ in their ratings of the same item. In this evaluation, each audio sample was rated independently by multiple annotators, and disagreement is quantified as the standard deviation of scores across annotators for each item. A standard deviation of 0 indicates perfect agreement, while higher values reflect greater divergence in judgment.

A high standard deviation  was observed in the absolute human evaluation scores, which can be attributed to the highly granular nature of the 0 -- 100 Likert scale, as it naturally introduces subjective calibration differences across annotators. Although annotators consistently agreed on the relative ranking of the systems (i.e., AudioLLMs outperform cascaded systems), their individual criteria for which constitute an ``acceptable" absolute score varied significantly. Despite the variance in absolute scoring, the aggregate trends robustly support our comparative conclusions.

\begin{figure*}[h]
    \centering
\includegraphics[width=1\textwidth]{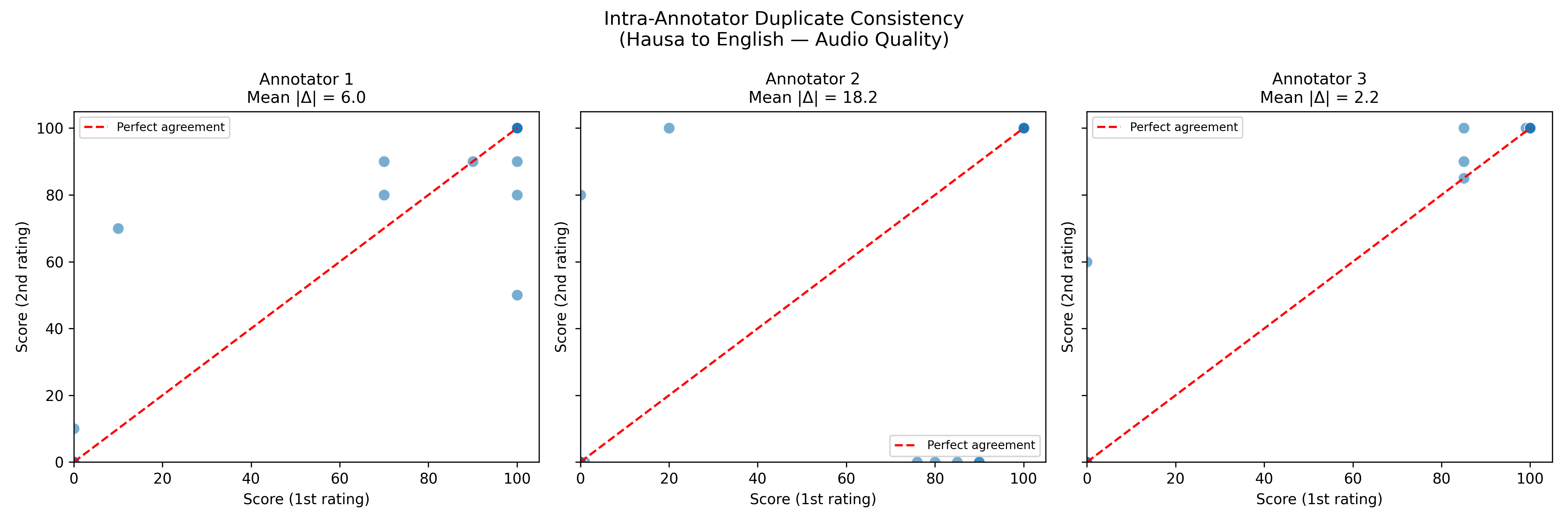}
\caption{Intra-Annotator Consistency for S2ST translation quality for Hausa$\rightarrow$English human evaluation. Points closer to the diagonal indicate higher self-consistency across different rating intervals.}
\label{fig:ibo_eng_reliability2}
\end{figure*}

\begin{figure*}[h]
    \centering
\includegraphics[width=1\textwidth]{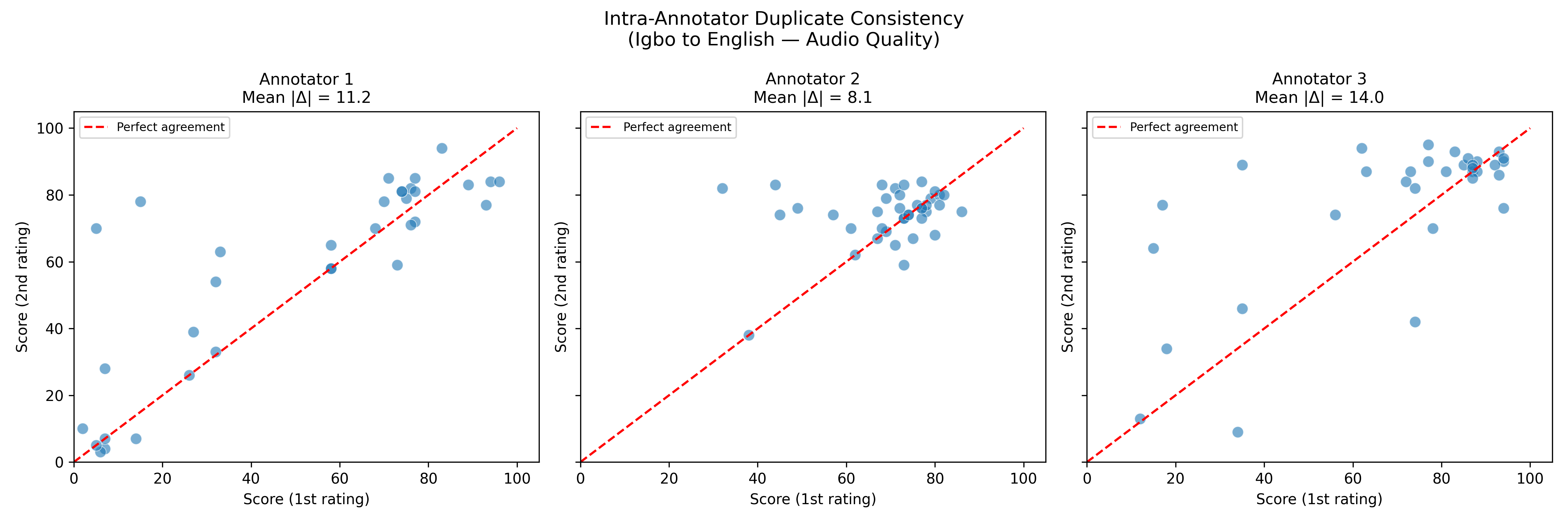}
\caption{Intra-Annotator Consistency for S2ST translation quality for Igbo$\rightarrow$English human evaluation. Points closer to the diagonal indicate higher self-consistency across different rating intervals.}
\label{fig:ibo_eng_reliability1}
\end{figure*}

\begin{figure*}[h]
    \centering
\includegraphics[width=1\textwidth]{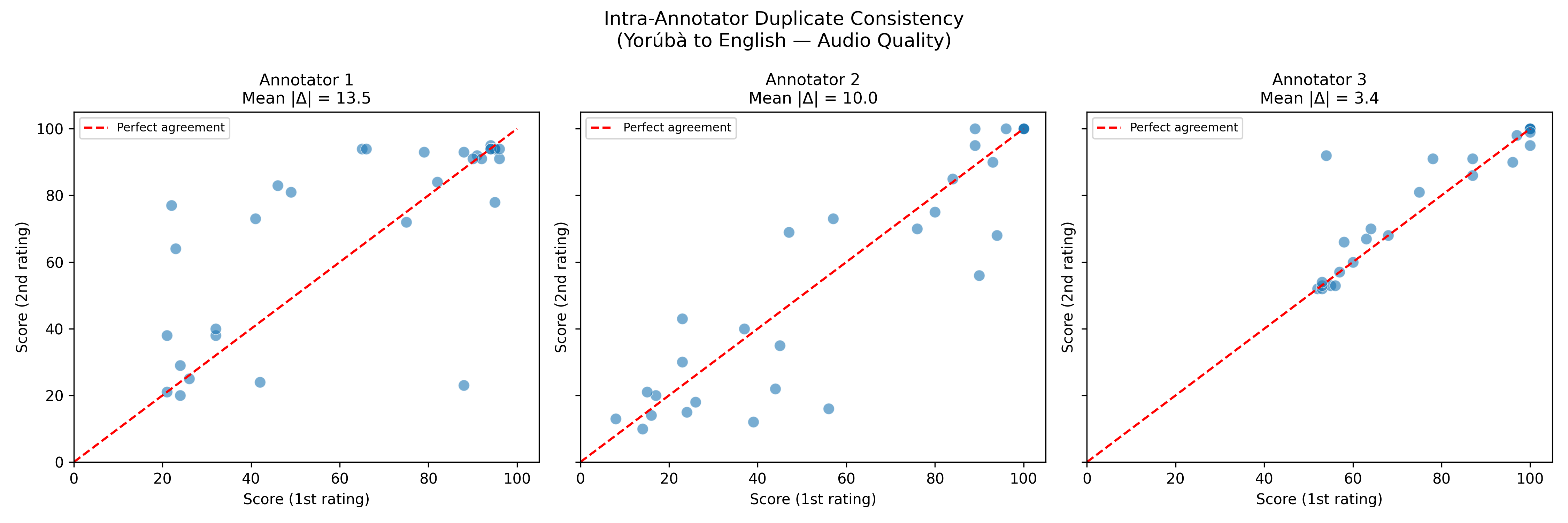}
\caption{Intra-Annotator Consistency for S2ST translation quality for \yoruba$\rightarrow$English human evaluation. Points closer to the diagonal indicate higher self-consistency across different rating intervals.}
\label{fig:yor_eng_reliability}
\end{figure*}

\begin{figure*}[h]
    \centering
\includegraphics[width=1\textwidth]{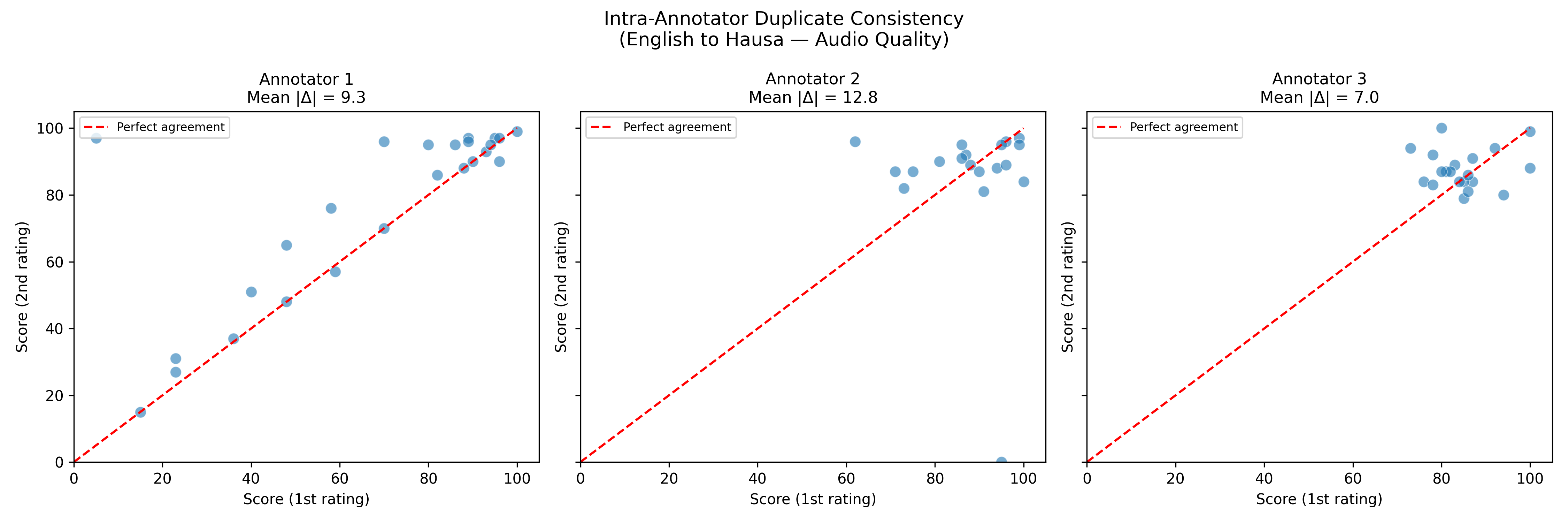}
\caption{Intra-Annotator Consistency for S2ST translation quality for English$\rightarrow$English human evaluation. Points closer to the diagonal indicate higher self-consistency across different rating intervals.}
\label{fig:eng_hau_reliability}
\end{figure*}

\begin{figure*}[h]
    \centering
\includegraphics[width=1\textwidth]{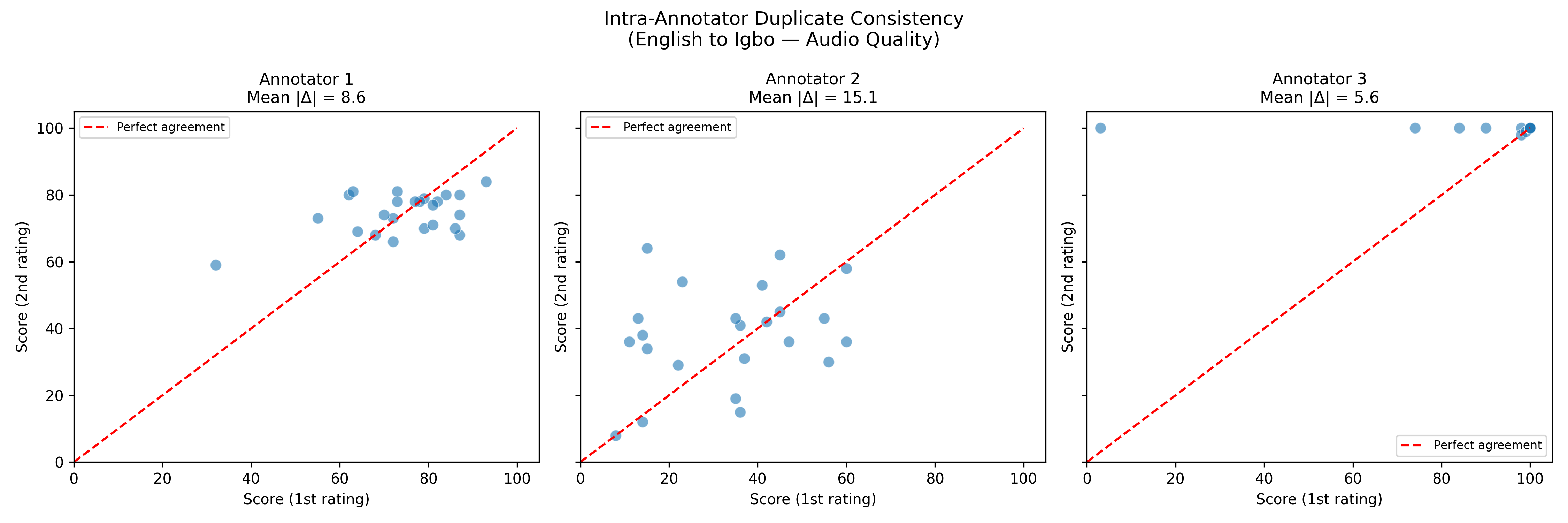}
\caption{Intra-Annotator Consistency for S2ST translation quality for English$\rightarrow$Igbo human evaluation. Points closer to the diagonal indicate higher self-consistency across different rating intervals.}
\label{fig:eng_ibo_reliability}
\end{figure*}

\begin{figure*}[h]
    \centering
\includegraphics[width=1\textwidth]{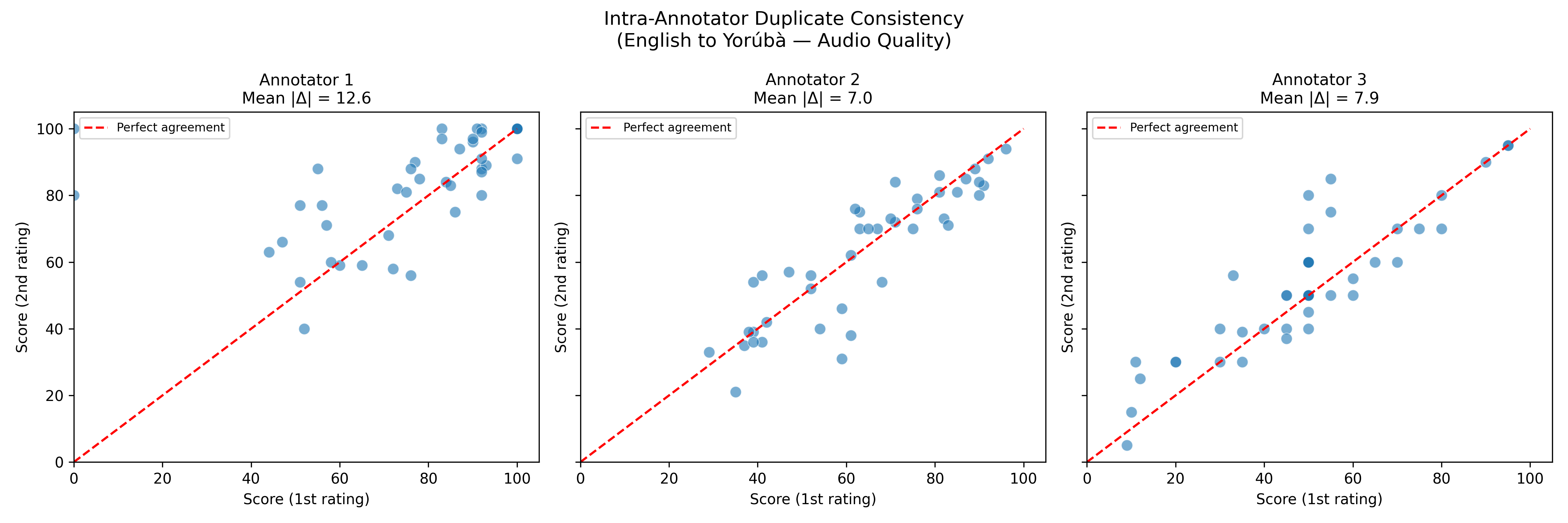}
\caption{Intra-Annotator Consistency for S2ST translation quality for English$\rightarrow$\yoruba human evaluation. Points closer to the diagonal indicate higher self-consistency across different rating intervals.}
\label{fig:eng_yor_reliability}
\end{figure*}

\begin{figure*}[htbp]
  \centering
  \begin{subfigure}[b]{0.5\linewidth}
    \centering
    \includegraphics[width=\linewidth]{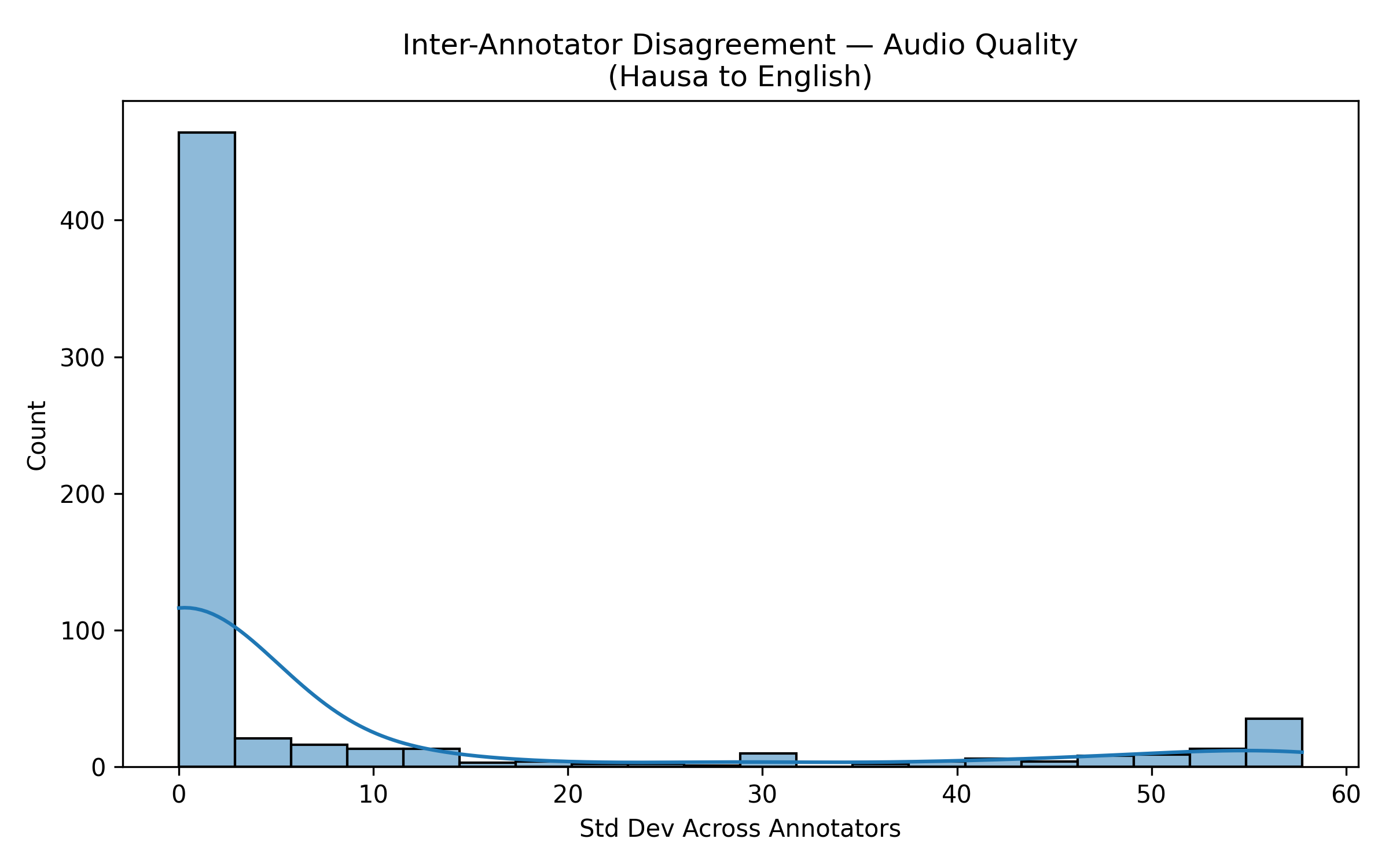}
    \caption{Hausa$\rightarrow$English}
    \label{fig:hau_eng_disagreement}
  \end{subfigure}\hfill
  \begin{subfigure}[b]{0.5\linewidth}
    \centering
    \includegraphics[width=\linewidth]{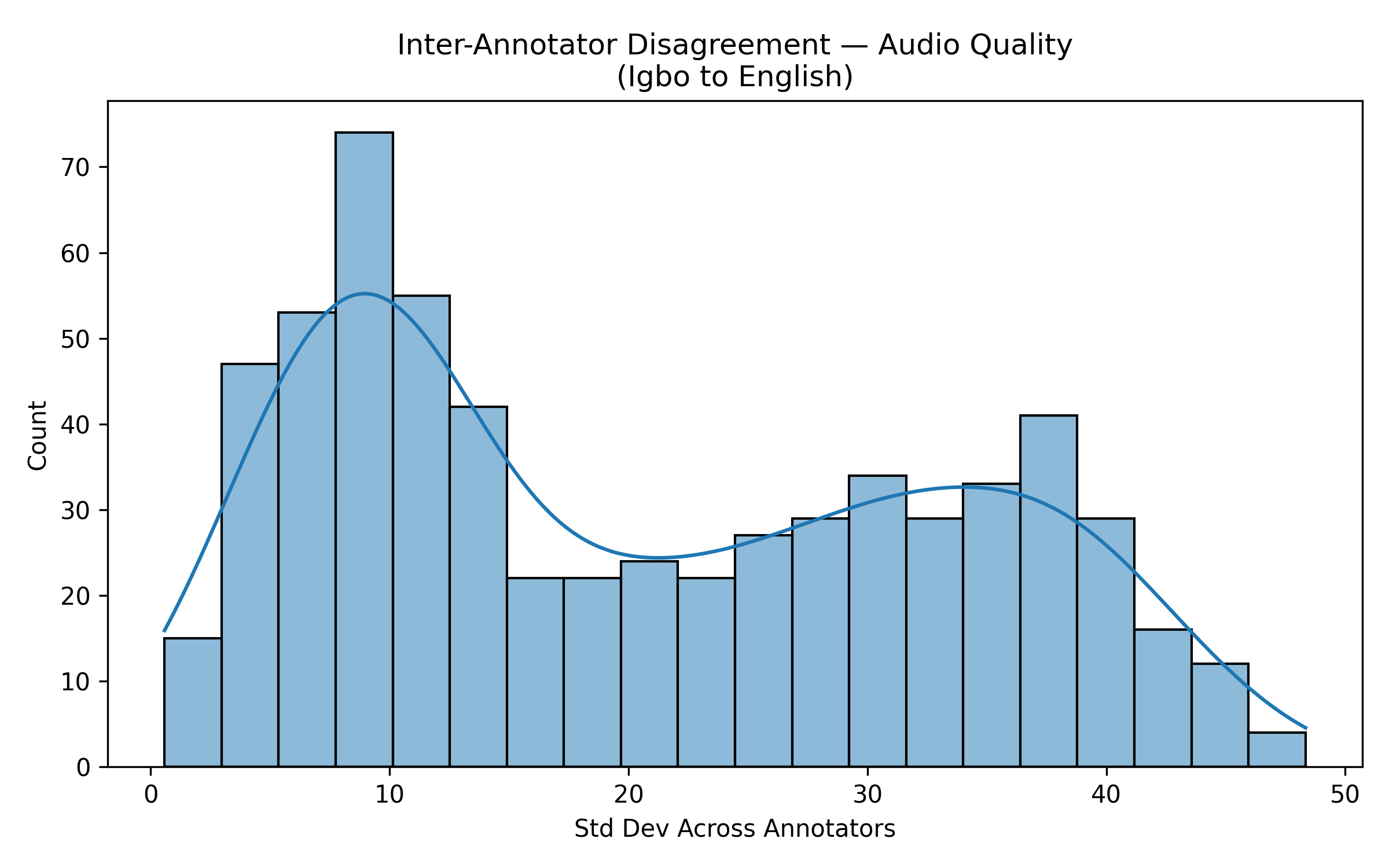}
    \caption{Igbo$\rightarrow$English}
    \label{fig:ibo_eng_disagreement}
  \end{subfigure}

  \vspace{2em}

  \begin{subfigure}[b]{0.5\linewidth}
    \centering
    \includegraphics[width=\linewidth]{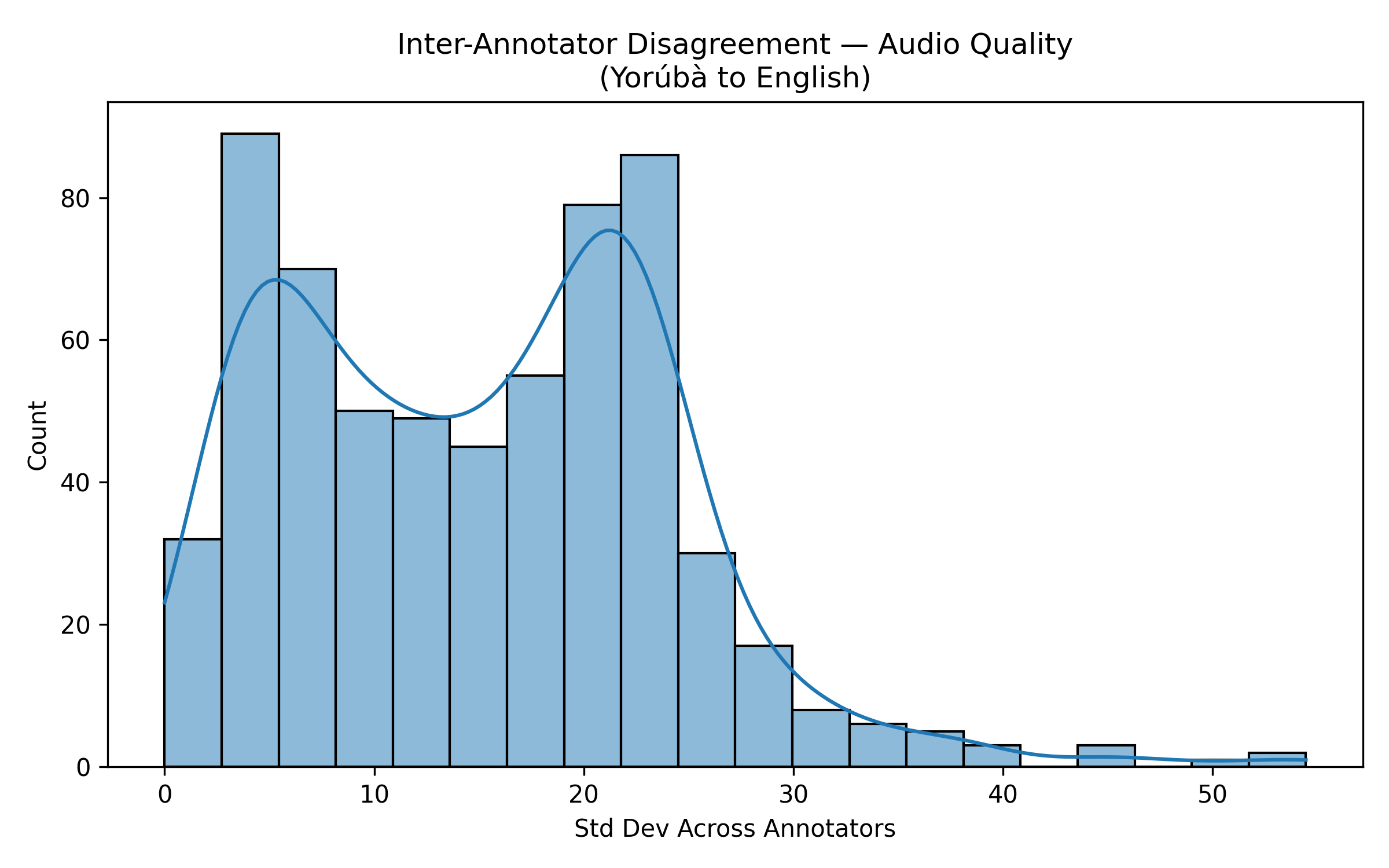}
    \caption{\yoruba$\rightarrow$English}
    \label{fig:yor_eng_disagreement}
  \end{subfigure}\hfill
  \begin{subfigure}[b]{0.5\linewidth}
    \centering
    \includegraphics[width=\linewidth]{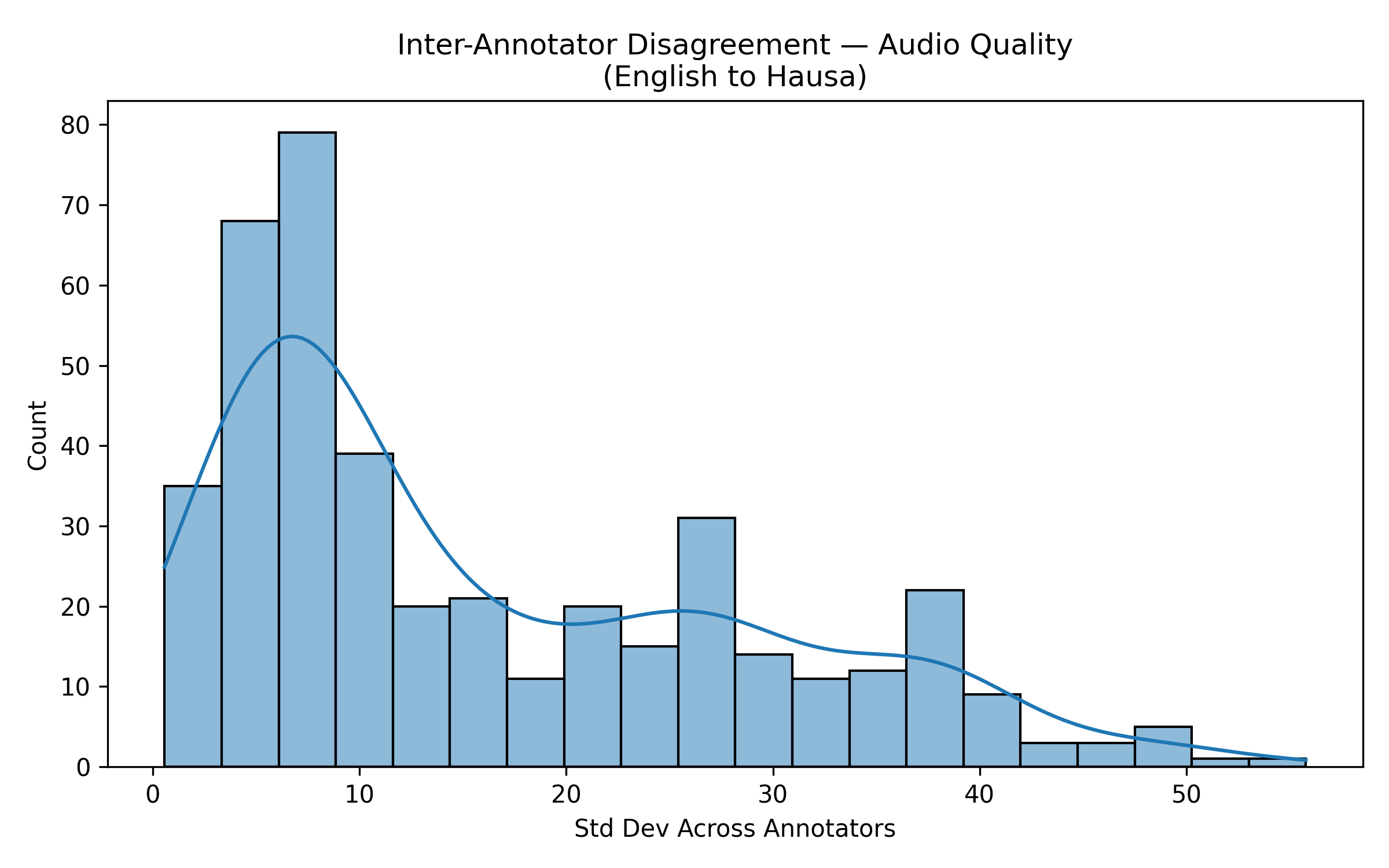}
    \caption{English$\rightarrow$Hausa}
    \label{fig:eng_hau_disagreement}
  \end{subfigure}

  \vspace{2em}

  \begin{subfigure}[b]{0.5\linewidth}
    \centering
    \includegraphics[width=\linewidth]{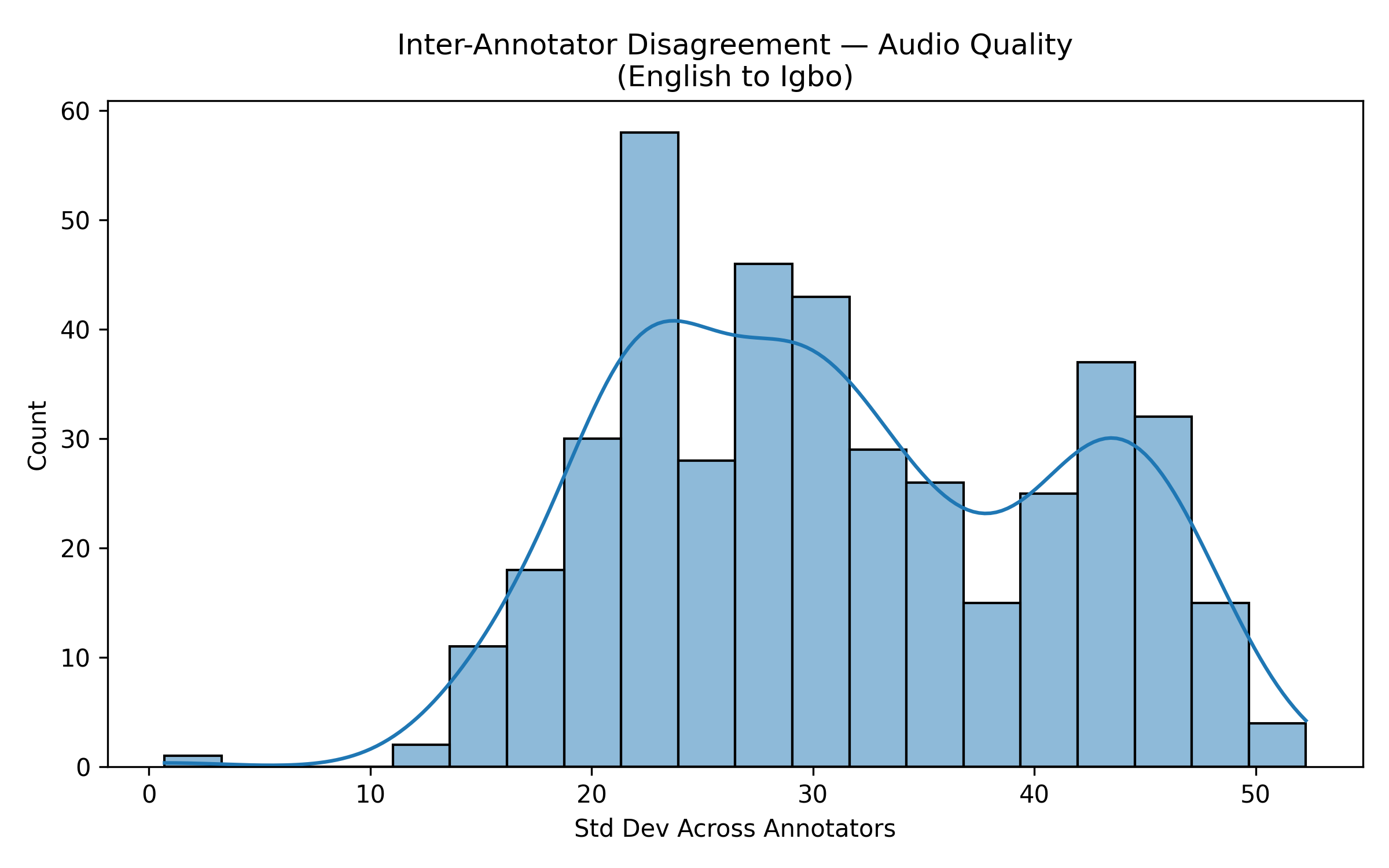}
    \caption{English$\rightarrow$Igbo}
    \label{fig:eng_ibo_disagreement}
  \end{subfigure}\hfill
  \begin{subfigure}[b]{0.5\linewidth}
    \centering
    \includegraphics[width=\linewidth]{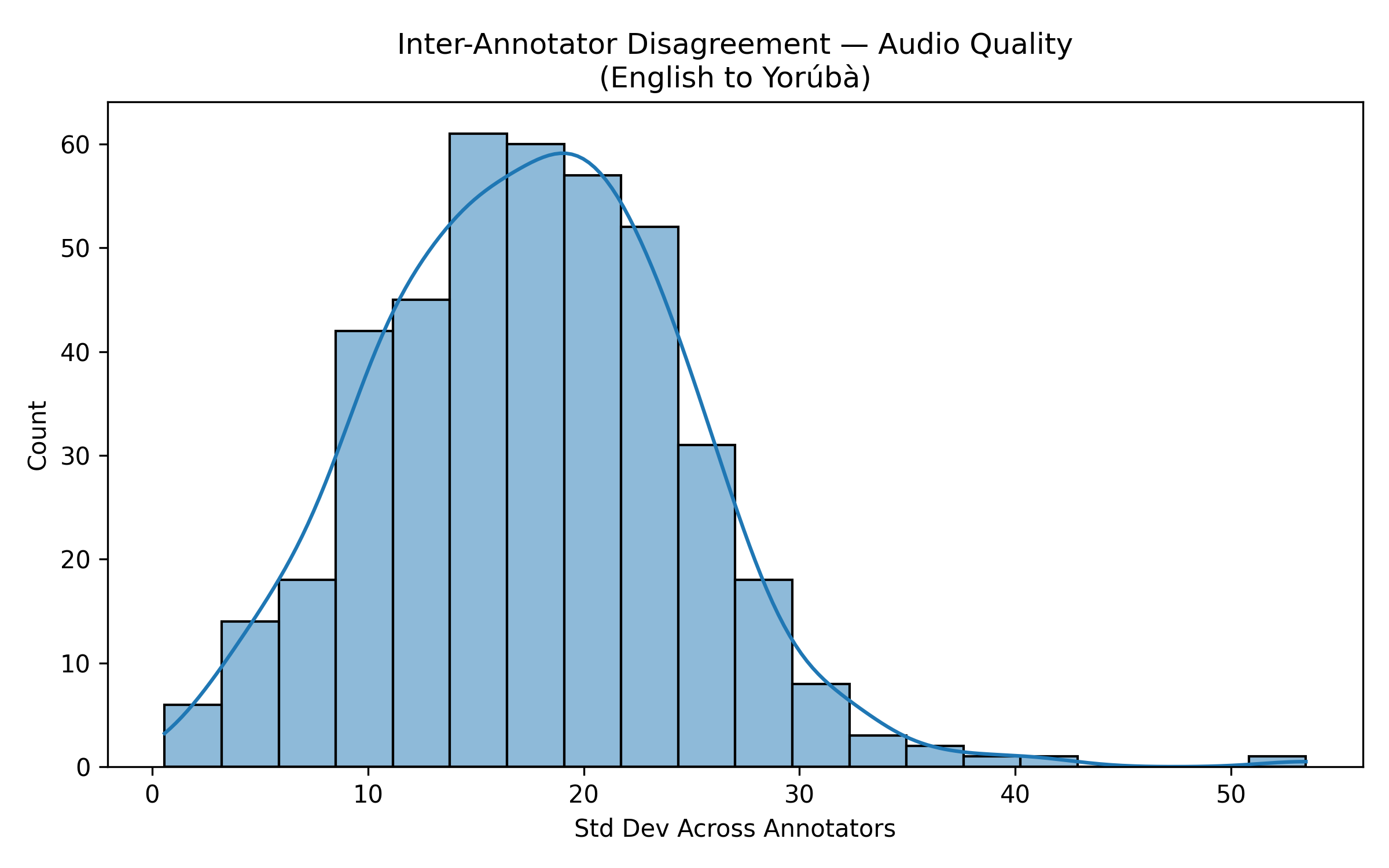}
    \caption{English $\rightarrow$ \yoruba}
    \label{fig:eng_yor_disagreement}
  \end{subfigure}

  \caption{Inter-Annotator Variance for S2ST translation quality human evaluation across all language pairs.}
  \label{fig:all_disagreement}
\end{figure*}

\end{document}